\documentclass{aa}  
\usepackage{graphicx}
\usepackage{txfonts}

\usepackage{natbib,twoopt}
\bibpunct{(}{)}{;}{a}{}{,} 

\usepackage{hyperref}
\usepackage{placeins}
\hypersetup
{
    colorlinks= true,
    linkcolor = blue,
    filecolor = blue,      
    urlcolor  = blue,
    citecolor = blue
}

\newcommand{\Hmol}{\mbox{H$_{\rm 2}$}}

\def\approxlt{\lower.2em\hbox{$\buildrel < \over \sim$}}
\def\approxgt{\lower.2em\hbox{$\buildrel > \over \sim$}}

\def\gtrsim{\mathrel{\hbox{\rlap{\hbox{\lower4pt\hbox{$\sim$}}}\hbox{$>$}}}}

\def\lesssim{\mathrel{\hbox{\rlap{\hbox{\lower4pt\hbox{$\sim$}}}\hbox{$<$}}}}

\def\la{\mathrel{\hbox{\rlap{\hbox{\lower4pt\hbox{$\sim$}}}\hbox{$<$}}}}
\def\ga{\mathrel{\hbox{\rlap{\hbox{\lower4pt\hbox{$\sim$}}}\hbox{$>$}}}}

\begin{document}

\title{Shock excitation of H$_2$ in the \textit{James Webb} Space Telescope era\thanks{Tables B.1 -- B.7 are only available in electronic form
at the CDS via anonymous ftp to cdsarc.cds.unistra.fr (\url{130.79.128.5})
or via \url{https://cdsarc.cds.unistra.fr/cgi-bin/qcat?J/A+A/675/A86}}}

\author{L.E. Kristensen\inst{1}
\and B. Godard\inst{2,3} 
\and P. Guillard\inst{4,5} 
\and A. Gusdorf\inst{3,2} 
\and G. Pineau des For\^ets\inst{6,2}
}

\institute{Niels Bohr Institute, University of Copenhagen, {\O}ster Voldgade 5--7, 1350 Copenhagen K, Denmark \\ 
\email{lars.kristensen@nbi.ku.dk} 
\and
Observatoire de Paris, Universit{\'e} PSL, Sorbonne Universit{\'e}, LERMA, 75014 Paris, France \\
\email{benjamin.godard@obspm.fr}
\and
Laboratoire de Physique de l’{\'E}cole Normale Sup{\'e}rieure, ENS, Universit{\'e} PSL, CNRS, Sorbonne Universit{\'e}, Universit{\'e} Paris Cit{\'e}, 75005 Paris, France 
\and
Sorbonne Universit{\'e}, CNRS, UMR 7095, Institut d’Astrophysique de Paris, 98bis bd Arago, F-75014 Paris, France 
\and
Institut Universitaire de France, Minist{\`e}re de l’Enseignement Sup{\'e}rieur et de la Recherche, 1 rue Descartes, 75231 Paris Cedex F-05, France 
\and
Universit{\'e} Paris-Saclay, CNRS, Institut d’Astrophysique Spatiale, 91405 Orsay, France
}

\date{Received 27 February 2023; accepted 23 May 2023}

\abstract
{Molecular hydrogen, H$_2$, is the most abundant molecule in the Universe. Thanks to its widely spaced energy levels, it predominantly lights up in warm gas, $T \gtrsim 10^2$ K, such as shocked regions externally irradiated or not by interstellar UV photons, and it is one of the prime targets of \textit{James Webb} Space Telescope (JWST) observations. These may include shocks from protostellar outflows, supernova remnants impinging on molecular clouds, all the way up to starburst galaxies and active galactic nuclei. }
{Sophisticated shock models are able to simulate H$_2$ emission from such shocked regions. We aim to explore H$_2$ excitation using shock models, and to test over which parameter space distinct signatures are produced in H$_2$ emission. }
{We here present simulated H$_2$ emission using the Paris-Durham shock code over an extensive grid of $\sim$ 14,000 plane-parallel stationary shock models, a large subset of which are exposed to a semi-isotropic external UV radiation field. The grid samples six input parameters: the preshock density, shock velocity, transverse magnetic field strength, UV radiation field strength, the cosmic-ray-ionization rate, and the abundance of polycyclic aromatic hydrocarbons, PAHs. Physical quantities resulting from our self-consistent calculations, such as temperature, density, and width, have been extracted along with H$_2$ integrated line intensities. These simulations and results are publicly available on the Interstellar Medium Services platform.}
{The strength of the transverse magnetic field, as quantified by the magnetic scaling factor, $b$, plays a key role in the excitation of H$_2$. At low values of $b$ ($\lesssim$ 0.3, J-type shocks), H$_2$ excitation is dominated by vibrationally excited lines; whereas, at higher values ($b$ $\gtrsim$ 1, C-type shocks), rotational lines dominate the spectrum for shocks with an external radiation field comparable to (or lower than) the solar neighborhood. Shocks with $b$ $\ge$ 1 can potentially be spatially resolved with JWST for nearby objects. H$_2$ is typically the dominant coolant at lower densities ($\lesssim$ 10$^4$ cm$^{-3}$); at higher densities, other molecules such as CO, OH, and H$_2$O take over at velocities $\lesssim$ 20 km s$^{-1}$ and atoms, for example, H, O, and S, dominate at higher velocities. Together, the velocity and density set the input kinetic energy flux. When this increases, the excitation and integrated intensity of H$_2$ increases similarly. An external UV field mainly serves to increase the excitation, particularly for shocks where the input radiation energy is comparable to the input kinetic energy flux. These results provide an overview of the energetic reprocessing of input kinetic energy flux and the resulting H$_2$ line emission.}
{}

\keywords{Shock waves --- Methods: numerical --- ISM: general --- Galaxies: ISM}

\maketitle
%

\section{Introduction}\label{sect:intro}

Shocks are inherently out-of-equilibrium time-dependent phenomena that permeate space. They appear over a wide range of scales, ranging from, for example, accretion onto stars or protoplanetary disks, winds and jets driven by accreting (proto)stars, planetary nebulae, supernova remnants, starburst galaxies, jets from active galactic nuclei (AGN), and to galaxy-galaxy collisions \citep[physical sizes ranging from subastronomical unit to kiloparsec scales; e.g.,][]{bally16, Wright1993, Mouri1994, Goldader1997, Appleton2006}. Common to all these phenomena is that the input kinetic energy flux dissipated by the shock accelerates, heats, and compresses the medium. When the medium cools down, radiation is emitted, which we observe. To understand the physical origin of emission (e.g., preshock density, shock velocity) and the energetic processing taking place in shocks, it is thus necessary to reverse engineer the observed light. Doing so requires models. 

One of the often-used tracers of shocks is molecular hydrogen, H$_2$ \citep[e.g.,][]{hollenbach89, kaufman96b, rosenthal00}. This is the most abundant molecule in the interstellar medium by some four orders of magnitude over CO and H$_2$O. The molecule is the lightest, and so it has the most widely spaced rotational levels ($J$ = 1 has $E_{\rm up} / k_{\rm B}$ = 170 K and $J$ = 2 has $E_{\rm up} / k_{\rm B}$ = 510 K). As such, it is predominantly excited in warm ($T \gtrsim 10^2$~K) and hot ($T \gtrsim 10^3$~K) molecular gas. This molecule has no permanent dipole moment, and only forbidden electric quadrupole transitions occur, although at low probability. The main reason H$_2$ emission is still bright is because of its high abundance.

\begin{figure*}
    \centering
    \includegraphics[width=\textwidth]{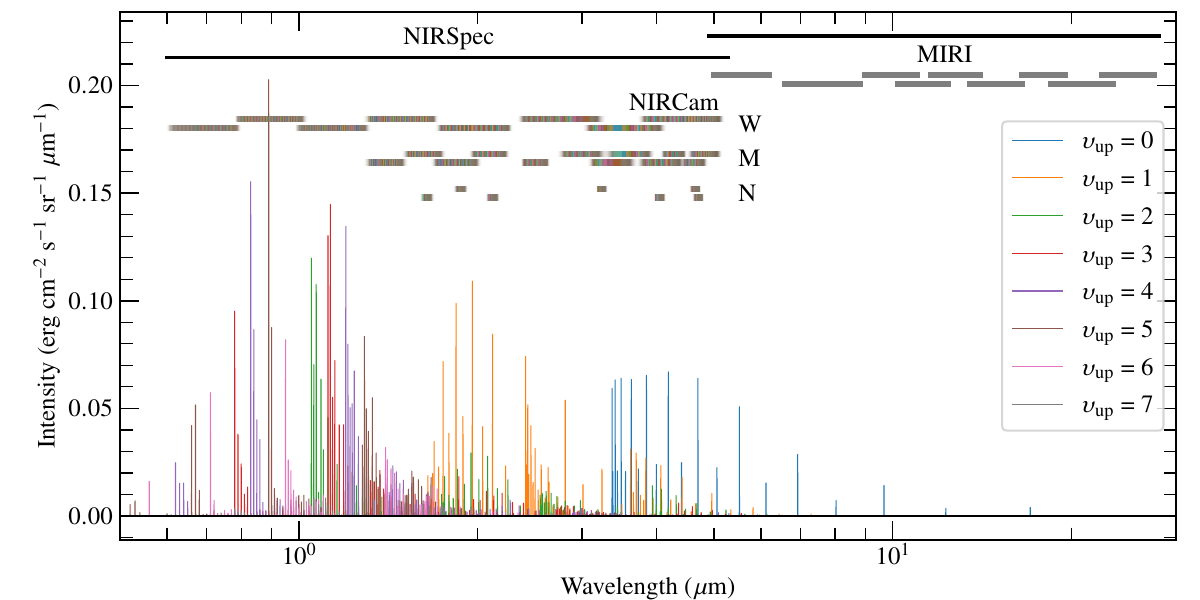}
    \caption{
    Synthetic H$_2$ spectrum produced with a shock model with velocity 30 km s$^{-1}$, preshock density 10$^4$ cm$^{-3}$, a transverse magnetic field strength of 10 $\mu$G, and no external UV radiation. Wavelength ranges of the NIRSpec and MIRI spectrographs, as well as the wide-, medium-, and narrow-band filters for NIRCam and the MIRI filters on JWST are indicated as black and gray horizontal bars.  The colors are for lines with different vibrational upper levels. The resolving power is assumed to be uniform across the wavelength range at $\lambda/\Delta\lambda$=2500.}
    \label{fig:spec_filter}
\end{figure*}

H$_2$ emission is readily observed from the ground, particularly in higher-excited rovibrational transitions at near-infrared wavelengths \citep[e.g.,][]{froebrich15}. The brightest of these is typically the $\varv$ = 1--0 S(1) line at 2.12 $\mu$m. A few pure rotational lines are also accessible from the ground, and the line profiles may even be velocity resolved on telescopes such as the Very Large Telescope \citep[VLT,][]{santangelo14}. However, it is necessary to go above the atmosphere to observe the lower-excited pure rotational transitions of H$_2$. Space-based telescopes such as the Infrared Space Observatory (ISO) and the \textit{Spitzer} Space Telescope (\textit{Spitzer}) both observed these transitions toward numerous shocked regions \citep[e.g.,][]{rosenthal00, neufeld06, Valentijn1999, Lutz2003, Verma2005}, as did the Stratospheric Observatory For Infrared Astronomy \citep[SOFIA,][]{reach19, neufeld19}. Now the \textit{James Webb} Space Telescope (JWST) is doing the same \citep[e.g.,][]{garcia-bernete22, berne22, yang22, Appleton2023, Alvarez-Marquez2022}. Particularly, the MIRI instrument is observing the rotational \Hmol\ transitions with a gain in sensitivity and spatial resolution of two orders of magnitude compared with \textit{Spitzer}, and an increase in spectral resolution of a factor five \citep[e.g., Fig. 7 and 8 of][]{rigby22}. Similar improvements are reached with the NIRSpec instrument compared with the VLT-SINFONI integral-field unit, allowing deep observations of the rovibrational lines of \Hmol. The wavelength coverage of NIRSpec, NIRCam, and MIRI are illustrated in Fig. \ref{fig:spec_filter}, which shows a simulated H$_2$ spectrum with the instrument wavelength coverages displayed. 

Planning and interpreting the abovementioned observations is often done by use of models. With models, it is possible to constrain, for example, the shock velocity and preshock density, which together give the input kinetic energy flux, 1/2 $\rho$ $\varv_{\rm s}^3$, where $\rho$ is the mass density and $\varv_{\rm s}$ is the shock velocity. In molecular shocks, a comparison reveals that up to 50\% of the input energy is radiated away in H$_2$ emission \citep{kaufman96b}, depending on shock conditions, making H$_2$ the dominant coolant in these shocks. \textit{Spitzer} particularly opened up for characterization of the pure rotational H$_2$ lines. Observations and subsequent modeling revealed that most H$_2$ emission could be reproduced by shock models \citep[e.g., in protostellar outflows;][]{maret09, dionatos10}. However, when additional constraints, such as the H/H$_2$ ratio and the cooling length are included for protostellar outflows, a single shock model no longer reproduces observations \citep{nisini10}. Instead, as argued, the observational beam likely catches different shocks, or more complex shock geometries than 1D, which is to be expected; this is not just the case for protostellar outflows, but also observations of shocks in the diffuse gas of starburst and colliding galaxies \citep{kristensen08, gustafsson10, lesaffre13, tram18, lehmann22}. Irrespective of the specific science case, the first step in comparing observations to models is to have the models available. 

The Paris-Durham shock code \citep[e.g.,][ and references therein]{godard19} has been developed and maintained for more than 35 years \citep{flower85}. The code can either find jump (J-type shocks) or continuous (C-type shocks) solutions depending on the input physical parameters. Recent developments include the treatment of an external UV radiation field \citep{godard19}, and self-irradiation in high-velocity shocks \citep[$\varv_{\rm s} \gtrsim$ 30 km s$^{-1}$;][]{lehmann22}. Here we present the results of running a large grid of simulations of (externally irradiated) shocks with the goal of exploring how the input energy flux (kinetic and radiative) is reprocessed and ultimately results in H$_2$ emission. These model predictions can be used directly to interpret, for example, JWST observations of shock emission. 

The paper is organized as follows. Section \ref{sect:mod_desc} describes the shock model and the model grid, with a particular emphasis on H$_2$ excitation and emission. The section also describes which physical quantities were extracted from the models, and the methodology applied. Section \ref{sect:results} describes the results and provides a discussion of these results. Finally, the main points are summarized in Sect. \ref{sect:summary}.

\section{Model and grid description}\label{sect:mod_desc}

The current version of the multifluid shock code is extensively described in \citet{godard19} and references therein, and only the main relevant points will be described here. These points particularly relate to H$_2$ emission and other observable diagnostics, but also how the initial shock conditions are calculated. The code is publicly available\footnote{\url{http://ism.obspm.fr/shock.html}}, and the entire grid presented in this paper is also available on the ISM platform\footnote{ \url{https://app.ism.obspm.fr/ismdb/}}. In Appendix \ref{app:ismdb} we provide an introduction to this platform and demonstrate how it can be used.

\subsection{Initial conditions} \label{sect:initial}

\begin{figure*}
\centering
\hspace{38pt}
\includegraphics[width=0.8\textwidth]{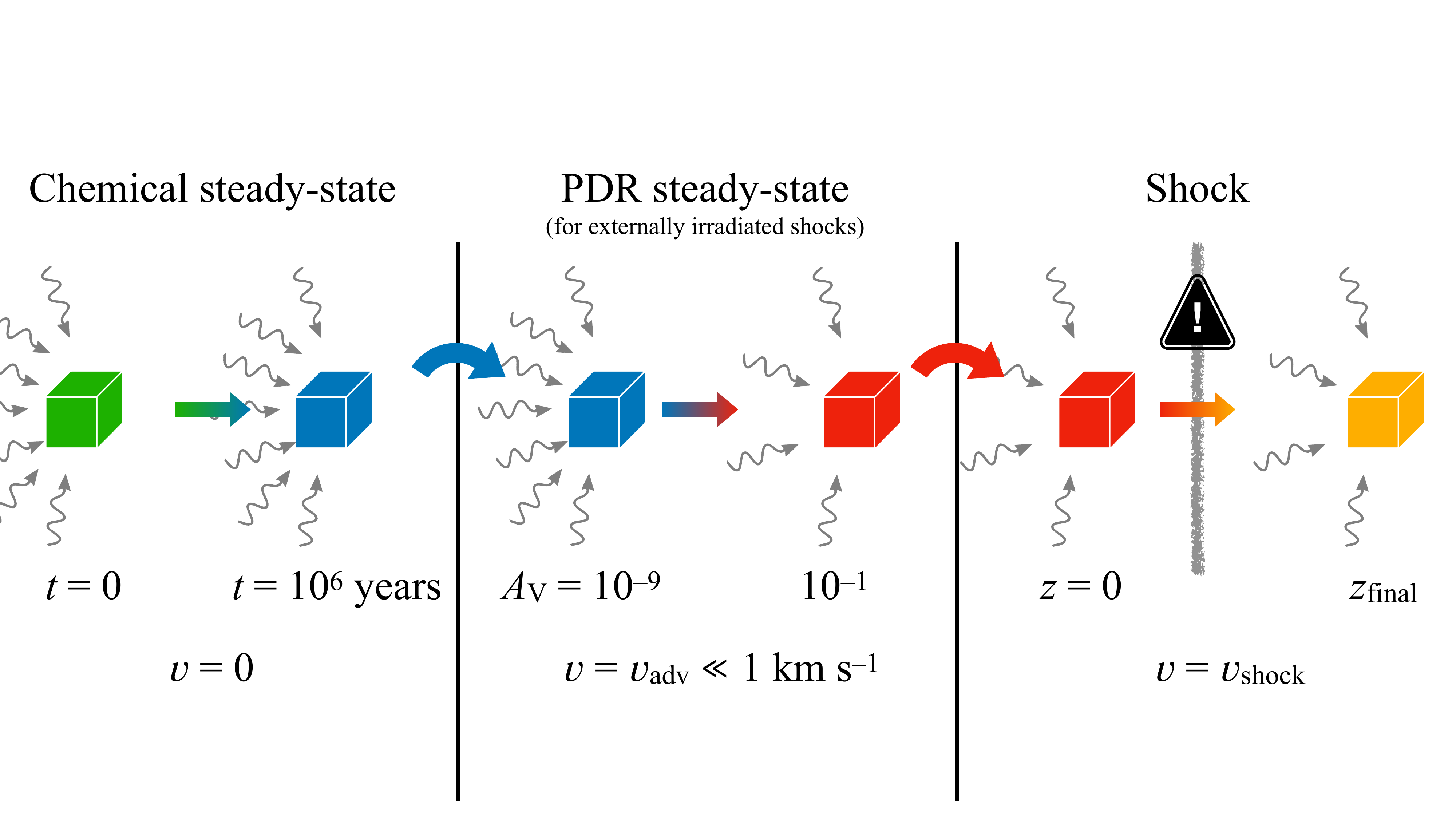}
\includegraphics[width=0.9\textwidth]{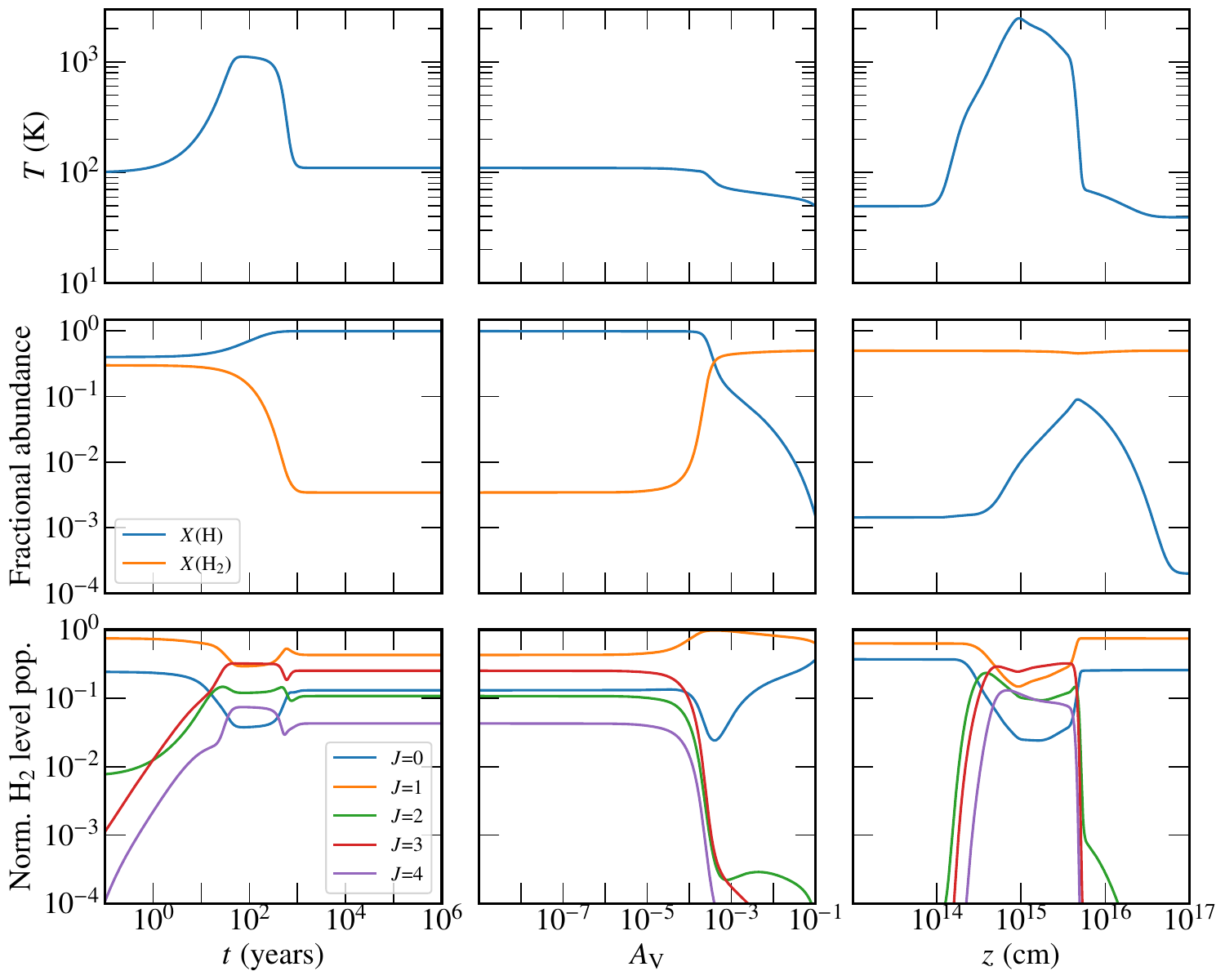}
\caption{Illustration of the three steps required for running an externally irradiated shock model. The shock model shown here has a preshock density of 10$^4$ cm$^{-3}$, shock velocity of 20 km s$^{-1}$, and it is irradiated by a UV radiation field with $G_0$ of 10. The strength of the transverse magnetic field is 100 $\mu$G. First a chemical steady-state model is run, and the thermal, chemical, and excitation output is used as input for a PDR model. The output of the PDR model is then used as input for the shock model. The top row shows the temperature evolution across the model run, the middle row the abundances of H and H$_2$, while the bottom row shows normalized populations of the first five rotational levels of H$_2$. The ``bump'' in the temperature profile at $t$ $\sim$ 10$^2$ years in the chemical steady-state model comes from reformation of a small fraction of H$_2$ on the grain, and the release of its binding energy. \label{fig:initial}}
\end{figure*}

\begin{table}
  \caption{Initial fractional elemental abundances, $n_{\rm X}/n_{\rm H}$.}
  \label{tab:ini_abun}
  \centering\small
  \begin{tabular}{l c c c c}
    \hline\hline
     & Fractional & Gas   & &  Grain \\
    Element & abundance  & phase & PAHs\tablefootmark{a} & cores\tablefootmark{b} \\
    \hline
    H  & 1.00     & 1.00     & 1.8(--5) &  \\
    He & 1.00(--1) & 1.00(--1) & &          \\
    C  & 3.55(--4) & 1.38(--4) & 5.4(--5) & 1.63(--4) \\
    N  & 7.94(--5) & 7.94(--5) & &  \\
    O  & 4.42(--4) & 3.02(--4) & & 1.40(--4) \\
    Mg & 3.70(--5) &          &         & 3.70(--5) \\
    Si & 3.67(-5) &   3.00(--6)  & & 3.37(--5) \\
    S  & 1.86(--5) & 1.86(--5) & &          \\
    Fe & 3.23(--5) & 1.50(--8) &          & 3.23(--5) \\
    \hline
  \end{tabular}
  \tablefoot{$a(b) = a \times 10^b$. 
  \tablefoottext{a}{The abundances of H and C in PAHs are given for an initial PAH abundance of 10$^{-6}$ here.}
  \tablefoottext{b}{The grain size distribution is an MRN distribution with a mass density of 2 g cm$^{-3}$ leading to a fractional grain abundance 6.9 $\times$ 10$^{-11}$.}
}
\end{table}

The main focus of this paper is on H$_2$, and so the chemistry considered in this paper and, more importantly, in the models run, is a gas-phase-only chemistry. That is, grain adsorption and desorption processes are not included. The only exceptions are the formation of H$_2$ on grains, and grain erosion for the release of elemental Si, Fe, etc. into the gas phase. Photochemistry is included in all steps of the calculation; readers can refer to the text below for more details.

Our assumption is that the initial conditions are in equilibrium, that is, thermal and chemical equilibrium with or without an incident radiation field. Running a shock model therefore requires multiple steps, all done using the Paris-Durham code \citep[see][for details]{godard19}. This code simulates steady-state gas equilibrium, photon-dominated regions (PDRs), or shocks. These steps are illustrated in Fig. \ref{fig:initial}. First, a chemical steady-state calculation is run with the given density and radiation field. For irradiated shocks, the next step is to take the final equilibrium conditions from the chemical steady-state calculation and use these as input for a PDR calculation, where a tracer particle is advected at a small velocity ($\le$ 0.01 km s$^{-1}$) from an $A_{\rm V}$ of 10$^{-9}$ to 10$^{-1}$. The advection speed is chosen such that the time it takes to cross the PDR front is long enough that equilibrium is reached; this timescale is 10$^5$--10$^9$ years for high to low densities. The choice of a final $A_{\rm V}$ of 0.1 is motivated by two considerations. First, the primary focus of this paper is H$_2$ and the $A_{\rm V}$ thus needs to be high enough that the preshock gas is substantially molecular (molecular fraction $\ge$ 0.1) for the majority of the $G_0$ values here, specifically the part of the grid where $G_0/n_{\rm H}$ $<$ 1. Second, the $A_{\rm V}$ should be low enough that H$_2$ is not fully self-shielded. These two conditions are met at an $A_{\rm V}$ of 0.1. The final conditions, in terms of steady-state abundances, temperature, and H$_2$ level populations, are then used as the input physical conditions of the shock calculation. The shock is run in the final step. 

The initial elemental abundances are provided in Table \ref{tab:ini_abun}. Of particular importance is the abundance of polycyclic aromatic hydrocarbons (PAHs). In the model, a representative PAH molecule is included, C$_{54}$H$_{18}$ and its singly charged ions. Table \ref{tab:ini_abun} reports the amount of H and C locked up in this PAH for a PAH abundance of $X$(PAH) = 10$^{-6}$. The grain temperature is kept fixed at 15 K.

We cover a 6D parameter space with preshock density ($n_{\rm H}$ = 2 $n$(H$_2$) + $n$(H)), shock velocity ($\varv_{\rm s}$), strength of the transverse magnetic field\footnote{The transverse magnetic field strength scales with the density as $B = b \times \sqrt{n_{\rm H}\ ({\rm cm}^{-3})}$~$\mu$G, where $b$ is a scaling factor.} ($b$), external UV radiation \citep[$G_0$ in units of the field from][]{mathis83}, H$_2$ cosmic-ray ionization rate ($\zeta_{\rm H2}$), and the fractional abundance of the PAHs ($X$(PAH)). The parameter space is presented in Table \ref{tab:grid}. Depending on the initial conditions, the code either finds a Jump (J-type) solution or a Continuous (C-type) solution (see below, Sect. \ref{sec:bparam} for more details). Throughout this paper, we use two shock models to illustrate differences when changing $b$ from 0.1 to 1.0; these are referred to as model A and B (Table \ref{tab:refmod}). For the given set of input parameters, model A gives rise to a J-type shock, and model B a C-type shock. 

\begin{table}[!t]
\caption{Shock grid parameters. \label{tab:grid}}
\begin{center}
\small
\begin{tabular}{l c}
\hline\hline
Parameter & Values \\ \hline
$n_{\rm H}$\tablefootmark{(a)} (cm$^{-3}$) & 10$^2$, 10$^3$, 10$^4$, 10$^5$, 10$^6$, 10$^7$, 10$^8$ \\
$b$\tablefootmark{(b)} & 0.1, 0.3, 1.0, 3.0, 10.0 \\
$\varv_{\rm s}$ (km s$^{-1}$), $b$=0.1 & 2, 3, 4, 5, 10, 15, 20, 25, 30 \\
$\varv_{\rm s}$ (km s$^{-1}$), $b$=0.3 & 2, 3, 4, 5, 10, 15, 20, 25, 30 \\
$\varv_{\rm s}$ (km s$^{-1}$), $b$=1.0 & 2, 3, 4, 5, 10, 15, 20, 25, 30 \\
$\varv_{\rm s}$ (km s$^{-1}$), $b$=3.0 & 10, 20, 30, 40, 50, 60 \\
$\varv_{\rm s}$ (km s$^{-1}$), $b$=10.0 & 20, 40, 60, 80, 90 \\
$G_0$\tablefootmark{(c)} & 0, 10$^{-1}$, 10$^{0}$, 10$^{1}$, 10$^{2}$, 10$^3$ \\
$\zeta_{\rm H2}$\tablefootmark{(d)} (s$^{-1}$) & 10$^{-17}$, 10$^{-16}$, 10$^{-15}$ \\ 
$X$(PAH) & 10$^{-8}$, 10$^{-7}$, 10$^{-6}$ \\ \hline
\end{tabular}
\end{center}
  \tablefoot{
\tablefoottext{a}{Proton density defined as 2 $n$(H$_2$) + $n$(H).}
\tablefoottext{b}{Scaling factor of the transverse magnetic field such that $B = b \times \sqrt{n_{\rm H}\ ({\rm cm}^{-3})}$~$\mu$G.}
\tablefoottext{c}{UV field strength in units of the \citet{mathis83} field.}
\tablefoottext{d}{H$_2$ cosmic-ray ionization rate.}}
\end{table}

\begin{table}[!t]
\caption{Reference models. \label{tab:refmod}}
\begin{center}
\small
\begin{tabular}{l c c}
\hline\hline
& Model A & Model B \\ \hline
$n_{\rm H}$ (cm$^{-3}$) & 10$^4$ & 10$^4$ \\
$b$ & 0.1 & 1.0 \\
$\varv_{\rm s}$ (km s$^{-1}$) & 20 & 20 \\
$G_0$ & 0 & 0 \\
$\zeta_{\rm H2}$ (s$^{-1}$) & 10$^{-17}$ & 10$^{-17}$ \\ 
$X$(PAH) & 10$^{-6}$ & 10$^{-6}$ \\ \hline
\end{tabular}
\end{center}
\tablefoot{These sets of input parameters lead to model A being a J-type shock, and model B a C-type shock.}
\end{table}

\subsection{Molecular hydrogen}\label{sect:h2}

Collisional excitation and de-excitation of H$_2$ is calculated for collisions with H, H$_2$, and He. The collisional rate coefficients for H$_2$-H$_2$ collisions are adopted from \citet{flowerroueff98} and for H$_2$-He collisions from \citet{flower98a}. In the case of H$_2$-H collisions, for the first 49 levels of H$_2$ the rates are from \citet{flower97} and \citet{flower98}, where the rates have been calculated using a full quantum mechanical approach. For the remaining levels, the rates from \citet{martin95} are used. They were calculated using a quasi-classical approach. The reactive reaction rates of H$_2$ with H are from \citet{lebourlot99}. 

The number of levels has been set to 150 here, and the highest level is $\varv$ = 8, $J$ = 3 ($E/k_{\rm B}=39,000$~K). The model assumes that there are no levels between the user-set value and the dissociation level. This may be important when calculating the dissociation rate of H$_2$, since molecules that are already excited have internal energies that are closer to the dissociation limit, and thus require less energy to dissociate. For the models run here, we find that there is no significant difference in H$_2$ emission by increasing the number of levels. 

Depending on the initial conditions, H$_2$ may dissociate in the shock through collisions. As the post-shock gas cools, H$_2$ reforms on the grains \citep[Appendix A of][]{flower13} and it is necessary to account for the bond energy released (4.5 eV $\sim$ 5.1 $\times$ 10$^4$ K). We assume that approximately one third of the energy goes to internal energy of the molecule. This internal energy distribution follows a Boltzmann distribution with a temperature corresponding to $\sim$ 17,000 K. The remaining energy is equally split between kinetic energy of the newly formed H$_2$ molecule, and heating of the grain.

The H$_2$ level populations are used for calculating the local H$_2$ line emissivities. This is done under the assumption of optically thin emission, which typically applies to H$_2$ emission because of its lack of a permanent dipole moment. Of these lines, 1000 are output explicitly and stored as emissivity profiles in this grid. About 900 of these H$_2$ lines are covered by the JWST instruments MIRI and NIRSpec. These two instruments together cover the wavelength range of 0.6 -- 28 $\mu$m, that is the $\varv$ = 0--0 S(0) ground-state line at 28.3 $\mu$m (Fig. \ref{fig:spec_filter}) is not covered. 

\subsection{Grid}\label{sect:grid}

The total set of grid parameters is presented in Table \ref{tab:grid}; covering this range of parameter space resulted in $\sim$ 14,000 simulations in total. Each simulation produces a number of outputs that are all stored in human-readable ASCII files and an HDF5 file for easy extraction\footnote{The full model outputs are provided on the ISM platform: \url{https://app.ism.obspm.fr/ismdb/}}. These include physical properties of the shock (e.g., temperature, density, velocity) as a function of distance and time through the shock, and chemical properties (e.g., local densities, charge state, column densities), excitation of H$_2$ (level populations and local emissivities). In this case, the time is calculated as the neutral flow time, $t_{\rm n} = \int {\rm d}z / \varv_{\rm n}$.  In total, more than 2600 quantities are stored as profiles through each shock, and 1400 quantities are stored as integrated values. 

The model integrates the gas state far downstream in order to ensure that a steady-state solution is contained within the simulation. Therefore, special care needs to be taken when extracting integrated quantities such as column densities or line intensities. We here adopt a similar criterion for the size of the shock as in \citet{godard19} based on radiative energy dissipation. We here set that limit as the point where 99.9\% of the total radiation has been emitted (see Appendix \ref{app:tables}). Specifically, this means that the size, $z_{\rm s}$ is defined as: 
\begin{equation}
    \frac{\Upsilon(z_{\rm s}) - \Upsilon(0)}{\Upsilon(\infty)-\Upsilon(0)} = 99.9 \% \ , 
    \label{eq:width}
\end{equation}
where $\Upsilon$ is the sum of the kinetic, magnetic, and thermal energy fluxes. 

For ease of use, we provide a number of tables containing already-extracted results at the Centre de Donn{\'e}es astronomiques de Strasbourg (CDS\footnote{Add link to CDS archive at publication stage.}). Example tables are provided in Appendix \ref{app:tables} in Tables \ref{tab:phys} --  \ref{tab:atom}. These tables include: 
\begin{itemize}
    \item[\ref{tab:phys}] Physical parameters such as peak temperature, density, width, and age of the shock; 
    \item[\ref{tab:coldens}] Column densities of selected species, particularly H, H$_2$, O, OH, H$_2$, C$^+$, C, and CO; 
    \item[\ref{tab:exc}] Data required for creating H$_2$ excitation diagrams, i.e., ln($N$/$g$) and $E$ for each of the 150 levels; 
    \item[\ref{tab:int}] H$_2$ integrated intensities of the 1000 lines extracted, along with their wavelength; 
    \item[\ref{tab:width}] Width of the H$_2$ emitting zone for the $\varv$ = 0--0 S(1), 1--0 S(1), 0--0 S(9), 1--0 O(5), and 2--1 S(1) lines; 
    \item[\ref{tab:opr}] H$_2$ $o/p$ ratios determined both locally and integrated through the shock; 
    \item[\ref{tab:atom}] Integrated line intensities of 29 transitions arising from C$^+$, Si$^+$, H, C, Si, O, S$^+$, N$^+$, N, and S. 
\end{itemize}
On occasion, the model does not converge for numerical reasons; this happens in $\sim$5\% of cases. This convergence-failure occurs often in C$^*$-type shocks, when the flow crosses the first sonic point \citep[see Appendix C in][]{godard19}. In these cases, the model output is ignored but the input parameters are still recorded in the tables.

\begin{figure*}
\centering
\includegraphics[width=0.49\textwidth]{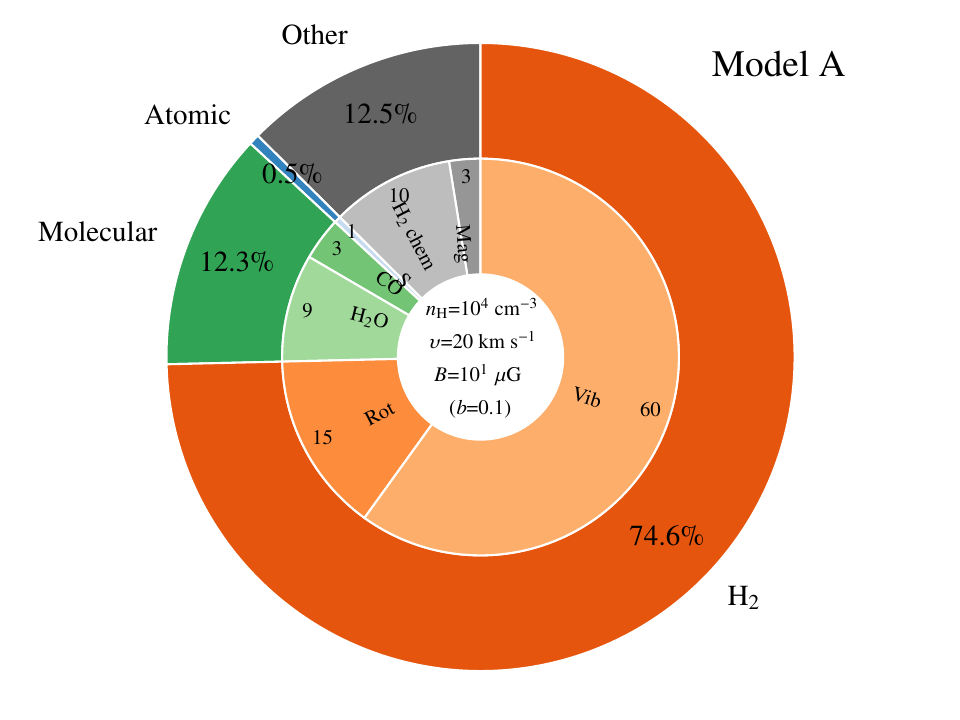}
\includegraphics[width=0.49\textwidth]{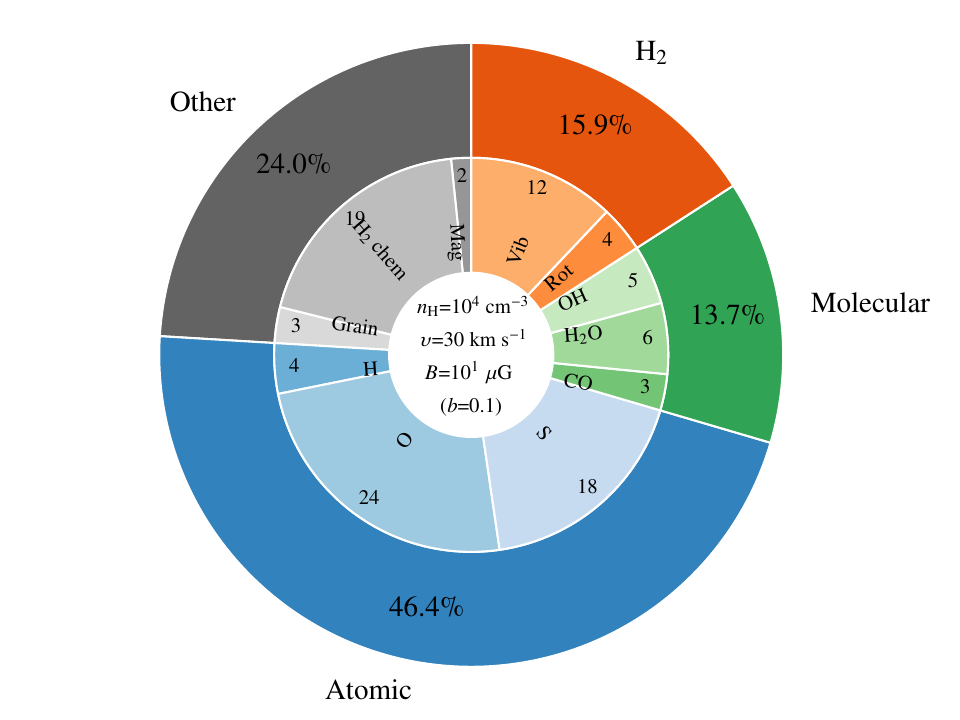}
\includegraphics[width=0.49\textwidth]{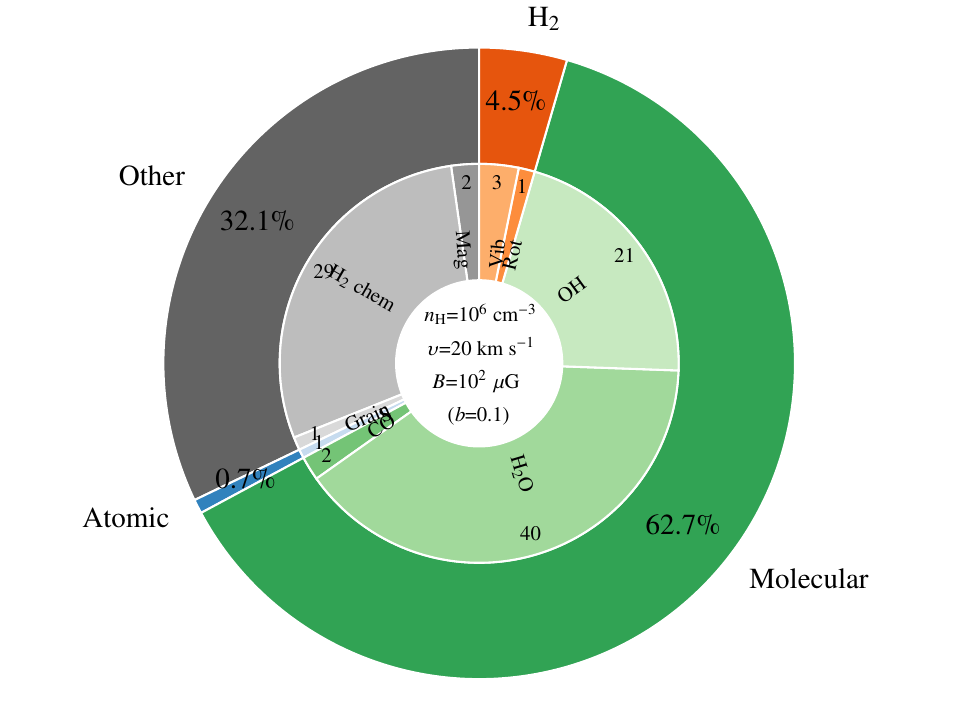}
\includegraphics[width=0.49\textwidth]{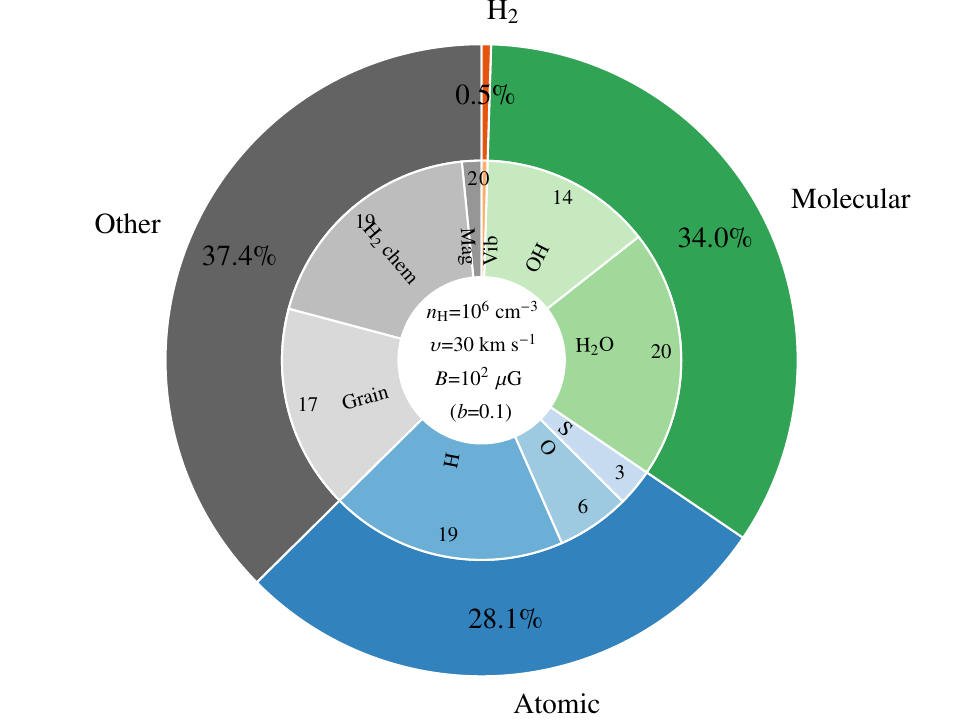}
\caption{Energetic reprocessing for four shocks with $b$ = 0.1. The pie charts show the percentage of energy lost relative to the input kinetic energy flux. The kinetic energy flux is primarily converted to heat, which goes to exciting the atoms and molecules that then radiate the energy away. This radiation is either from H$_2$ (rotational and vibrational emission), other molecules (primarily CO, OH and H$_2$O), or atoms (primarily H, O, and S). Some kinetic energy goes into compressing the magnetic field (``mag''), dissociating H$_2$ collisionally (``H$_2$ chem''), or atoms/molecules thermalizing with grains (``grain''). The percentages are shown in each pie slice, and the input shock parameters inside the pies. The input parameters all result in the shocks being J-type shocks, and model A is marked. \label{fig:pie_b01}}
\end{figure*}

\begin{figure*}
\centering
\includegraphics[width=0.49\textwidth]{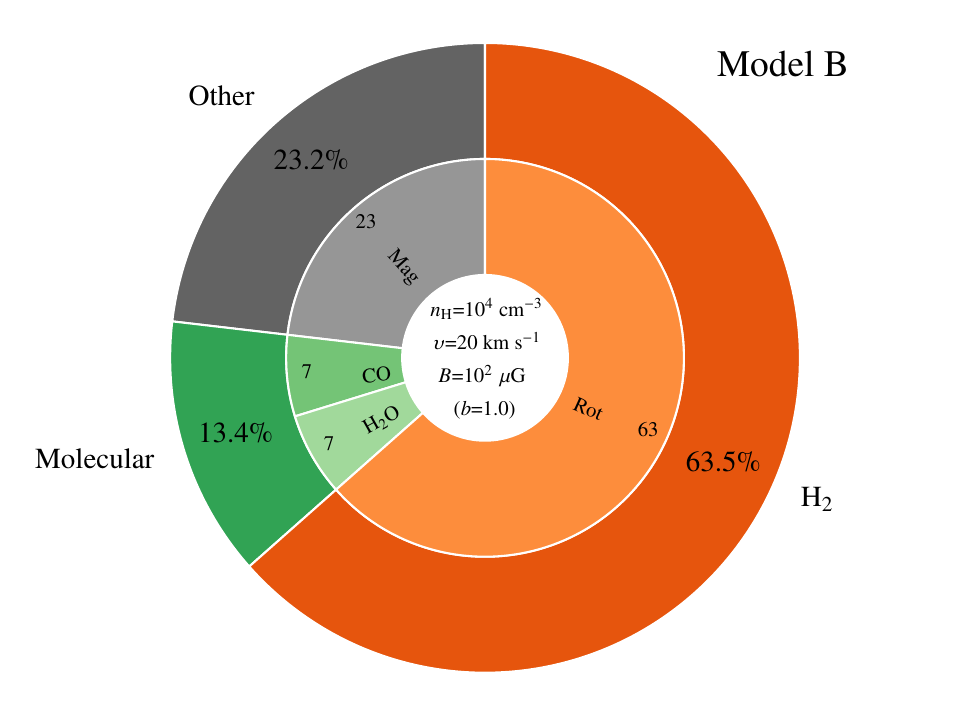}
\includegraphics[width=0.49\textwidth]{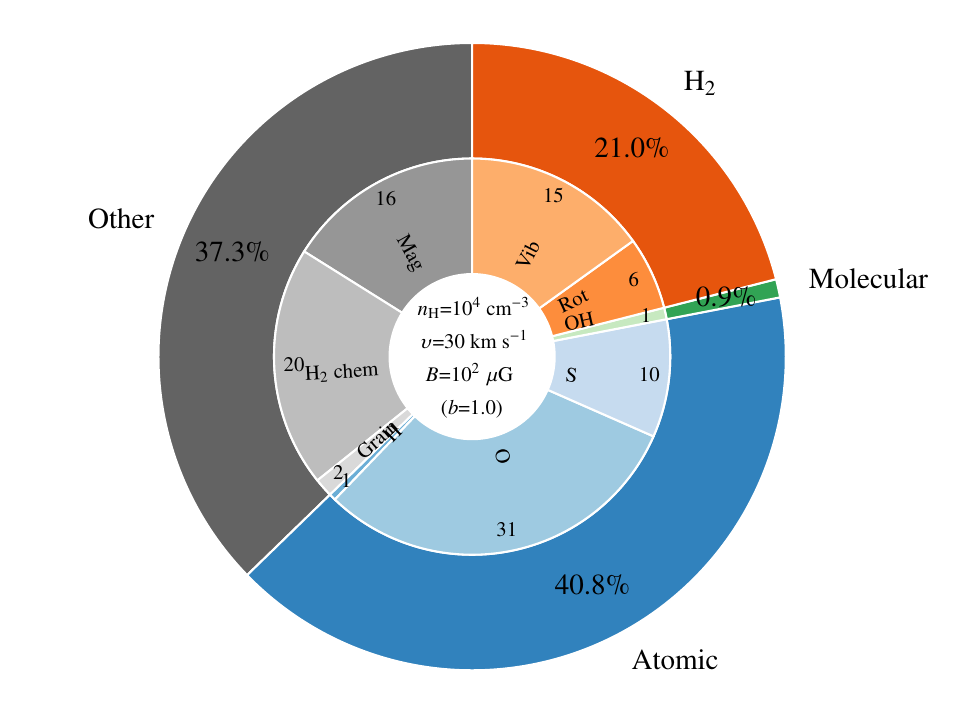}
\includegraphics[width=0.49\textwidth]{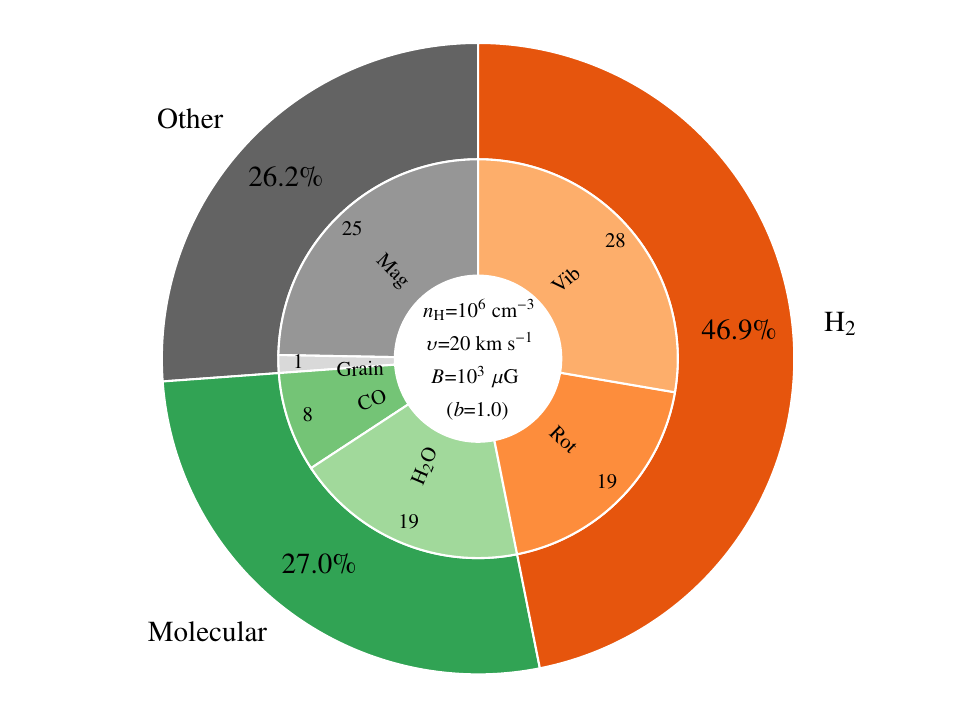}
\includegraphics[width=0.49\textwidth]{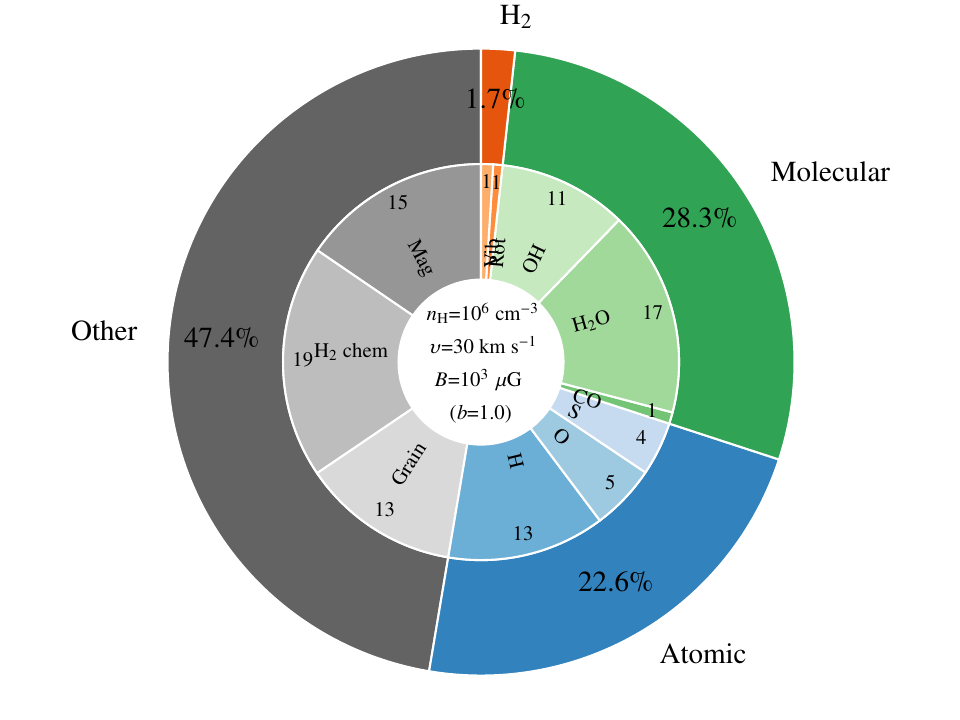}
\caption{As Fig. \ref{fig:pie_b01} but for $b$ = 1.0. For this change in $b$, shock models represented in the two left-most pie-charts are C-type, while the two right pie charts are J-type shocks; model B is marked.\label{fig:pie_b10}}
\end{figure*}

\subsection{Model limitations}\label{sect:limits}

The model has a number of inherent assumptions, which are discussed in the following. The include the shock geometry, magnetic field orientation, self-irradiation, stationary shocks, and grain chemistry. 

\textbf{Geometry.} The model treats a plane-parallel shock front, thus ignoring geometry. The lack of geometry is especially important in J-type shocks, where the gas may be compressed by four orders of magnitude or more. In nature, such a compression would quickly lead to a expansion of the high-pressure post-shock gas into the surrounding low-pressure medium, however, that is not possible in a 1D simulation. As a result, the post-shock density could be overestimated. For the case of H$_2$ emission, this is less important: most of the H$_2$ emission is generated in the warm parts of the shock where $T$ $>$ 100 K, prior to where significant post-shock expansion would occur. 

\textbf{Magnetic field orientation.} The magnetic field orientation is assumed to be perpendicular to the direction of motion. This may not always be the case in molecular clouds, in fact, there is no a priori reason to assume the shock wave and field orientation are well aligned. If the field is not perpendicular to the direction of motion, the compression will lead to a change in field geometry, as described and discussed in \citet{lehmann16}. These effects are not included here. 

\textbf{Self-irradiation.} The model is best suited for molecular shocks. In shocks where H$_2$ is dissociated and atomic H is excited, the shocks become self-irradiated. While this self-irradiation can be solved iteratively \citep{lehmann20, lehmann22}, it is not included in the present version of the grid. This limits J-type shocks to $\varv_{\rm s} \lesssim 30$~km\,s$^{-1}$. 

\textbf{Stationary shocks.} All the shocks in this paper are stationary shocks. This implies there needs to be enough time for the stationary structure to fully develop. While the code can mimic non-stationary shocks, an additional free parameter, the age of the shock, is needed, and it is deemed beyond the scope of this work to explore the effects of that parameter \citep[e.g.,][]{lesaffre04a, lesaffre04b, gusdorf08b}. 

\textbf{Grain chemistry.} Grain-grain interactions are omitted in this grid. For conditions where the velocity is below $\sim$ 25 km s$^{-1}$ and the density is below $\sim$ 10$^5$ cm$^{-3}$, this assumption is likely valid \citep{guillet09, guillet11}. At larger velocities or densities, grains may interact, leading to grain evaporation and fragmentation which changes the size distribution of grains. Finally, in this grid we do not include ice mantles on the grains.

\section{Results and discussion}\label{sect:results}

The shock has an initial kinetic energy flux of 1/2 $\rho$ $\varv_{\rm s}^3$, where $\rho$ = 1.4 $n_{\rm H}$ $m_{\rm H}$ is the mass density; most of this energy is radiated away in the shock. Figure \ref{fig:pie_b01} shows how the energy is lost in shocks with $b$ = 0.1, velocities of 20 and 30 km s$^{-1}$, and densities of 10$^4$ and 10$^6$ cm$^{-3}$. The pie charts are sorted by initial kinetic energy flux going from left to right, and top to bottom. The H$_2$ fraction decreases with increasing velocity and density because of dissociation. H$_2$ then reforms on the grains in the postshock gas introducing a heating term which counteracts the cooling of H$_2$. This is visible in the pie charts as the fraction of H$_2$ emission decreases monotonically with input kinetic energy flux, from 75\% to 0.5\%. 

Figure \ref{fig:pie_b10} is similar to Fig. \ref{fig:pie_b01}, but for a stronger magnetic field ($b$ = 1.0), i.e., the input kinetic energy fluxes are the same as above. Increasing $b$ to 1 has the consequence that the two 20-km s$^{-1}$ shocks become C-type shocks; the 30-km s$^{-1}$ shocks remain J-type shocks. The J-type shocks are dissociative, and the H$_2$ cooling fraction thus decreases significantly, as also illustrated in Fig. \ref{fig:pie_b01}. 

The distribution of energy flux into emission lines has been described previously \citep[e.g.,][]{kaufman96b, flower10, flower15, lehmann20}, and a comparison in H$_2$ cooling fractions of the total input kinetic energy flux reveals broad agreement between different models and previous versions of the Paris-Durham model. These pie charts provide a global view of the energetic reprocessing in these shocks. In the following, the role of the different input parameters on the energetic reprocessing will be discussed in more detail, with a specific emphasis on H$_2$ emission.

\subsection{Magnetic field}\label{sec:bparam}

\begin{figure*}
\centering
\includegraphics[width=0.55\textwidth]{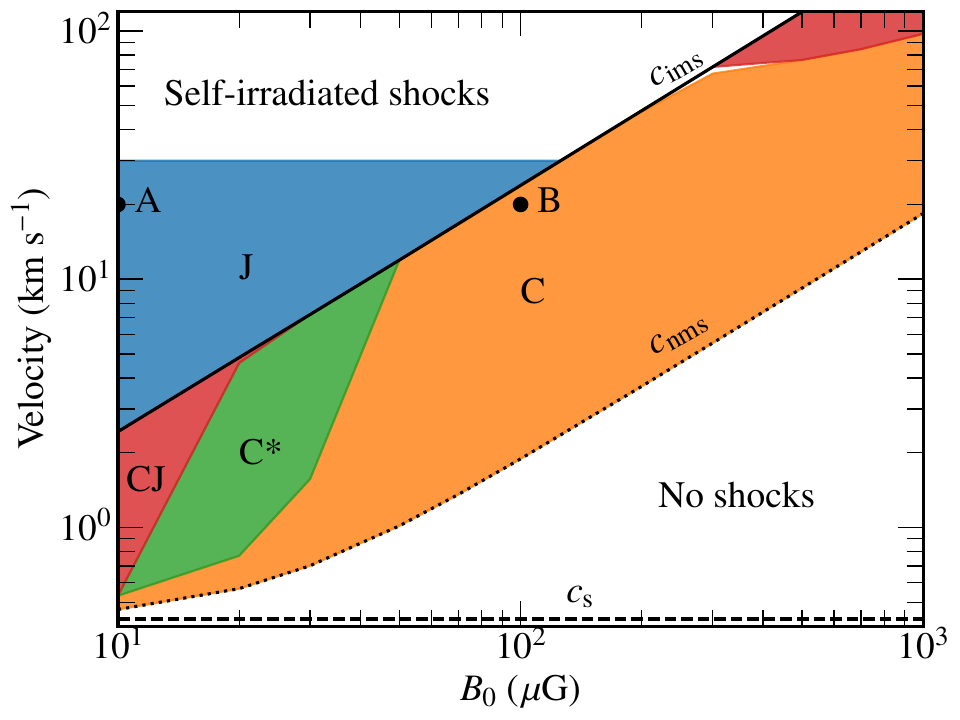}
\includegraphics[width=0.47\textwidth]{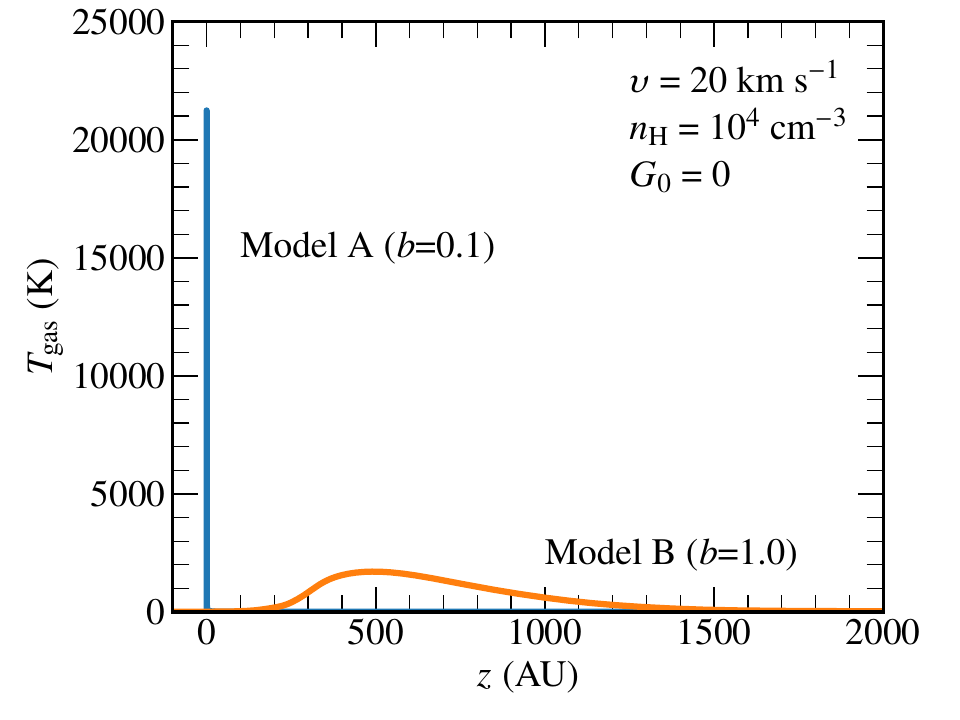}
\includegraphics[width=0.47\textwidth]{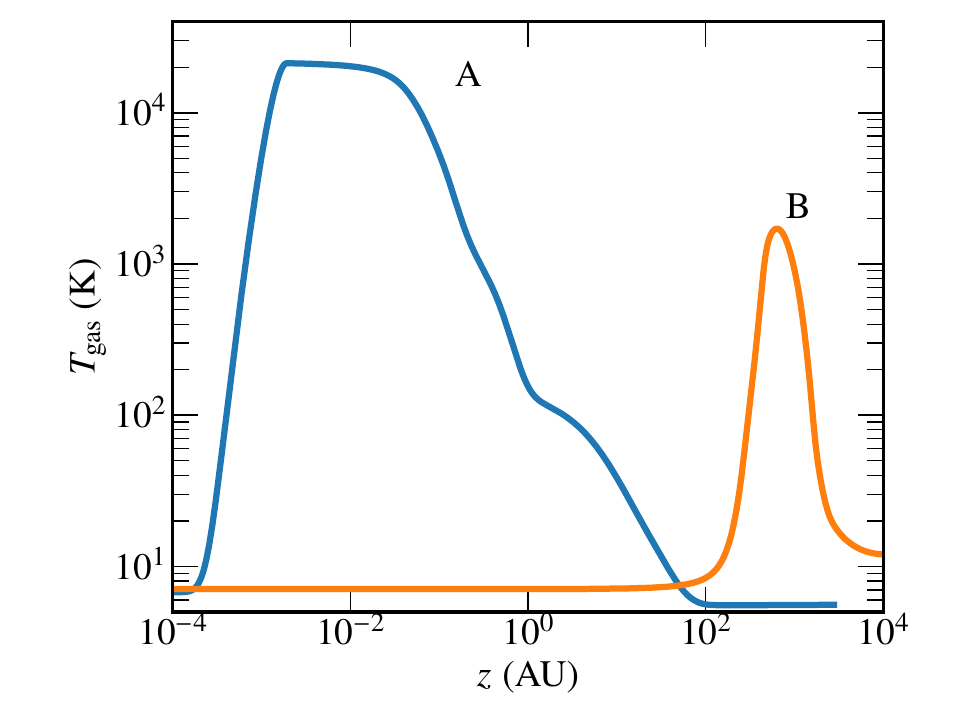}
\includegraphics[width=0.47\textwidth]{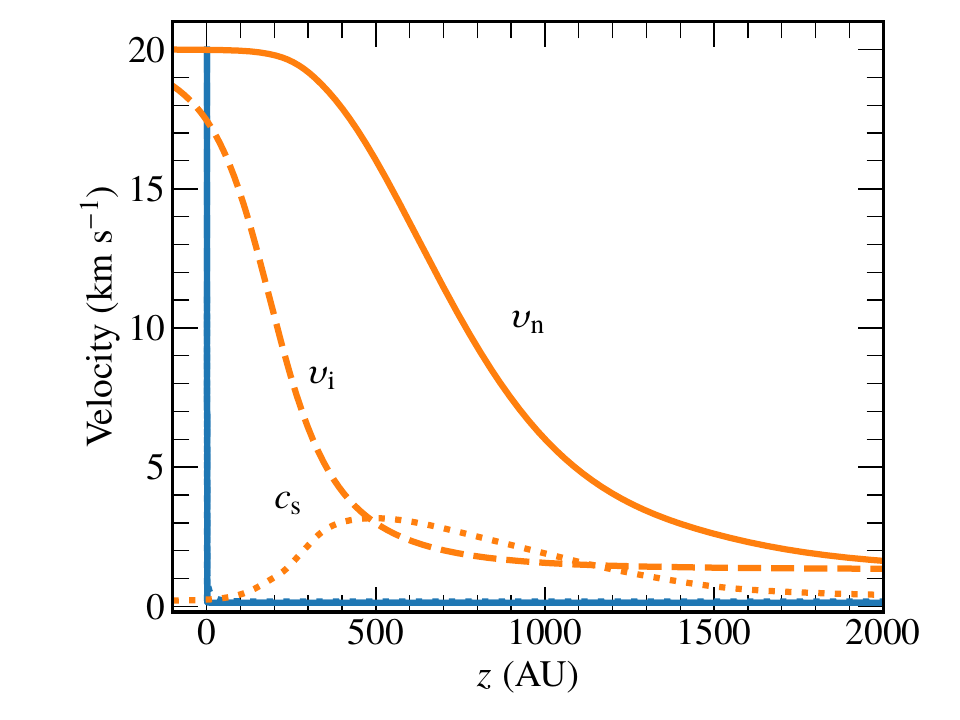}
\includegraphics[width=0.47\textwidth]{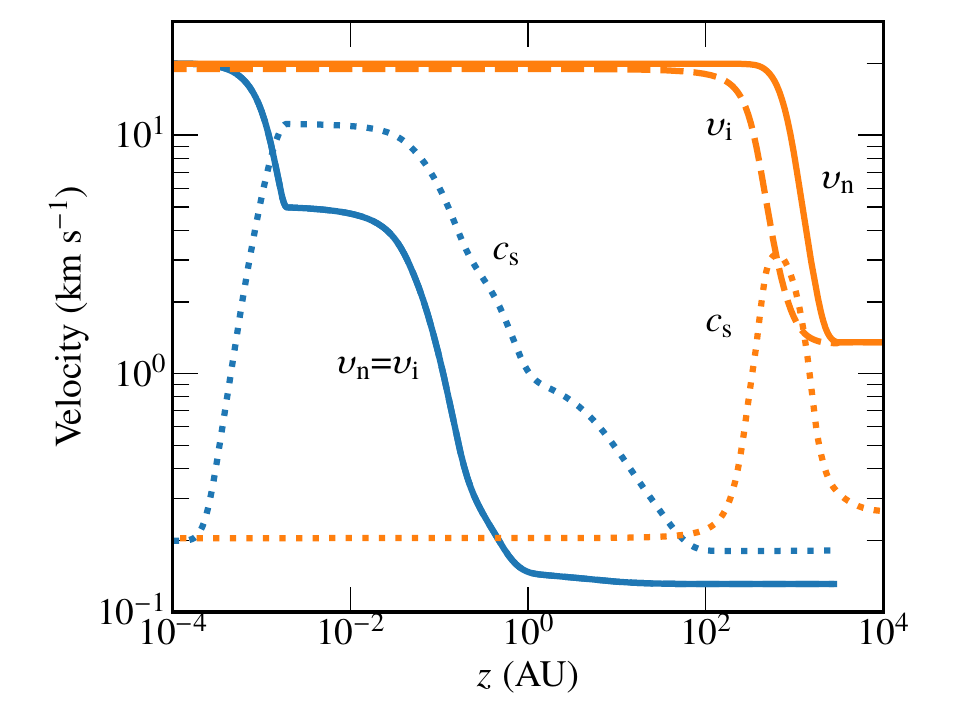}
\caption{Illustration of shock parameter space. \textit{Top.} Various shock type regimes as a function of transverse magnetic field strength and shock velocity for $G_0$ = 0 and preshock density of 10$^4$ cm$^{-3}$ \citep[adapted from Fig. 3,][]{godard19}. Here a field strength of 10 $\mu$G corresponds to $b$ = 0.1, and 10$^3$ $\mu$G is $b$ = 10. The blue region is for J-type shocks where the velocity exceeds the ion-magnetosonic velocity ($c_{\rm ims}$). Above 30 km s$^{-1}$ the shocks start producing UV-photons and become self-irradiated, which is not included in the current grid \citep{lehmann22}. The red, green and orange areas are for CJ, C*, and C-type shocks, respectively; these shocks fall between the ion- and neutral-magnetosonic velocities ($c_{\rm ims}$ and $c_{\rm nms}$, where the latter is calculated in a similar manner to $c_{\rm ims}$, Eq. \ref{eq:cims}, but with the neutral mass density). Models A and B are marked. \textit{Middle.} Gas temperature profiles for models A and B. \textit{Bottom.} Velocity profiles for models A and B. The dotted line shows the local sound speed, $c_{\rm s}$, the dashed is for the ion speed, $v_{\rm i}$, and the full is for the neutral speed, $v_{\rm n}$. All velocities are in the reference frame of the shock, and the line colors are for the same $b$ parameters as above. The difference between the left- and right-hand sides of the middle and bottom panels is that the left panel is on a linear-linear scale, while the right is on a log-log scale. \label{fig:paramspace}}
\end{figure*}

\begin{figure*}[!th]
\centering
\includegraphics[width=\textwidth]{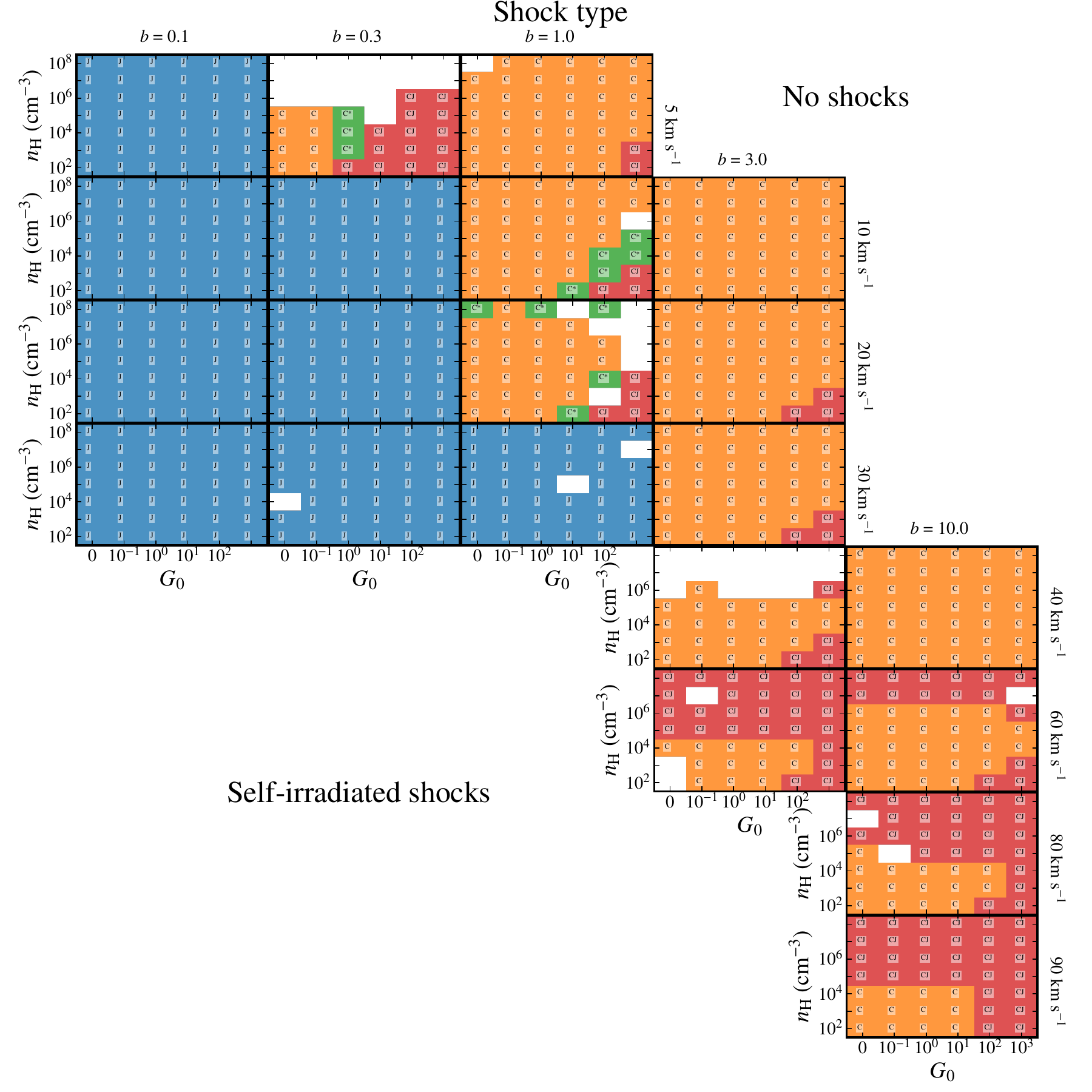}
\caption{Resulting shock type as a function of UV radiation field strength, preshock density, shock velocity, and magnetic field strength for a subset of the grid (for brevity, the shock velocities 2, 3, 4, 15, 25, and 50 km s$^{-1}$ are not shown). The cell color denotes the shock type, where blue is for J-type shocks, orange is for C-type shocks, red is for CJ-type shocks, and green is for C$^*$-type shocks, as is also written in each cell. The figure shows model results for $\zeta_{\rm H2}$=10$^{-17}$~s$^{-1}$, $X$(PAH)=10$^{-6}$. White cells are for models that did not converge numerically. \label{fig:type}}
\end{figure*}

The strength of the transverse magnetic field, $B$, sets the ion-magnetosonic speed, $c_{\rm ims}$, together with the ion mass density, $\rho_{\rm i}$: 
\begin{equation}
    c_{\rm ims} = \left(c_{\rm s} + B^2 / 4\pi\rho_{\rm i} \right)^{1/2},\end{equation}
where $c_{\rm s}$ is the sound speed. For $\varv_{\rm s} < c_{\rm ims}$, the ionized and neutral fluids are decoupled and a magnetic precursor is present \citep{mullan71, draine80}; the code treats these multiple fluids self-consistently. For $\varv_{\rm s} > c_{\rm ims}$, the ionized and neutral fluids are coupled, and there is no magnetic precursor (Fig. \ref{fig:paramspace}). We refer to Sect. 2.1 of \citet{lehmann22} for a more in-depth description of the differences between J- and C-type shocks. Figure \ref{fig:paramspace} shows where the different shock types are as a function of $b$ and $\varv_{\rm s}$ for a density of 10$^4$ cm$^{-3}$, Fig. \ref{fig:type} shows the shock type for a part of the grid presented in this paper. For low values of $b$ ($\lesssim$0.3), the resulting shocks are J-type shocks, while for $b$ $\gtrsim$ 1.0 the resulting shocks are predominantly C-type shocks.

\begin{figure}
\centering
\includegraphics[width=\columnwidth]{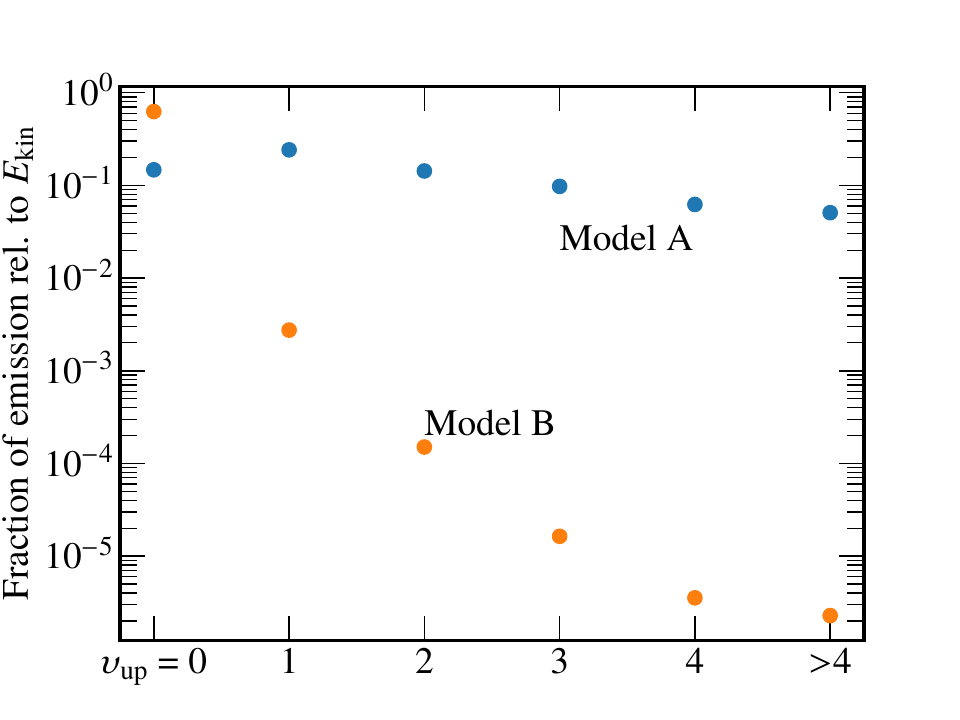}
\includegraphics[width=\columnwidth]{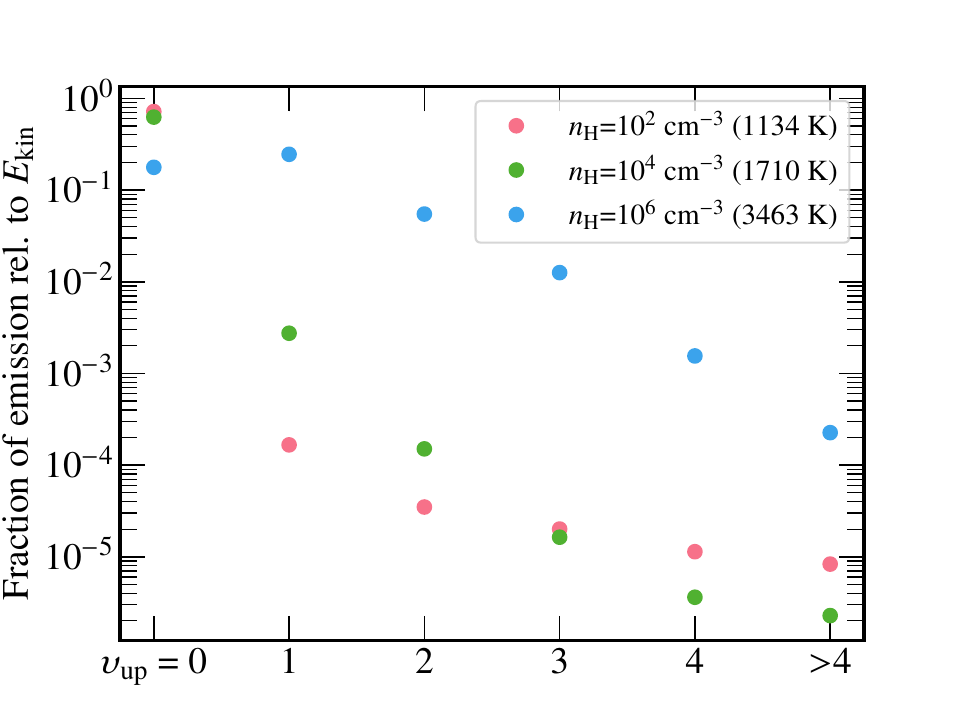}
\caption{\textit{Top.} H$_2$ integrated intensities integrated over 4$\pi$ steradian from each vibrational level compared to the input kinetic energy flux for model A and B. The total H$_2$ intensity  is similar in the two models, 5.5 $\times$ 10$^{-3}$ and 4.7 $\times$ 10$^{-3}$ erg s$^{-1}$ cm$^{-2}$ sr$^{-1}$ for model A and B, respectively. \textit{Bottom.} As the top panel, but for densities of 10$^2$, 10$^4$, and 10$^6$ cm$^{-3}$, $b$ = 1.0, and $\varv_{\rm s}$ = 20 km s$^{-1}$. The maximum neutral gas temperature is given in brackets. \label{fig:h2_emech}}
\end{figure}

The effects of the magnetic precursor is that the input kinetic energy flux is deposited over a much larger spatial range (Fig. \ref{fig:paramspace}), resulting in lower peak temperatures when compared to shocks with the same input kinetic energy flux but no magnetic precursor. This naturally affects the excitation of H$_2$, as illustrated in Fig. \ref{fig:h2_emech} in the form of the fraction of total integrated intensity to initial kinetic energy flux. The H$_2$ excitation is illustrated for the two reference shocks (Table \ref{tab:refmod}), both with the same input kinetic energy flux. The figure demonstrates that for both shocks, most of the kinetic energy is radiated away in H$_2$ emission (see Fig. \ref{fig:pie_b01} and \ref{fig:pie_b10}); the difference in total H$_2$ integrated intensity from the two shocks is $\sim$ 15\%. However, the integrated intensity from model B ($b$=1.0) is dominated by pure rotational emission ($>$ 99\% of H$_2$ emission), whereas it is spread over the vibrational levels in model A ($b$=0.1). 

\begin{figure*}
\centering
\includegraphics[width=0.41\textwidth]{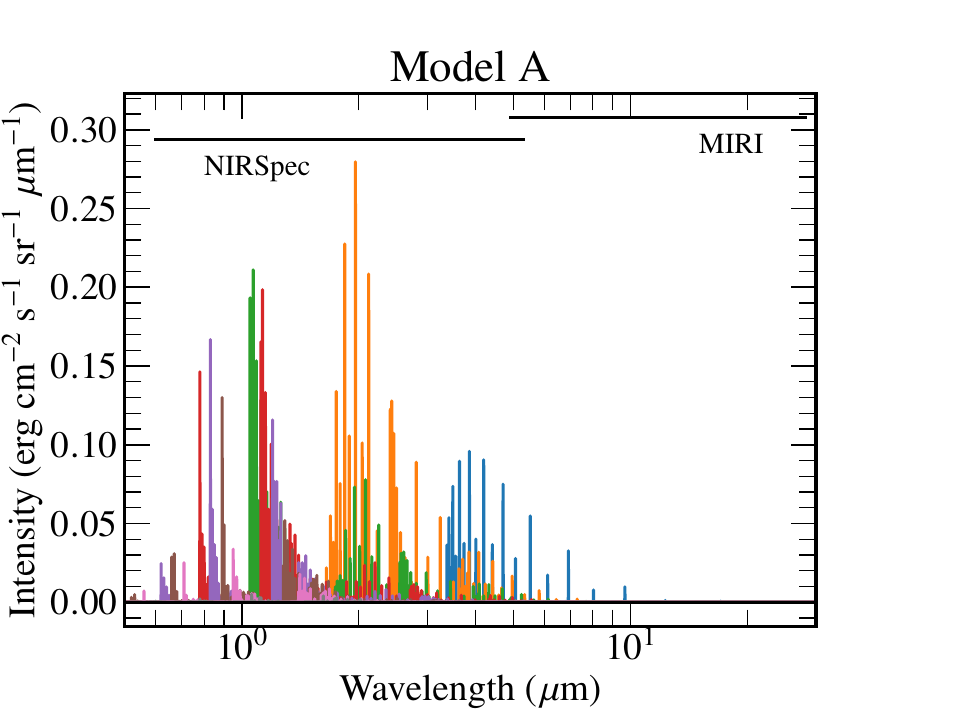}
\includegraphics[width=0.41\textwidth]{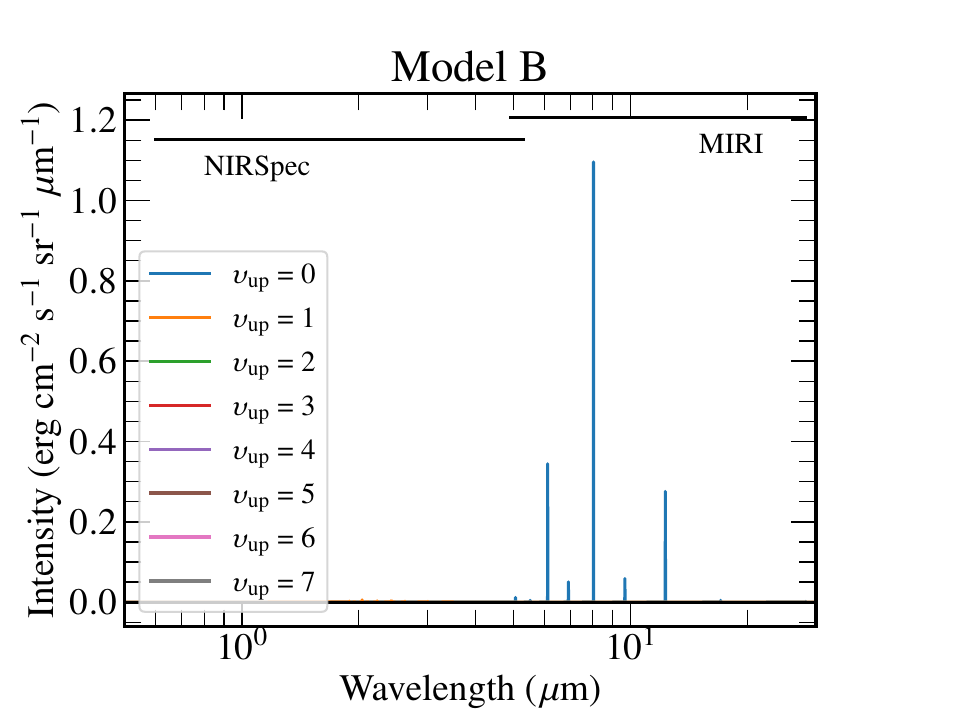}
\includegraphics[width=0.41\textwidth]{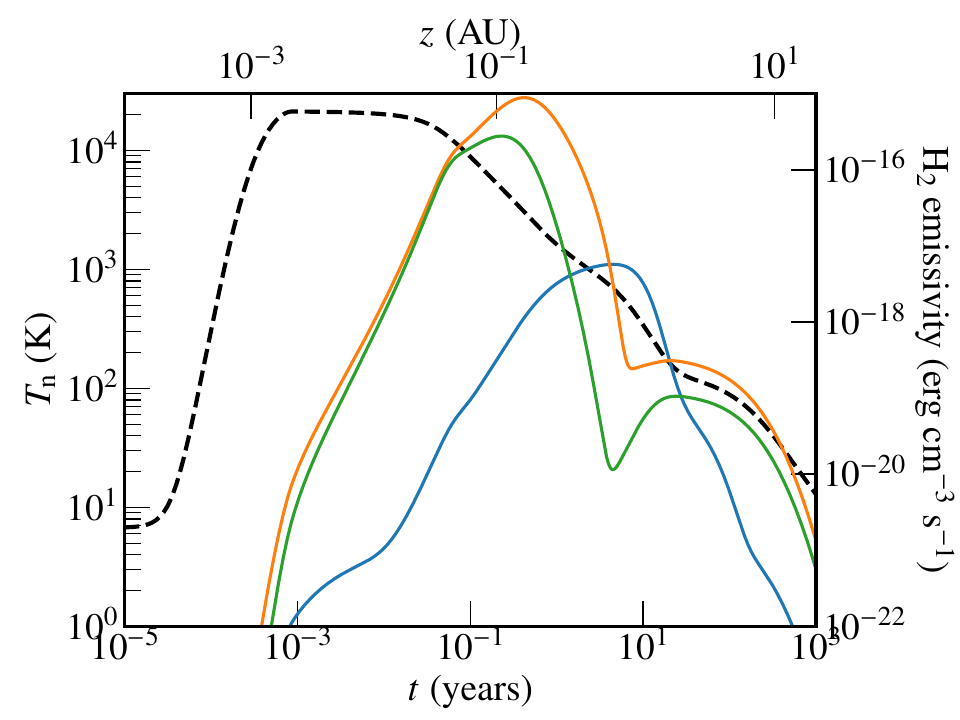}
\includegraphics[width=0.41\textwidth]{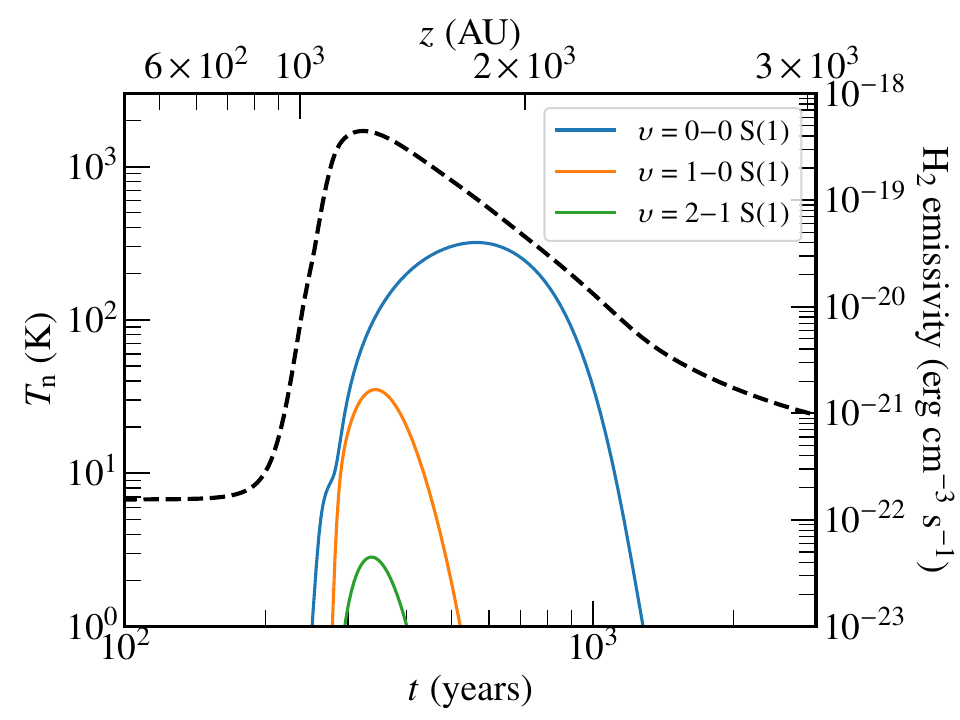}
\includegraphics[width=0.41\textwidth]{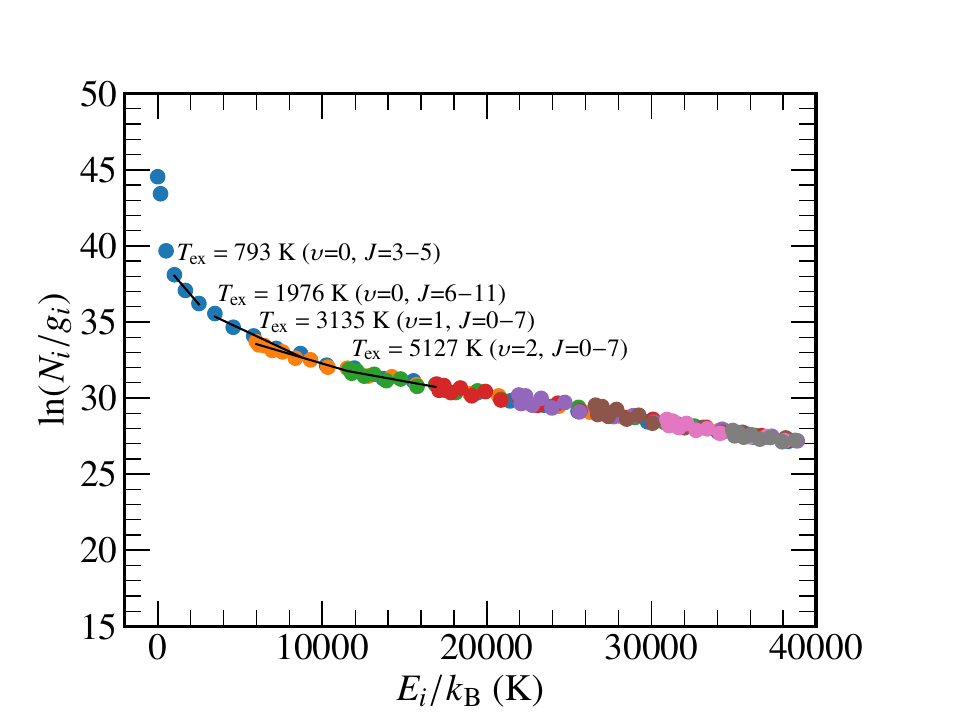}
\includegraphics[width=0.41\textwidth]{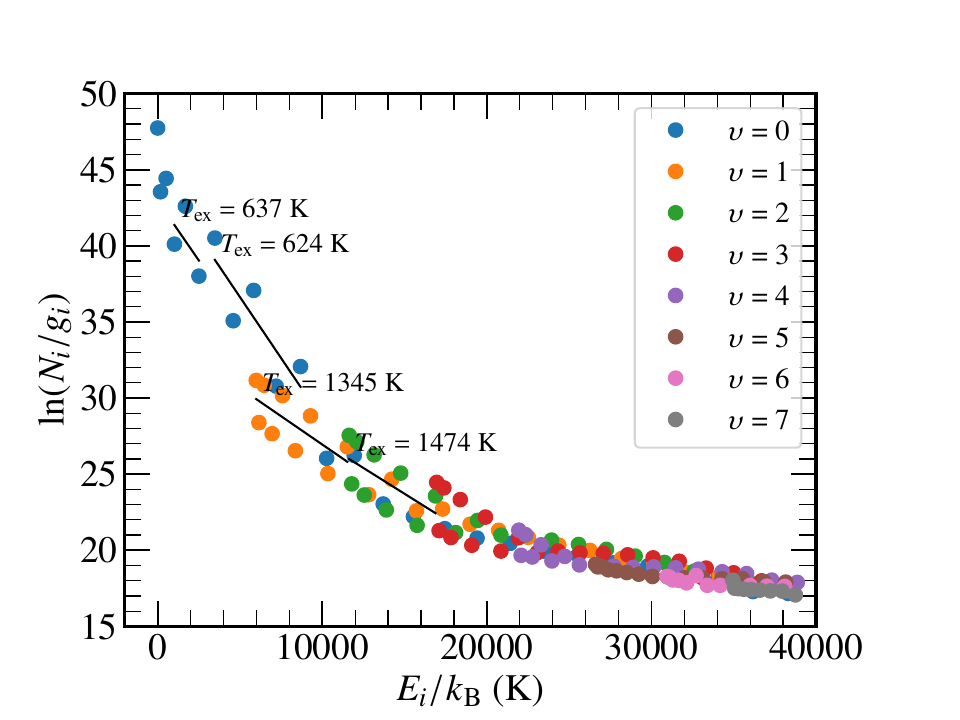}
\includegraphics[width=0.41\textwidth]{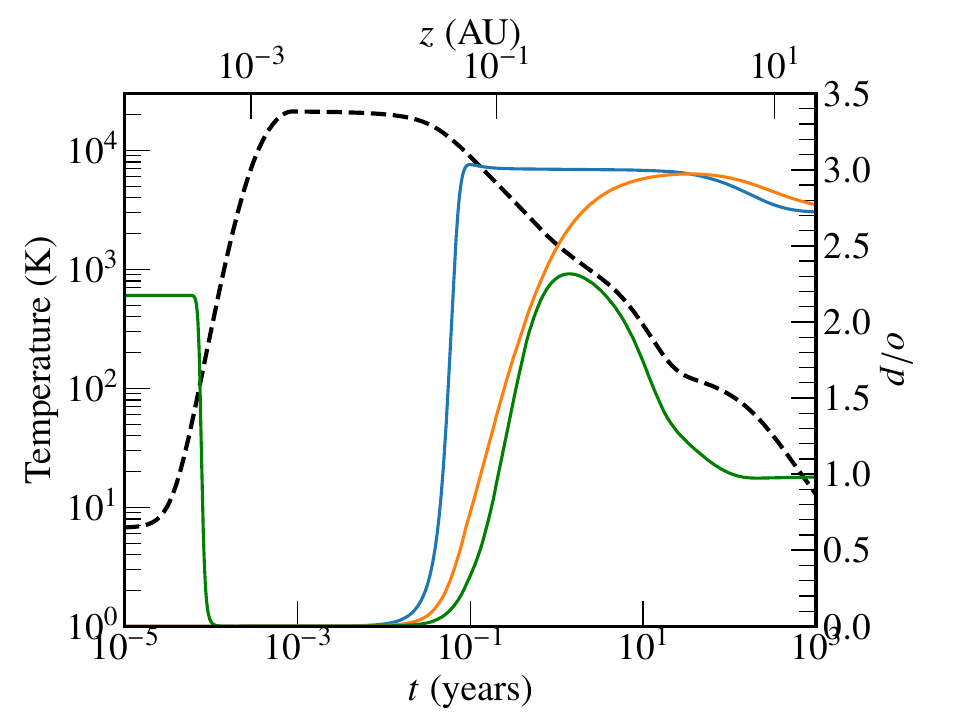}
\includegraphics[width=0.41\textwidth]{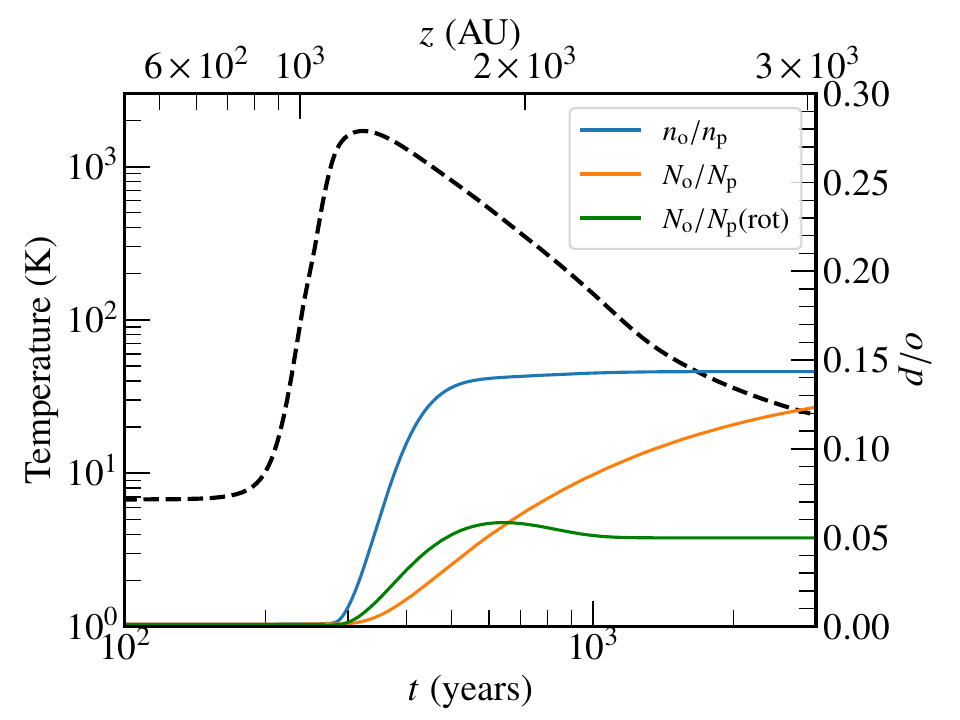}
\caption{
\textit{Top row.} Resulting H$_2$ spectrum in model A and B (Table \ref{tab:refmod}). The wavelength ranges of NIRSpec and MIRI on the JWST are labeled. The resolving power is uniform across the spectrum, implying that the linewidths increase with wavelength (not visible on this plot). 
\textit{Second row.} H$_2$ emissivities for three lines. The temperature profile is shown in black. 
\textit{Third row.} H$_2$ excitation diagrams color-coded according to vibrational level, and with four extracted excitation temperatures shown. The excitation temperatures are obtained from a linear fit to all the indicated levels.   
\textit{Fourth row.}  $o/p$ ratio measured from local densities ($n_{\rm o}/n_{\rm p}$), cumulative column densities ($N_{\rm o}/N_{\rm p}$), and from the column densities of the $\varv_{\rm up}$ = 0, $J$ = 2--9 levels ($N_{\rm o}/N_{\rm p}$(rot)). 
\label{fig:h2_cj}}
\end{figure*}

The differences in H$_2$ excitation and the origin thereof for different values of $b$ are further explored in Fig. \ref{fig:h2_cj} for models A and B in the left and right column, respectively. The first row shows the emerging H$_2$ spectrum from the two shocks. As was already clear from Fig. \ref{fig:h2_emech}, most of the H$_2$ emission in model A is spread over the vibrational transitions, whereas emission in model B predominantly is rotational. To make these artificial spectra, a uniform resolving power of $R = \lambda/\Delta\lambda$ = 2500 is assumed, similar to the resolving powers of the NIRSpec and MIRI instruments on JWST, and the line shapes are Gaussian. That is, the integrated intensity calculated in the models is $I_{\rm total} = \sqrt{\pi} I_{\rm peak} \Delta \lambda / (2 \sqrt{2 \ln 2})$. A uniform resolving power implies that the emission from longer-wavelength transitions is spread over a larger wavelength range, and thus the peak emission is lower. This stark difference in the H$_2$ spectra can be understood from the physical structure of the shock. 

The kinetic energy flux injected into the two shocks is the same, but the temperature structure is very different. For J-type shocks, such as model A, the maximum temperature can be approximated by \citep{lesaffre13}: 
\begin{equation}
    T_{\rm max} = 53~{\rm K} \left(\frac{\varv_{\rm s}}{1~{\rm km s}^{-1}} \right)^2.
    \label{eq:rh_temp}
\end{equation}
For model A, the maximum temperature is $\sim$ 2$\times$10$^4$ K (Fig. \ref{fig:h2_cj}, second row). This high temperature ensures that the vibrational H$_2$ levels are readily populated. For model B ($b$ = 1.0), on the other hand, the magnetic precursor causes the kinetic energy to be deposited over a much larger scale ($\sim$ 10$^3$ AU vs. $\sim$ 1 AU), and the resulting peak temperature is much lower ($\sim$ 2000 K). In this case, the temperature is so low that only the rotational levels are significantly excited. 

The third row of Fig. \ref{fig:h2_cj} shows excitation diagrams for the two shocks. For model A, all points fall on a single curved line, indicating that the levels are probing a range of excitation temperatures, $T_{\rm ex}$. Particularly, the higher-$J$ and rovibrational transitions probe hotter gas than the lower-$J$ transitions, and the slope is thus shallower (slope = --1/$T_{\rm ex}$). In this case, the excitation temperatures is similar to the gas temperature where the local emissivity peaks (second row of Fig. \ref{fig:h2_cj}). The excitation diagram for model B shows more scatter (caused by the low initial $o/p$ ratio, see below), but the excitation temperatures still match the gas kinetic temperature where the levels are excited.  In Appendix \ref{app:tex} we provide figures showing the extracted excitation temperatures sampling the full range of initial density and shock velocity for $b$ = 0.1 and 1.0, and $G_0$ = 0 and 1. 

Another feature of the excitation diagram for model B is that there is a clear difference between the ortho- and para-levels of H$_2$. Here the ortho-levels (odd $J$) are displaced downward compared to the corresponding para-levels (even $J$), and the resulting zigzag pattern indicates that the ortho/para ($o/p$) ratio is lower than the high-temperature statistical equilibrium value of 3 \citep{neufeld06}. 

There are no radiative or collisional transitions between ortho- and para-H$_2$ levels, only exchange reactions with H, H$_2$, and protonated ions (e.g., H$_3^+$, HCO$^+$) can change the spin state \citep[Sect. 2.1 of][]{lebourlot99}. The line emission and resulting excitation diagram is integrated through the shock, and thus does not provide information on the local $o/p$ ratio. This is calculated directly from the level populations as $n_{\rm o} / n_{\rm p}$, and it can be compared to the cumulative column density ratio, $N_{\rm o} / N_{\rm p}$. Both these values are shown in the bottom row of Fig. \ref{fig:h2_cj}. This column density ratio is often dominated by the column densities of H$_2$ in the two lowest rotational levels, $J$=0 and 1, which are not accessible in emission. Therefore, we also show the $o/p$ ratio as calculated from the column densities of the lowest observable rotational levels, in this case from the $J$ = 2--9 levels (S(0) to S(7) transitions). In model A, the temperature is high enough that the H exchange reaction H$_2^{\rm para}$ + H $\rightarrow$ H$_2^{\rm ortho}$ + H proceeds efficiently \citep[e.g.,][]{wilgenbus00}. The resulting $o/p$ ratios are thus close to 3, although the inferred rotational $o/p$ is somewhat lower than 3 ($\sim$ 1). For model B, the temperature never get high enough that the exchange reactions with H become dominant; instead, the ion-neutral proton-transfer reactions dominate, but they are limited by the low abundances of ions. Thus, the $o/p$ ratios remain at $\sim$ 0.1. In both models, the initial temperature is 10 K and the gas is dense, which leads to a steady-state $o/p$ ratio of 10$^{-3}$ \citep[see Fig. 1 of][]{flower06}. Had the initial temperature been higher or the gas not been in steady state, the initial $o/p$ ratio would have been higher, and the $o/p$ ratio through the shock also correspondingly higher. All in all, however, special care must be taken when interpreting $o/p$ ratios inferred from observations \citep[see also Fig. 4 of][]{wilgenbus00}. 

\begin{figure*}
\centering
\includegraphics[width=0.48\textwidth]{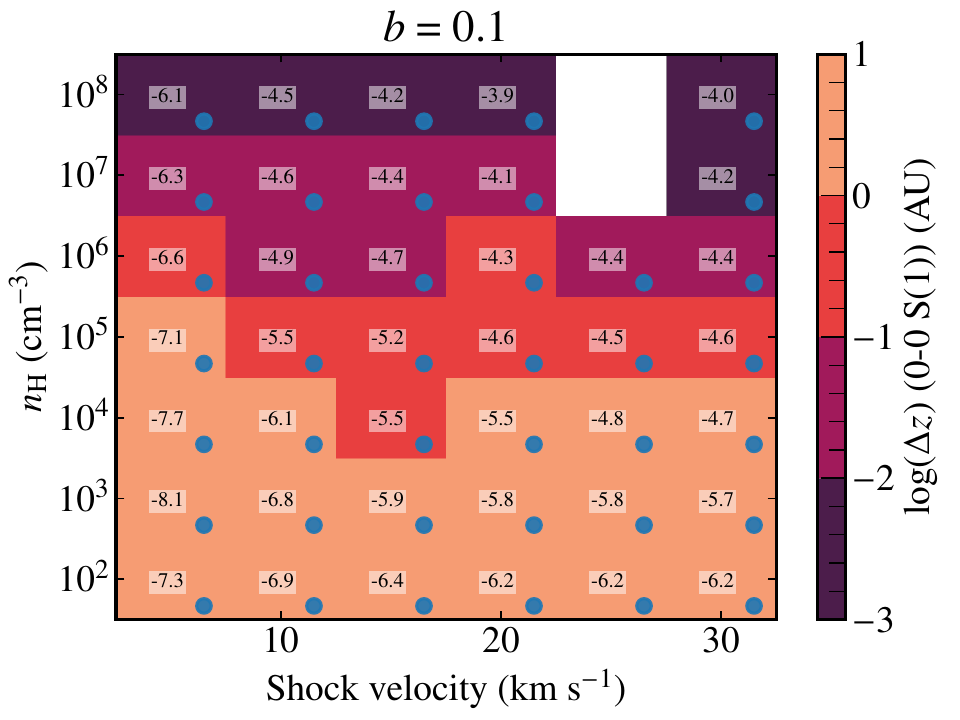}
\includegraphics[width=0.48\textwidth]{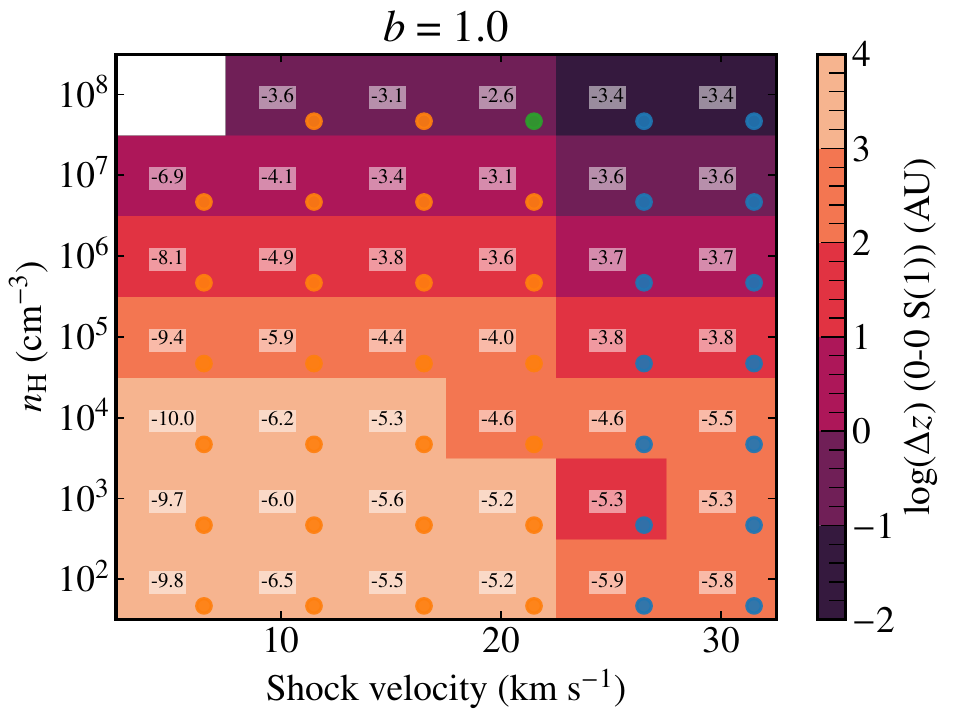}
\includegraphics[width=0.48\textwidth]{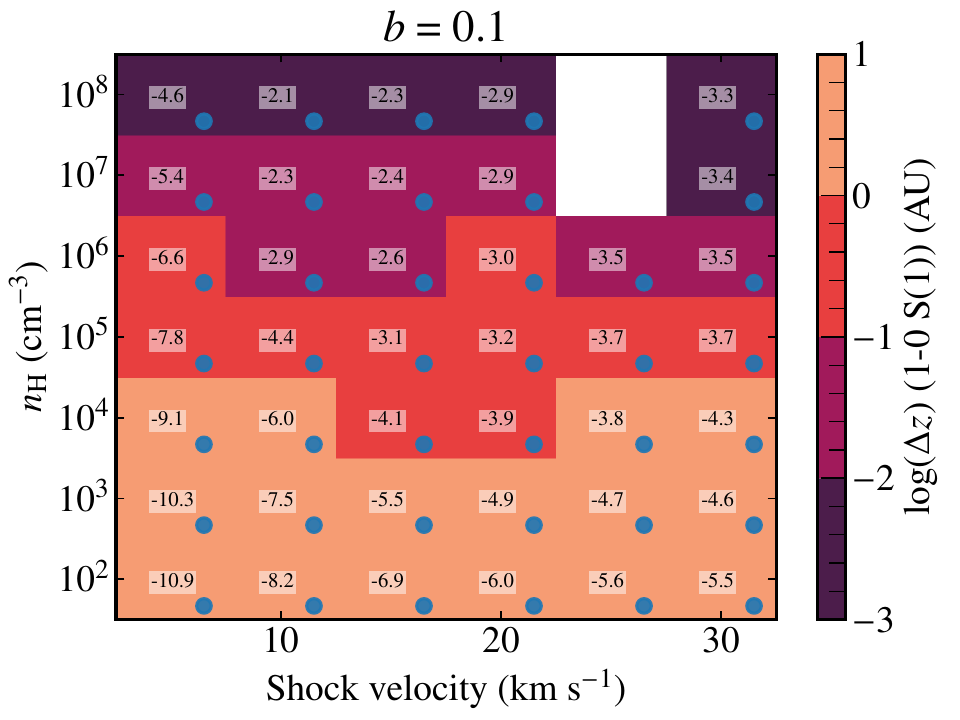}
\includegraphics[width=0.48\textwidth]{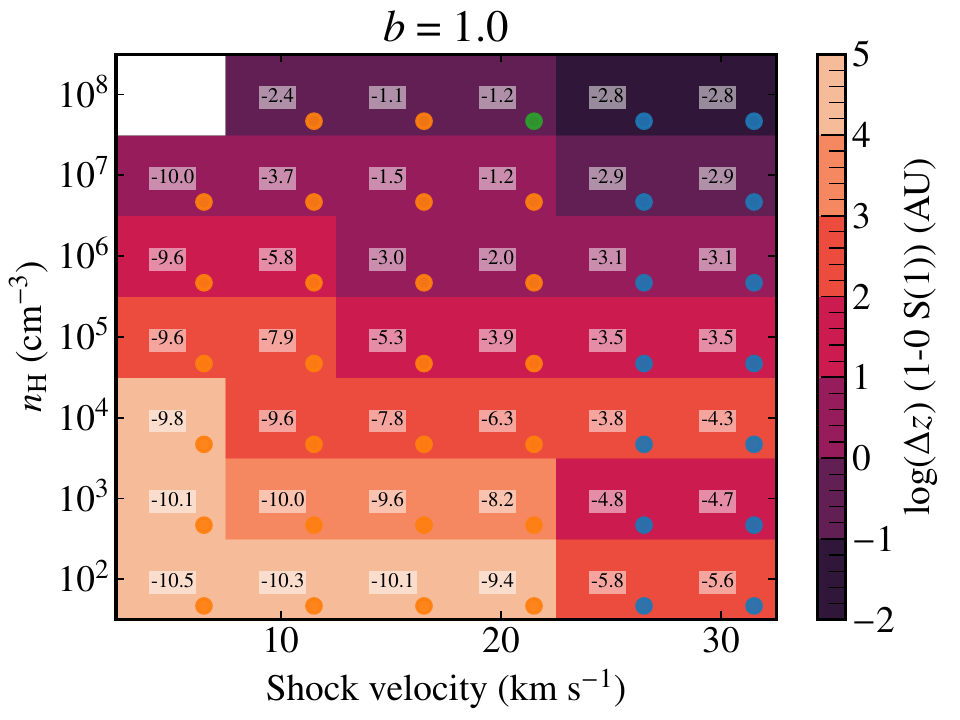}
\caption{
Width of the H$_2$ emitting zone for $b$ = 0.1 (left) and 1.0 (right) for the transitions $\varv$ = 0--0 S(1) (17.05 $\mu$m; top) and 1--0 S(1) (2.12 $\mu$m; bottom). The width is measured as the extent of the region where 80\% of the emission is radiated away. In each cell is written the log of the integrated intensity of the line in units of erg cm$^{-2}$ s$^{-1}$ sr$^{-1}$. The colored dots indicate the resulting shock type, where blue is for J-type shocks and orange for C-type shocks. \label{fig:h2_width}}
\end{figure*}

As mentioned above, the input kinetic energy flux is deposited over a larger spatial range for increasing values of $b$. Specifically, a ``phase transition'' occurs when the resulting shock type goes from being J- to C-type, and a magnetic precursor develops. This typically happens at higher values of $b$ or lower velocities (Fig. \ref{fig:type} shows which physical conditions lead to which shock type). Naturally the ionization fraction also plays a role in setting the shock type (Eq. \ref{eq:cims}), but the gas is primarily neutral for the conditions examined here, and effectively this fraction does not play a role here. To measure the width and to make it a usable observational constraint, we have extracted the scale over which 80\% of the H$_2$ emissivity is generated for a subset of lines: the $\varv$ = 0--0 S(1), 1--0 S(1), 0--0 S(9), 1--0 O(5), and 2--1 S(1) lines. These widths are shown in Fig. \ref{fig:h2_width} together with the integrated intensity of the lines; here we show the widths of the $\varv$ = 0--0 S(1) and 1--0 S(1) emitting regions. The shocks with $b$ = 0.1 all have widths less than 10 AU, whereas the $b$ = 1 shocks have widths up to $\sim$ 10$^5$ AU or $\sim$ 1 pc. For these shocks, there is an anticorrelation between the width and the integrated intensity: the wider shocks have lower integrated intensities. The J-type shocks occurring for $b$ = 1 and $\varv_{\rm s} \ge$ 25 km s$^{-1}$ have larger widths than their $b$ = 0.1 counterparts by one order of magnitude. Even though these are J-type shocks, the magnetic field still plays a significant role.

\subsection{Velocity and density}

\begin{figure*}
\centering
\includegraphics[width=0.48\textwidth]{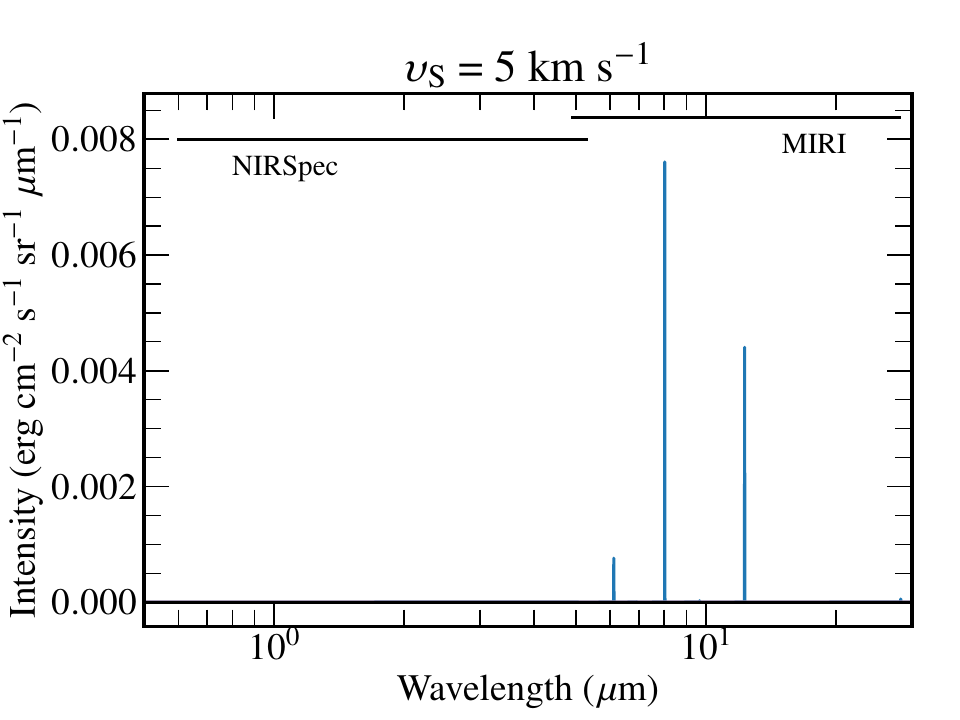}
\includegraphics[width=0.48\textwidth]{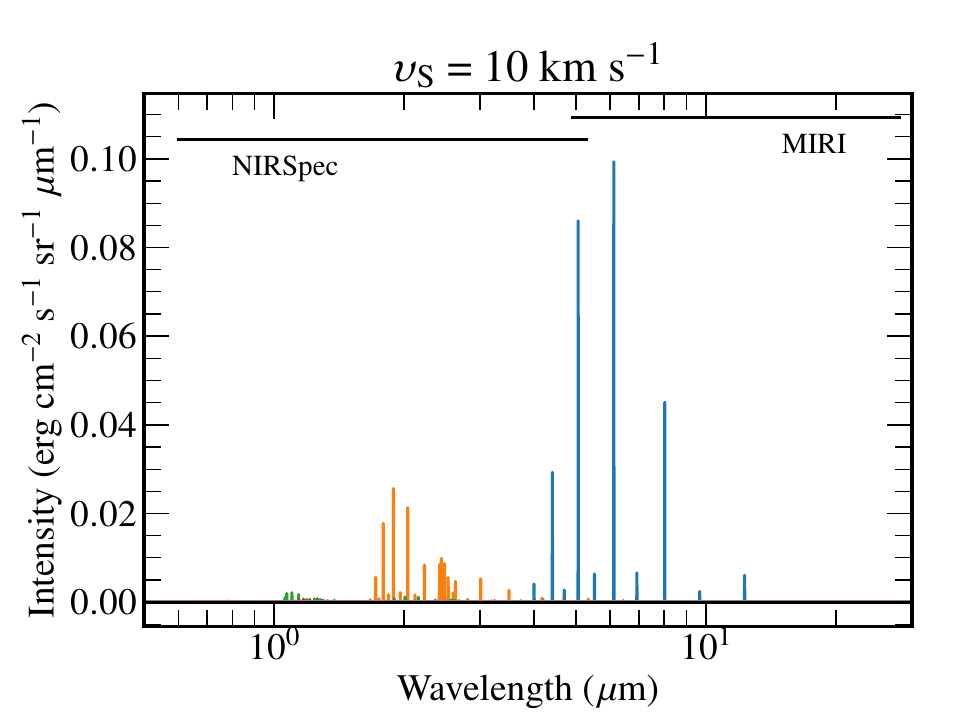}
\includegraphics[width=0.48\textwidth]{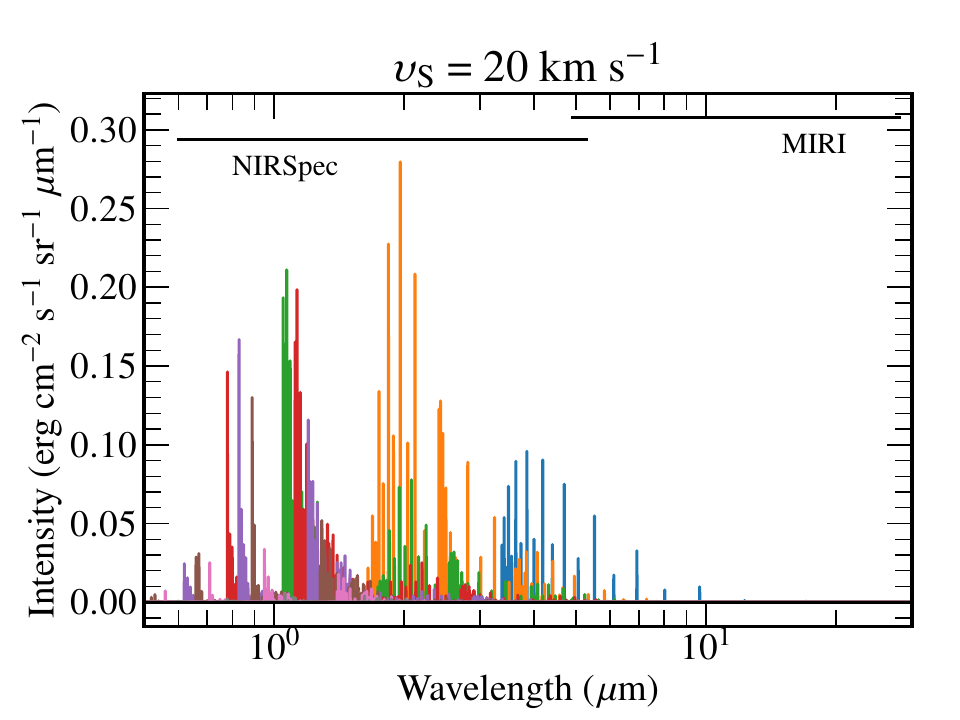}
\includegraphics[width=0.48\textwidth]{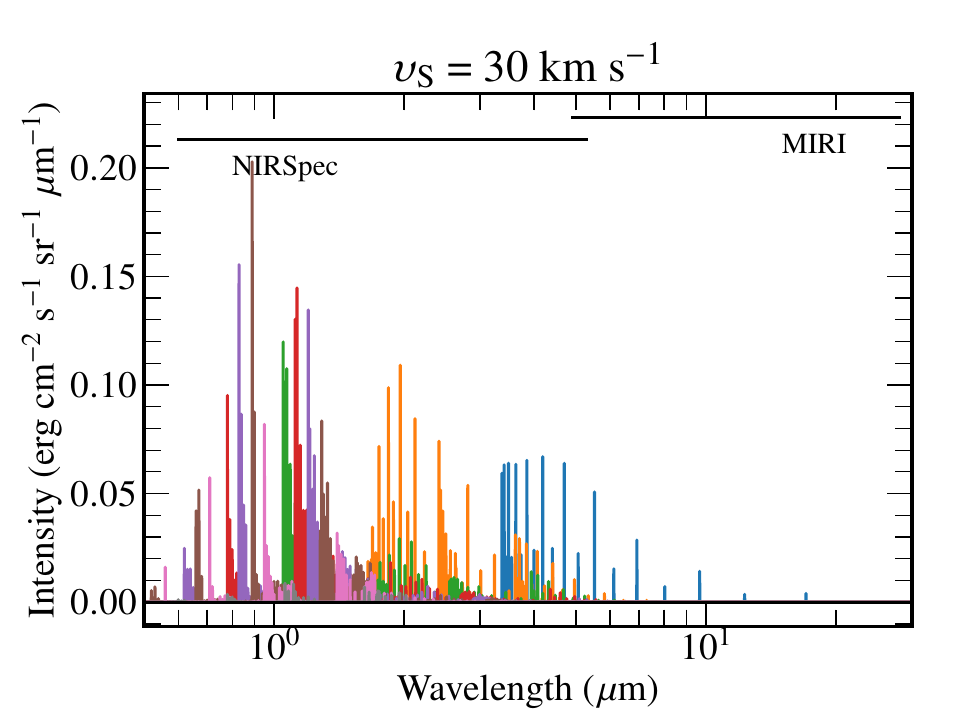}
\caption{H$_2$ spectra for velocities of 5, 10, 20, and 30 km s$^{-1}$ and a density of 10$^{4}$ cm$^{-3}$, $b$ = 0.1, $G_0$ = 0; the shock with $\varv_{\rm s}$ = 20 km s$^{-1}$ is model A, and the 30 km-s$^{-1}$ shock is shown in Fig. \ref{fig:spec_filter}. The colors are for different vibrational levels as in Fig. \ref{fig:h2_cj}. The complete coverage of NIRSpec and MIRI are shown; we refer to Fig. \ref{fig:spec_filter} for the NIRCam and MIRI filter coverage.  \label{fig:spec_vs}}
\end{figure*}

The shock velocity, $\varv_{\rm s}$ sets the maximum temperature in J-type shocks (Eq. \ref{eq:rh_temp}). H$_2$ excitation is sensitive to temperature, and so the velocity effectively sets the excitation. This is seen in the simulated spectra (Fig. \ref{fig:spec_vs}). At the lowest velocity (5 km s$^{-1}$), the integrated intensity is low and only a few rotational lines are seen in the spectrum. On the contrary, at velocities $\gtrsim$ 20 km s$^{-1}$, we see rich vibrational H$_2$ spectra. At the same time the peak specific intensity increases by a factor of $\sim$10, until the velocity reaches 30 km s$^{-1}$ and the shock becomes dissociative. In this case, H$_2$ only contributes to the cooling once it has reformed on the grains. Thus, to a first order, the excitation is set primarily by the velocity in J-type shocks, and the density plays a role in setting the total integrated intensity. 

\begin{figure*}
\centering
\includegraphics[width=0.48\textwidth]{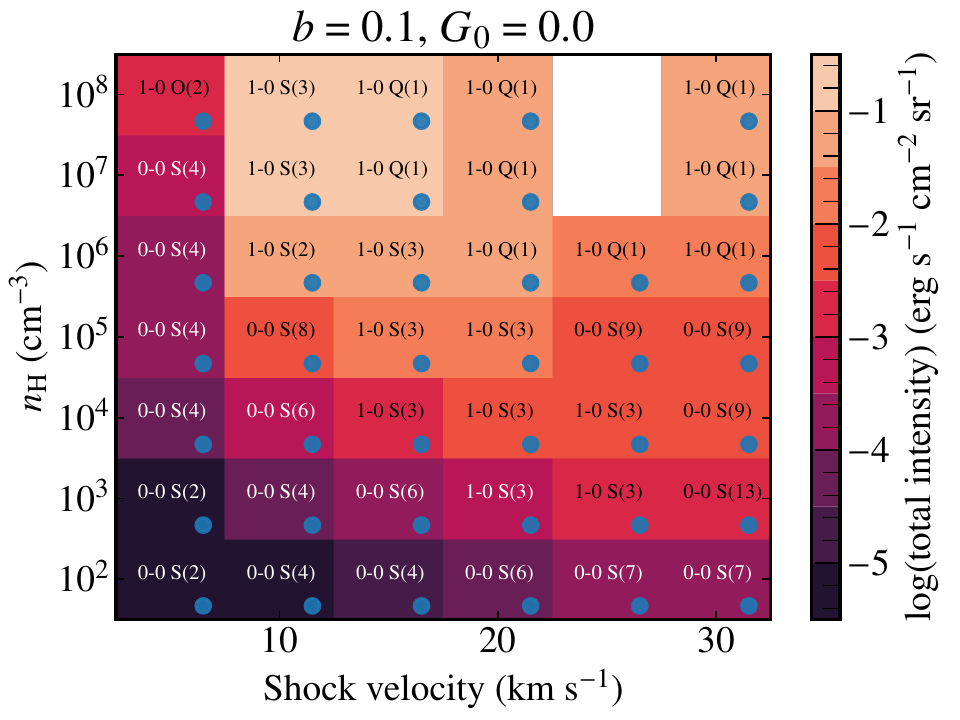}
\includegraphics[width=0.48\textwidth]{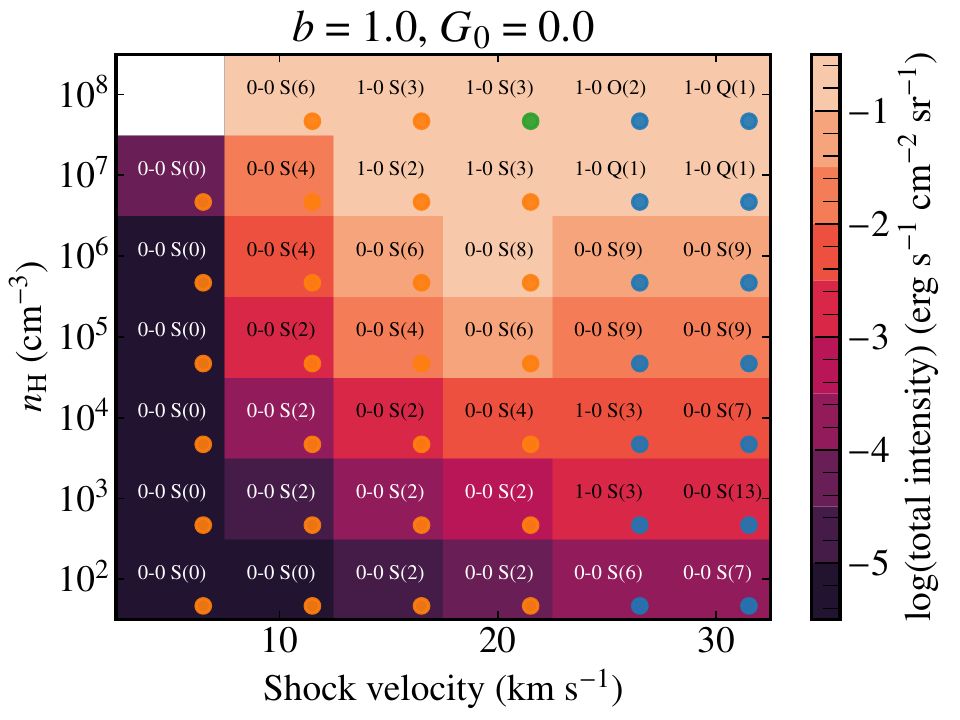}
\includegraphics[width=0.48\textwidth]{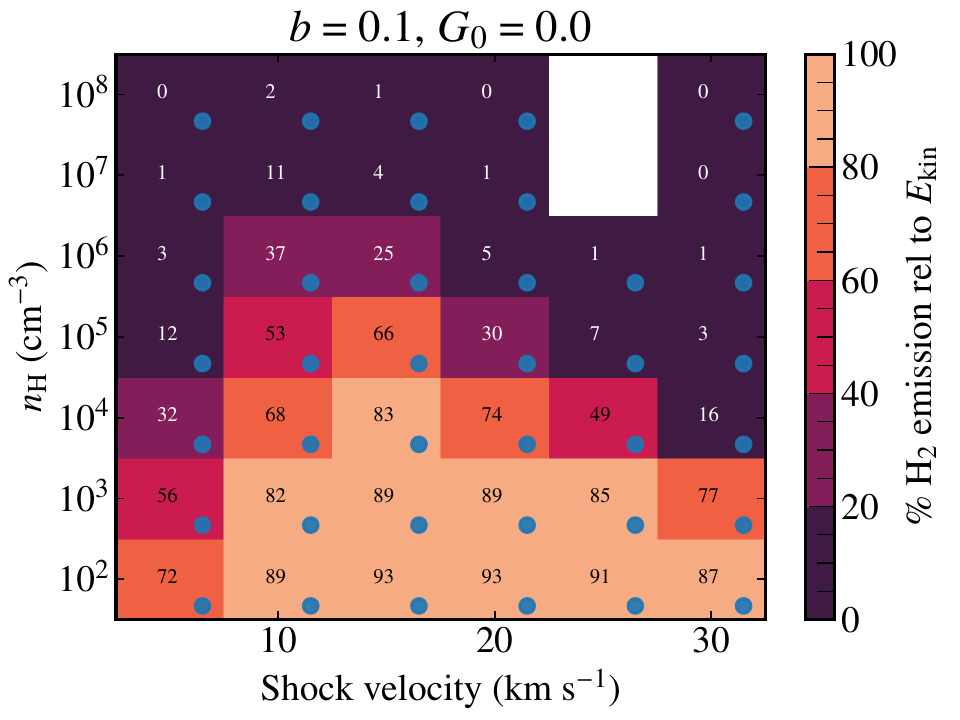}
\includegraphics[width=0.48\textwidth]{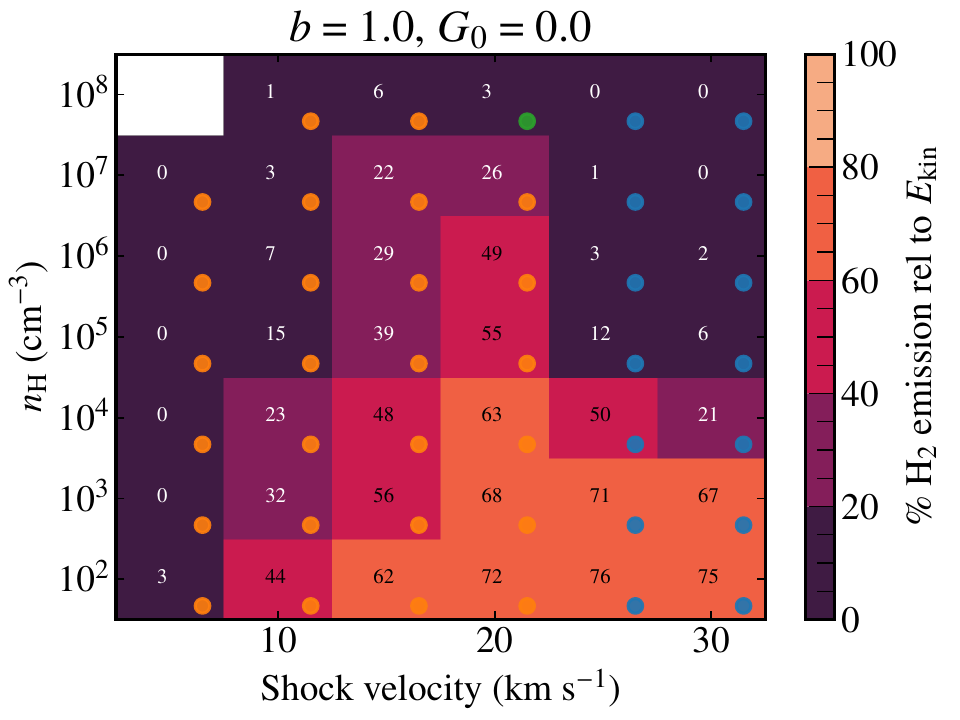}
\caption{
\textit{Top row.} Total H$_2$ integrated intensity in two subgrids of models with $b$ = 0.1 (left) and 1.0 (right), $G_0$ = 0. The colored dots indicate the resulting shock type, where blue is for J-type shocks and orange for C-type shocks (the single green point is a C$^*$ shock). The brightest line is written in each cell. 
\textit{Bottom row.} As the top row, but for the total H$_2$ integrated intensity radiated away compared to the input kinetic energy flux. The actual percentage is provided in each cell. \label{fig:h2_total}}
\end{figure*}

In C-type shocks, the combination of density and velocity is what affects the excitation and the integrated intensity (Fig. \ref{fig:h2_emech}, bottom panel). This is illustrated in the top row of Fig. \ref{fig:h2_total}, which shows the total H$_2$ integrated intensity emitted as well as the brightest line. Here, the brightest line serves as a proxy for the excitation in the sense that the higher excited the brightest line is, the higher the excitation is. For the C-type shocks (orange dots), there is a clear intensity and excitation gradient which depends on both density and velocity. The brightest lines are rotational over the bulk of parameter space (from 0--0 S(0) to S(6)), and they are typically para-H$_2$ transitions (even $J$). For the case of J-type shocks (blue dots), the intensity gradient is dominated by the density, as discussed above. However, the brightest lines quickly become vibrational; the $\varv$ = 1--0 Q(1) line (2.41 $\mu$m) is predicted to be particularly bright, as is the $\varv$ = 1--0 S(3) line (1.96 $\mu$m). Thus, identifying the brightest line in the H$_2$ spectrum provides constraints on where in parameter space the shock is located. Appendix \ref{app:cool} provides an overview of the dominant cooling lines across the grid. 

The H$_2$ fraction in the gas is highest at the lower densities and lower velocities where H$_2$ does not dissociate. However, for a given velocity, the total H$_2$ integrated intensity increases monotonically with density, as shown in Fig. \ref{fig:h2_total}. This is in spite of the fraction of input kinetic energy flux radiated by H$_2$ is monotonically decreasing. Thus, for the shocks with the brightest H$_2$ emission, other molecules and atoms are needed to trace the bulk deposition of kinetic energy. Examples include emission from CO and H$_2$O at lower velocities, and O, S, and H at higher velocities. 

\subsection{UV radiation field}

\begin{figure}
\centering
\includegraphics[width=0.9\columnwidth]{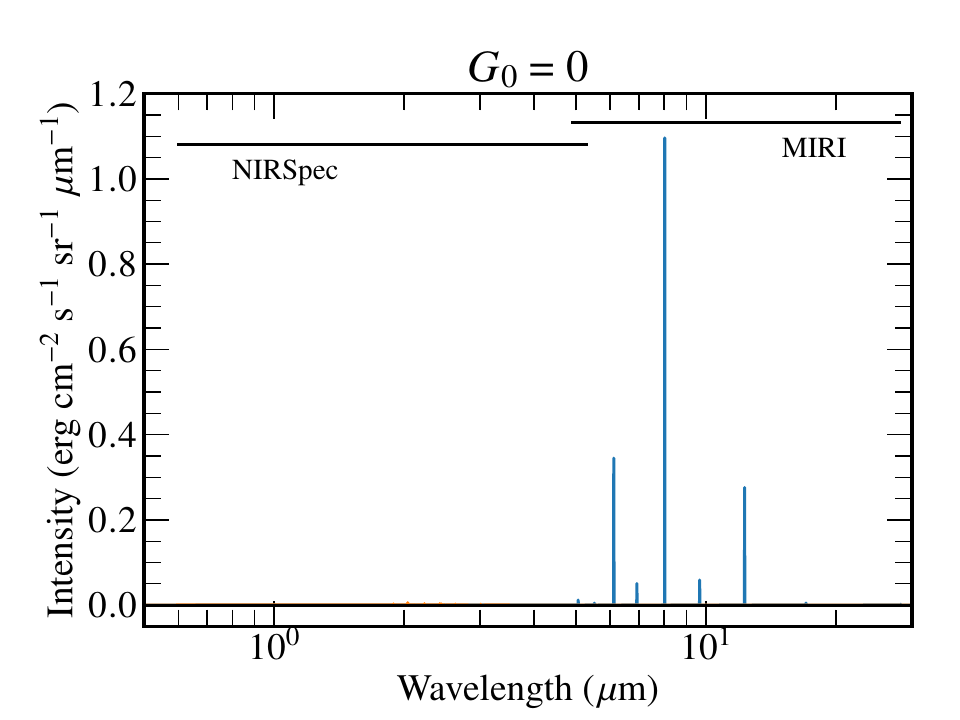}
\includegraphics[width=0.9\columnwidth]{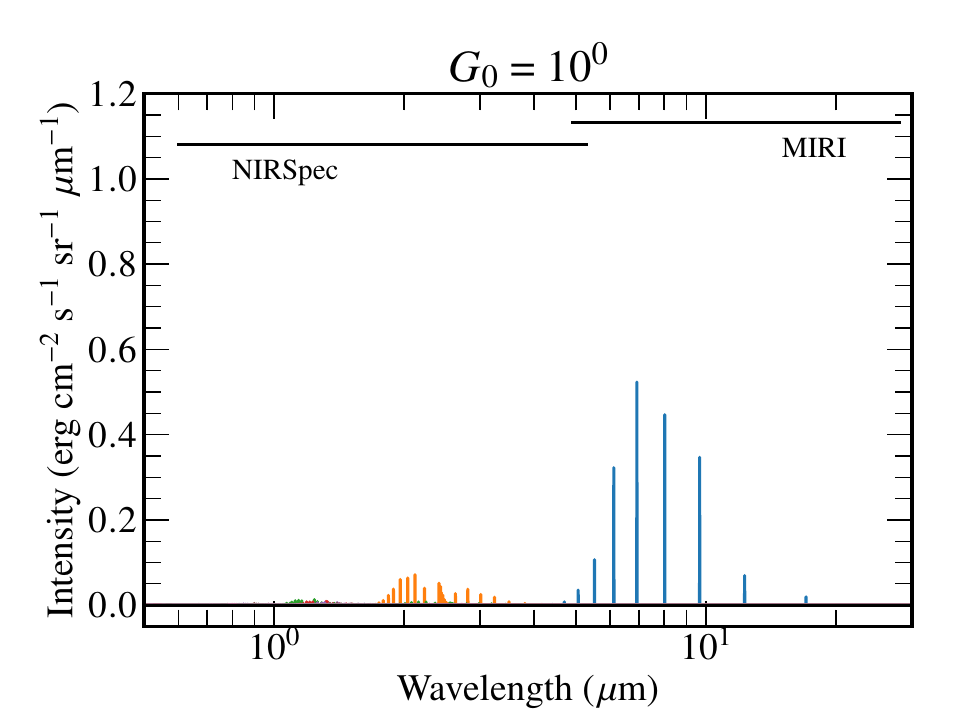}
\includegraphics[width=0.9\columnwidth]{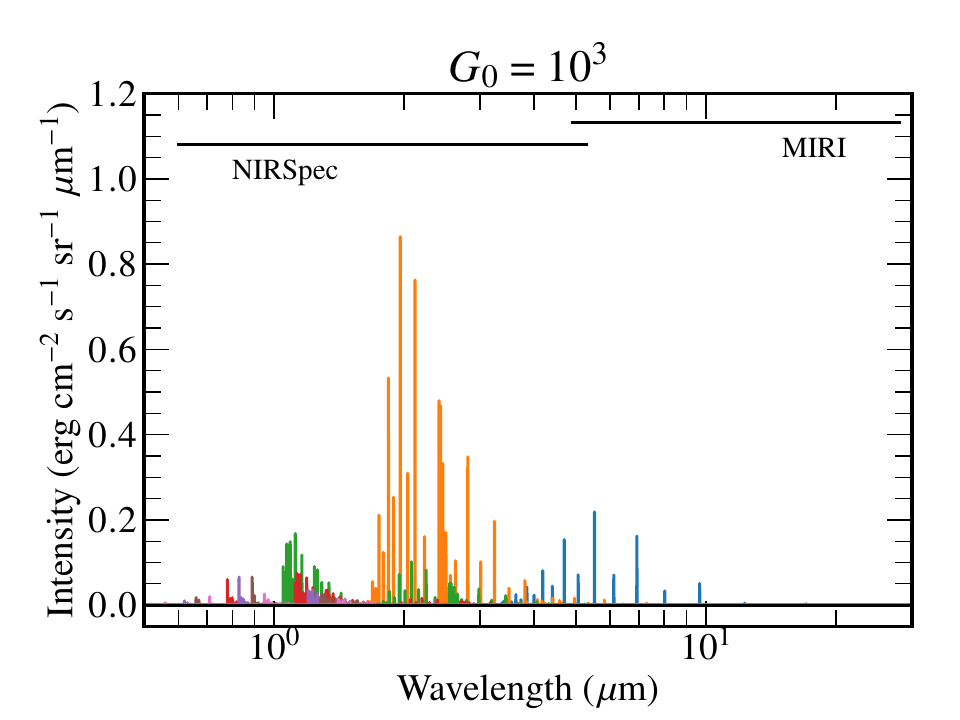}
\caption{H$_2$ spectra for model B but with $G_0$ of 0 (Model B), 1, and 10$^3$. The colors are for different vibrational levels as in Fig. \ref{fig:h2_cj}. The complete coverage of NIRSpec and MIRI are shown; we refer to Fig. \ref{fig:spec_filter} for the NIRCam and MIRI filter coverage. \label{fig:spec_g0}}
\end{figure}

In an externally UV-irradiated shock, the UV photons lead to increased gas ionization and thus higher density of the charged fluid. This increase causes a tighter coupling between the neutral and charged fluids, which in turn leads to the kinetic energy typically being deposited over shorter scales compared to in the absence of external UV radiation. Thus, the temperature typically increases and the shocks become narrower \citep[see Fig. 6 in][]{godard19}. The increased temperature naturally causes higher excitation of H$_2$, as is illustrated in the H$_2$ spectra in Fig. \ref{fig:spec_g0}. Here, the shock in model B, showing pure rotational excitation of H$_2$, is exposed to increasing strengths of an external UV-field, from $G_0$ = 0 to 10$^3$. The increase in temperature (from 1700 K to 2800 K) leads to an increase in excitation, and the vibrational levels start to become populated. 

\begin{figure}
\centering
\includegraphics[width=0.48\textwidth]{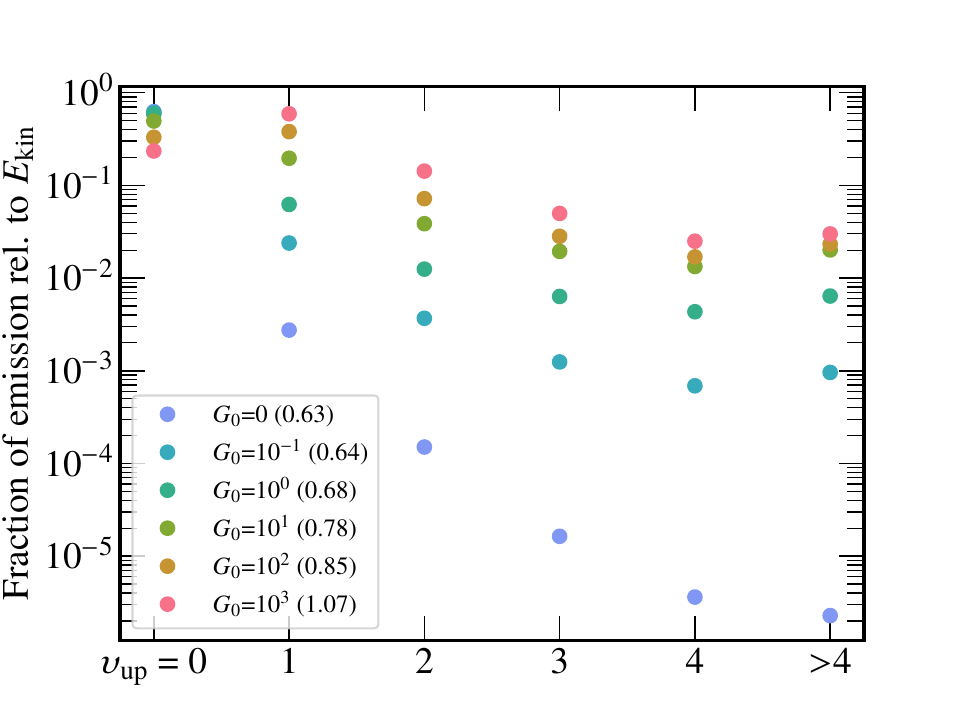}
\caption{Emission distribution for model B with increasing $G_0$ (similar to Fig. \ref{fig:h2_emech}). The total fraction of H$_2$ emission to input kinetic energy flux is provided in the legend. \label{fig:h2_g0}}
\end{figure}

The second effect of the UV field is to deposit additional energy into the shock \citep[Fig. 12 in][]{godard19}. Either this energy deposition is indirect in the form of ionization followed by recombination and release of binding energy, or the energy deposition is direct, where UV photons excite H$_2$ electronically, from which the molecules can de-excite radiatively. It is clear that for the highest values of $G_0$, the additional energetic input is significant. This is illustrated in Fig. \ref{fig:h2_g0}. Here, the energy radiated away by H$_2$ as a function of vibrational level is shown for model B, similar to Fig. \ref{fig:h2_emech}. In this case, model B is exposed to stronger UV fields, and the higher vibrational levels are excited, as also seen in Fig. \ref{fig:spec_g0}. The total fraction of energy lost in H$_2$ emission increases almost monotonically from 0.63 to 1.07 of the input kinetic energy flux. Thus, at least 7\% of the excitation is caused by the UV field, and likely more as there are other channels of energy loss (Fig. \ref{fig:pie_b10}). For a quantitative description of the role of UV pumping on the H$_2$ level populations, we refer to Fig. 8 of \citet{godard19}. 

\begin{figure*}
\centering
\includegraphics[width=0.48\textwidth]{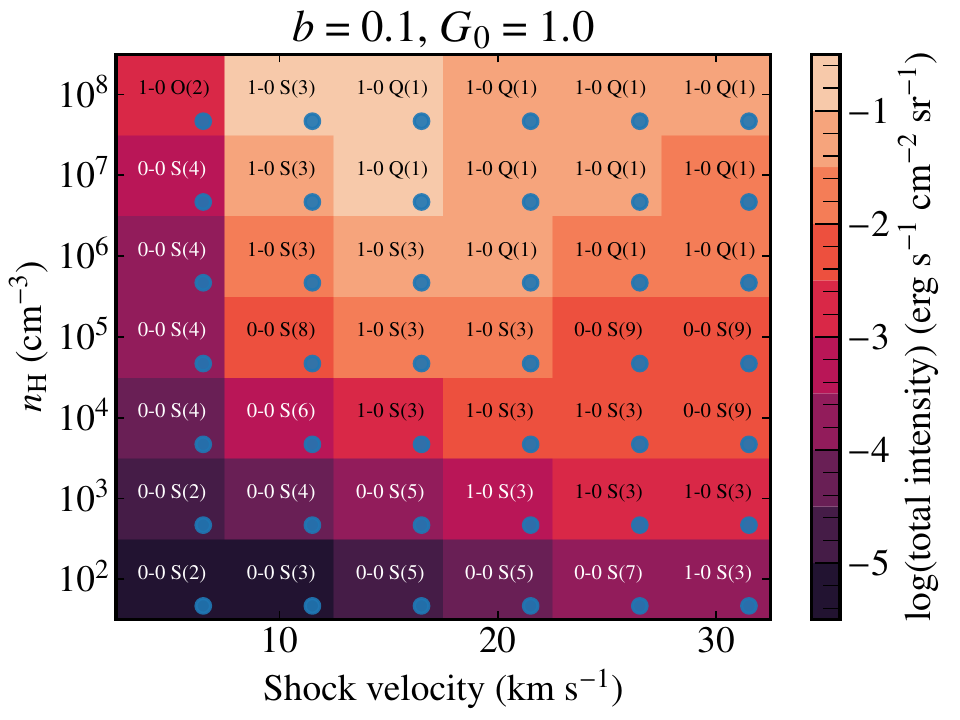}
\includegraphics[width=0.48\textwidth]{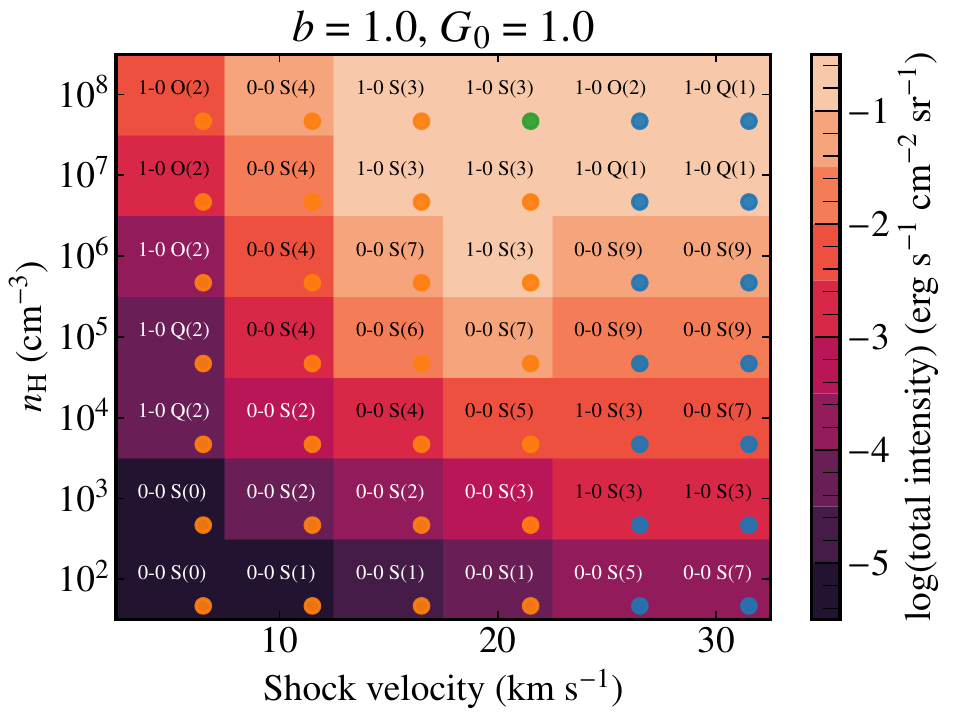}
\includegraphics[width=0.48\textwidth]{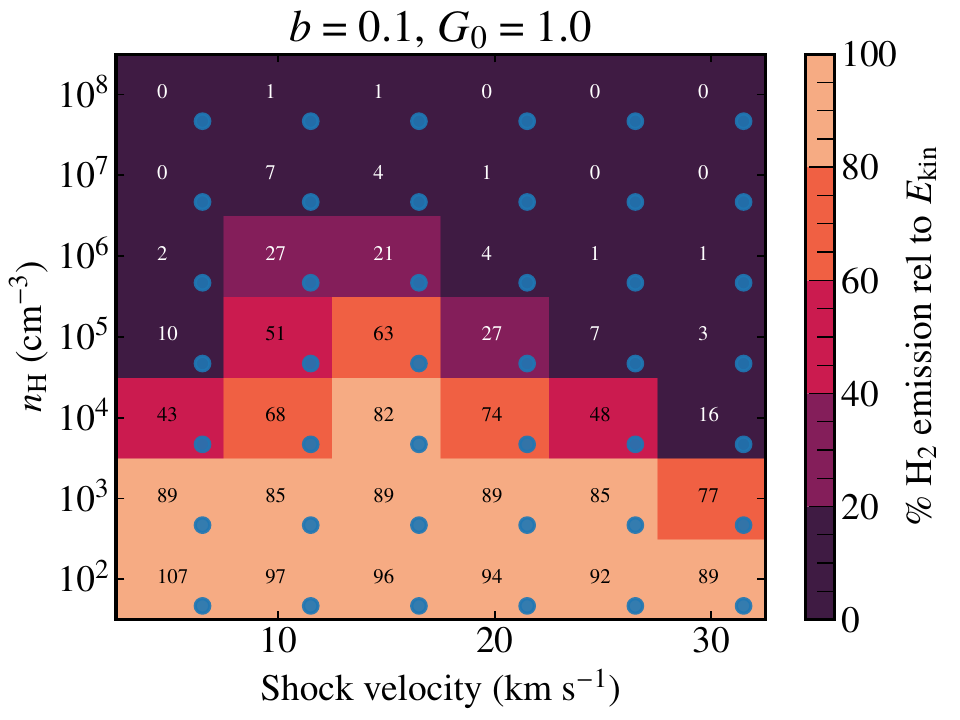}
\includegraphics[width=0.48\textwidth]{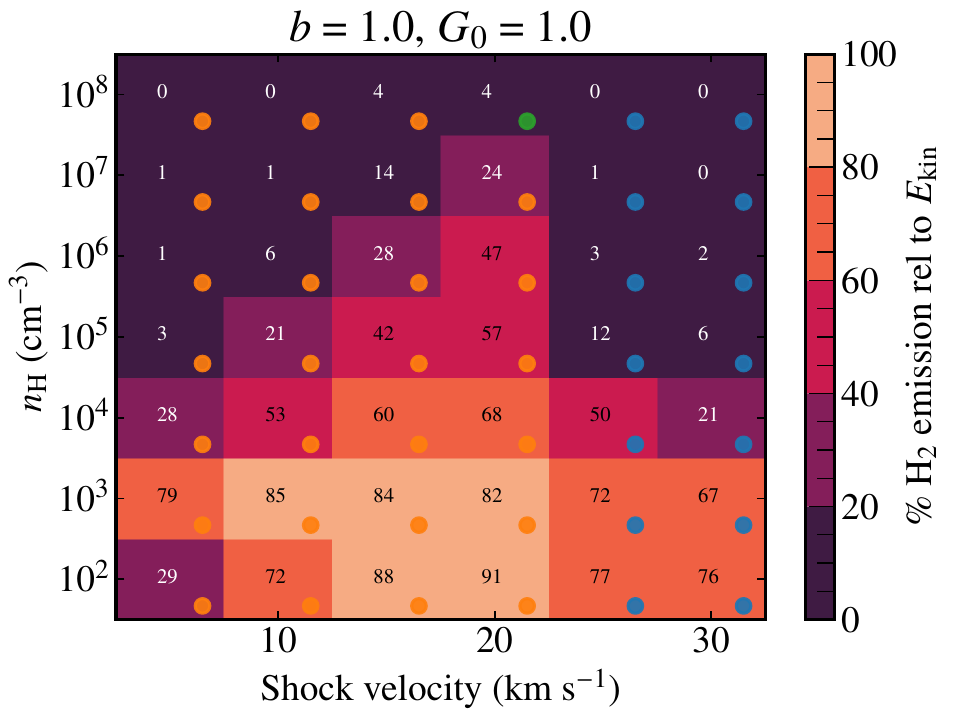}
\caption{As Fig. \ref{fig:h2_total}, but for $G_0$ = 1. \label{fig:h2_total_g0}}
\end{figure*}

Even for relatively weak UV field strengths (e.g., $G_0$ = 1), the UV photons may play a significant role. Figure \ref{fig:h2_total_g0} is similar to Fig. \ref{fig:h2_total} in that the top panels show the total amount of H$_2$ emission and the strongest H$_2$ line. For the weak shocks (low density, low velocity), one major difference is seen when the UV field is turned on: in the absence of external UV radiation, the brightest lines are all para-H$_2$ lines (even $J$) because there is no significant para- to ortho-H$_2$ conversion. For the weak UV field, the strongest lines are predominantly ortho-lines (odd $J$), which is consistent with observations of the diffuse gas in colliding galaxies \citep{Ingalls2011, Guillard2012, Appleton2017}. This suggests that interstellar shocks in general are not fully shielded, but exposed to some UV radiation.

\subsection{H$_2$ excitation for JWST observers} 

JWST represents an increase in sensitivity, spatial and spectral resolution by more than an order of magnitude over previous infrared space-based telescopes \citep{rigby22}. We here outline some of the ways in which the models may be used to plan and interpret the JWST observations of shocked regions, keeping in mind the model limitations listed in Sect. \ref{sect:limits}. 

\textbf{H$_2$ spectroscopy.} The spectroscopic capabilities of NIRSpec and MIRI make them perfectly suited for observing H$_2$ line emission. The excitation of H$_2$ is the result of a complex interplay between various input parameters, as discussed above, with some degeneracies, especially between the density and shock velocity. This is for example illustrated in Fig. 13 of \citet{kristensen07}, where observations of H$_2$ emission from the explosive Orion-KL protostellar outflow are analyzed. With high enough spectral resolution, independent constraints can be made on the shock velocity, thus directly breaking the degeneracy \citep{santangelo14}. 

It will likely not be possible to strongly constrain shock conditions from H$_2$ observations alone, unless the observers only consider subgrids of physical parameters relevant to their studies. An example could be that if shocks in diffuse clouds are studied, only the lowest densities in the grid would be relevant. Furthermore, in a large number of cases, $G_0$ can be independently constrained, for example, by studying ionized gas lines, UV continuum observations, or PAH features at infrared wavelengths. Observers should also be aware that, in shock-dominated environments, the total H$_2$ line emission in a given beam is likely the product of a distribution of shocks arising from a multiphase medium with different conditions. Such an example of shock probability distributions convolved with the use of grids of shock models have been used to interpret H$_2$ observations in the intragroup shocked diffuse gas in colliding galaxies \citep[e.g.,][]{Guillard2009, lesaffre13}. 

\textbf{Shock width.} The NIRCam instrument on JWST is well-suited for observing H$_2$ emission. The instrument contains three categories of filters, narrow-, medium-, and wide-band filters. Their wavelength coverages are illustrated in Fig. \ref{fig:spec_filter}. Of the narrowband filters, three center on H$_2$ lines: F212N ($\varv=1-0$ S(1)), F323N ($\varv=1-0$ O(5)), and F470N ($\varv=0-0$ S(9)). The spatial resolution ranges from 0\farcs07 to 0\farcs16, corresponding to linear scales of 14 and 32 AU at a distance of 200 pc, a typical distance to nearby star-forming regions. As illustrated in Fig. \ref{fig:h2_width}, the width of shocks with $b$ = 1.0 is typically resolvable if the shock is observed close to edge on, except at the highest densities ($\gtrsim$10$^7$ cm$^{-3}$ for C-type shocks, and $\gtrsim$10$^6$ cm$^{-3}$ for J-type shocks). Shocks with $b$ = 0.1 are not resolvable at a distance of 200 pc. Having a measured shock width puts additional constraints on the shock models: the width is sensitive to the strength of the transverse magnetic field and thus serves as an independent constraint of this parameter \citep[Fig. 8 and 9 in][for observations of a spatially resolved bow shock in Orion-KL]{kristensen08}. Besides NIRCam, the MIRI IFU offers the possibility of producing spectral line maps of H$_2$ emission at 160 AU (0."5) spatial resolution at a distance of 200 pc of the 0--0 S(1) line at 17 $\mu$m. Emission from this line traces colder gas, and so is typically more extended than the higher-excited lines shown in Fig. \ref{fig:h2_width}. This resolution is therefore still enough to resolve shock-dominated line emission from dissipative regions in nearby star-forming clouds \citep{richard22}.

\textbf{H$_2$ photometry.} As shown in Fig.~\ref{fig:spec_filter}, the NIRCAM and MIRI imaging filters includes multiple ro-vibrational and rotational H$_2$ lines, so the use of a those filters may prove to be efficient as far as exposure time and mapping area are concerned. Such observations may be used for constraining shock conditions. As an example, Figs. \ref{fig:h2_total} and \ref{fig:h2_total_g0} show the brightest lines for a given set of initial conditions. Thus, if an observed region is dominated by shocked H$_2$ emission, then it might be possible to broadly constrain the range of parameter space where the emission is generated. That is, with the model results in hand, the user can construct ``H$_2$ photometry'' which can be compared to observations, assuming H$_2$ emission dominates the spectrum and the contribution from, e.g., PAH emission is negligible, or assuming that a combination of filters can be used to remove the contribution of the continuum emission. A similar approach has been shown to work efficiently for the wideband MIRI filters for observations of the colliding galaxies in Stephan's Quintet \citep{Appleton2023}.

\textbf{H$_2$ summary.} Table \ref{tab:lessons} summarizes what sets the H$_2$ integrated intensity and the excitation. This table is by no means exhaustive, but may be used as an overview guide of H$_2$ emission in shocks. To constrain the excitation properly, it is necessary to cover as large a wavelength range as possible, and to cover both rotational and rovibrational lines. The former are predominantly excited in C-type shocks, and the latter in J-type shocks. Once a solution has been found that approximately reproduces observations, we recommend the user to fine-tune the grid further for more precise solutions. This can be done either by interpolating the grid values; in this case care must be taken when going from one shock type to another. Alternatively the user can download the model and run their own shock models, in which case we recommend benchmarking their results against the models presented here in a first step. Finally, we recommend that the total integrated intensity of the H$_2$ lines is compared to the total available mechanical energy output from a given source, to ensure that the best-fit shock model is physical \citep[see][for the methodology]{lehmann22}. 

\textbf{Atomic lines.} Apart from H$_2$ emission, the model calculates line emission from several other atomic and ionic species. As an example, JWST-MIRI will observe the [\ion{S}{i}] line at 25 $\mu$m \citep[e.g., toward the nearby protostellar outflow from IRAS15398,][]{yang22}, and the integrated line intensity of this line is calculated and tabulated from the grid. The same applies to lines from other species, e.g., O, and C. Naturally, these lines light up in different parts of parameter space compared to H$_2$, and thus provide complementary information. 

\textbf{Other emission lines.} The abundances of some 140 other species have been calculated through the shock. Examples of particular relevance to JWST and shocks include Fe$^+$, OH and H$_2$O, because these species have a number of transitions visible in the NIRSpec and MIRI wavelength ranges and these species are some of the dominant coolants  (e.g., Fig. \ref{fig:pie_b01} and \ref{fig:pie_b10}). The abundance, temperature, and density profiles are calculated through the shock, which means that the profiles can be post-processed to calculate integrated line intensities using for example a large velocity gradient (LVG) radiative transfer code \citep[e.g.,][]{gusdorf11}, which has not been done for this grid of models. Just as for the atomic lines, these will provide complementary observational constraints. 

\begin{table*}[!t]
    \caption{Summary of H$_2$ excitation in shocks. \label{tab:lessons}}
    \centering
    \begin{tabular}{l l}
\hline \hline
Parameter & Role \\ \hline
$b$ & With $\varv_{\rm s}$, determines if a shock is J- or C-type \\
    & J-type: $T$ $\propto$ $\varv_{\rm s}^2$, and the shock is narrow \\
    & C-type: energy is dissipated over large scales, lower $T_{\rm max}$ \\ \hline
$\varv_{\rm s}$ & With $n_{\rm H}$ sets $E_{\rm kin} = 1/2\ \rho \varv_{\rm s}^3$ ($\rho = 1.4 m_{\rm H} n_{\rm H}$) \\
    & J-type: velocity sets $T$leading to excitation \\
    & C-type: sets $T$ together with $n_{\rm H}$, spectrum dominated by rotational lines, but intensity increases with $\varv_{\rm s}$ ($E_{\rm kin}$) \\ \hline
$n_{\rm H}$ 
    & J-type: sets intensity (up to the point where H$_2$O and OH take over, or H$_2$ is dissociated) \\ 
    & C-type: sets intensity, and at the highest $n_{\rm H}$ ($\ge$10$^6$ cm$^{-3}$) also excitation \\ \hline
$G_0$ & 
    J-type: little to no effect, except at lowest $\varv_{\rm s}$ where intensity increases \\
    & C-type: Changes excitation when $E_{\rm UV} \gtrsim E_{\rm kin}$ \\ \hline
    \end{tabular}
\end{table*}

\section{Summary}\label{sect:summary}

Here we present the results of an extensive grid of plane-parallel steady-state shock models. The grid was constructed by varying six parameters: the preshock density, shock velocity, strength of the transverse magnetic field, strength of the UV field impinging on the shock, the cosmic-ray-ionization rate, and the PAH abundance. This is the first time such an extensive grid of shock models has been run and made publicly available. 

The purpose of running this grid of models was to examine under which shock conditions H$_2$ is efficiently excited, and how shock conditions affect the H$_2$ excitation and integrated line intensities. H$_2$ is already being extensively observed with JWST, and the coming years will see a flood of H$_2$ observations. At the moment it is therefore critical for planning and interpreting JWST observations. 

We find that the strength of the transverse magnetic field, as quantified by the magnetic scaling factor, $b$, plays a key role in the excitation of H$_2$. At low values of $b$ ($\lesssim$ 0.3, J-type shocks), H$_2$ excitation is dominated by vibrationally excited lines; whereas, at higher values ($b$ $\gtrsim$ 1, C-type shocks), rotational lines dominate the spectrum for shocks without an external radiation field. Shocks with $b$ $\ge$ 1 can potentially be spatially resolved with JWST for nearby objects, which serves as an additional constraint. 

H$_2$ is typically the dominant coolant at lower densities ($\lesssim$ 10$^4$ cm$^{-3}$); at higher densities, other molecules such as CO, OH, and H$_2$O take over at velocities $\lesssim$ 20 km s$^{-1}$ and atoms, for example, H, O, and S, dominate at higher velocities. Together, the velocity and density set the input kinetic energy flux. When this increases, the excitation and integrated intensity of H$_2$ increases similarly. 

An external UV field mainly serves to increase the excitation, particularly for shocks where the input radiation energy is comparable to or greater than the input kinetic energy flux. Together, these results provide an overview of the energetic reprocessing of input energy and the resulting H$_2$ line emission observable by JWST.

\begin{acknowledgements}
We would like to thank F. Boulanger and S. Cabrit for simulating discussions, particularly at the beginning of this project, as well as J. A. Villa V{\'e}lez. The research leading to these results has received funding from the European Research Council, under the European Community’s Seventh framework Programme, through the Advanced Grant MIST (FP7/2017–2022, No. 742719). The grid of simulations used in this work has been run on the computing cluster Totoro of the ERC MIST, administered by MesoPSL. We would also like to acknowledge the support from the Programme National ``Physique et Chimie du Milieu Interstellaire'' (PCMI) of CNRS/INSU with INC/INP co-funded by CEA and CNES. The research of LEK is supported by a research grant (19127) from VILLUM FONDEN. PG would like to thank the Sorbonne University, the Institut Universitaire de France, the Centre National d'Etudes Spatiales (CNES), the ``Programme National de Cosmologie and Galaxies'' (PNCG). This work has made use of the Paris-Durham public shock code V1.1, distributed by the CNRS-INSU National Service ``ISM Platform'' at the Paris Observatory Data Center\footnote{\url{http://ism.obspm.fr}}. 
\end{acknowledgements}

\bibliographystyle{aa} 
\bibliography{bibliography} 

\begin{thebibliography}{60}
\expandafter\ifx\csname natexlab\endcsname\relax\def\natexlab#1{#1}\fi

\bibitem[{{{\'A}lvarez-M{\'a}rquez} {et~al.}(2022){{\'A}lvarez-M{\'a}rquez},
  {Labiano}, {Guillard}, {Dicken}, {Argyriou}, {Patapis}, {Law}, {Kavanagh},
  {Larson}, {Gasman}, {Mueller}, {Alberts}, {Brandl}, {Colina},
  {Garc{\'\i}a-Mar{\'\i}n}, {Jones}, {Noriega-Crespo}, {Shivaei}, {Temim}, \&
  {Wright}}]{Alvarez-Marquez2022}
{{\'A}lvarez-M{\'a}rquez}, J., {Labiano}, A., {Guillard}, P., {et~al.} 2022,
  arXiv e-prints, arXiv:2209.01695

\bibitem[{{Appleton} {et~al.}(2023){Appleton}, {Guillard}, {Emonts},
  {Boulanger}, {Togi}, {Reach}, {Alatalo}, {Cluver}, {Diaz Santos}, {Duc},
  {Gallagher}, {Ogle}, {O'Sullivan}, {Voggel}, \& {Xu}}]{Appleton2023}
{Appleton}, P.~N., {Guillard}, P., {Emonts}, B., {et~al.} 2023, arXiv e-prints,
  arXiv:2301.02928

\bibitem[{{Appleton} {et~al.}(2017){Appleton}, {Guillard}, {Togi}, {Alatalo},
  {Boulanger}, {Cluver}, {Pineau des For{\^e}ts}, {Lisenfeld}, {Ogle}, \&
  {Xu}}]{Appleton2017}
{Appleton}, P.~N., {Guillard}, P., {Togi}, A., {et~al.} 2017, \apj, 836, 76

\bibitem[{Appleton {et~al.}(2006)Appleton, Xu, Reach, Dopita, Gao, Lu, Popescu,
  Sulentic, Tuffs, \& Yun}]{Appleton2006}
Appleton, P.~N., Xu, K.~C., Reach, W., {et~al.} 2006, \apjl, 639, L51

\bibitem[{{Bally}(2016)}]{bally16}
{Bally}, J. 2016, \araa, 54, 491

\bibitem[{{Bern{\'e}} {et~al.}(2022){Bern{\'e}}, {Habart}, {Peeters},
  {Abergel}, {Bergin}, {Bernard-Salas}, {Bron}, {Cami}, {Dartois}, {Fuente},
  {Goicoechea}, {Gordon}, {Okada}, {Onaka}, {Robberto}, {R{\"o}llig},
  {Tielens}, {Vicente}, {Wolfire}, {Alarc{\'o}n}, {Boersma}, {Canin}, {Chown},
  {Dicken}, {Languignon}, {Le Gal}, {Pound}, {Trahin}, {Simmer}, {Sidhu}, {Van
  De Putte}, {Cuadrado}, {Guilloteau}, {Maragkoudakis}, {Schefter}, {Schirmer},
  {Cazaux}, {Aleman}, {Allamandola}, {Auchettl}, {Baratta}, {Bejaoui}, {Bera},
  {Bilalbegovi{\'c}}, {Black}, {Boulanger}, {Bouwman}, {Brandl}, {Brechignac},
  {Br{\"u}nken}, {Burkhardt}, {Candian}, {Cernicharo}, {Chabot}, {Chakraborty},
  {Champion}, {Colgan}, {Cooke}, {Coutens}, {Cox}, {Demyk}, {Donovan Meyer},
  {Engrand}, {Foschino}, {Garc{\'\i}a-Lario}, {Gavilan}, {Gerin}, {Godard},
  {Gottlieb}, {Guillard}, {Gusdorf}, {Hartigan}, {He}, {Herbst}, {Hornekaer},
  {J{\"a}ger}, {Janot-Pacheco}, {Joblin}, {Kaufman}, {Kemper}, {Kendrew},
  {Kirsanova}, {Klaassen}, {Knight}, {Kwok}, {Labiano}, {Lai}, {Lee},
  {Lefloch}, {Le Petit}, {Li}, {Linz}, {Mackie}, {Madden}, {Mascetti},
  {McGuire}, {Merino}, {Micelotta}, {Misselt}, {Morse}, {Mulas}, {Neelamkodan},
  {Ohsawa}, {Omont}, {Paladini}, {Palumbo}, {Pathak}, {Pendleton},
  {Petrignani}, {Pino}, {Puga}, {Rangwala}, {Rapacioli}, {Ricca},
  {Roman-Duval}, {Roser}, {Roueff}, {Rouill{\'e}}, {Salama}, {Sales},
  {Sandstrom}, {Sarre}, {Sciamma-O'Brien}, {Sellgren}, {Shannon}, {Shenoy},
  {Teyssier}, {Thomas}, {Togi}, {Verstraete}, {Witt}, {Wootten}, {Ysard},
  {Zettergren}, {Zhang}, {Zhang}, \& {Zhen}}]{berne22}
{Bern{\'e}}, O., {Habart}, {\'E}., {Peeters}, E., {et~al.} 2022, \pasp, 134,
  054301

\bibitem[{{Dionatos} {et~al.}(2010){Dionatos}, {Nisini}, {Cabrit},
  {Kristensen}, \& {Pineau des For{\^e}ts}}]{dionatos10}
{Dionatos}, O., {Nisini}, B., {Cabrit}, S., {Kristensen}, L., \& {Pineau des
  For{\^e}ts}, G. 2010, \aap, 521, A7

\bibitem[{{Draine}(1980)}]{draine80}
{Draine}, B.~T. 1980, \apj, 241, 1021

\bibitem[{{Flower}(1997)}]{flower97}
{Flower}, D.~R. 1997, \mnras, 288, 627

\bibitem[{{Flower} {et~al.}(1985){Flower}, {Pineau des For{\^ e}ts}, \&
  {Hartquist}}]{flower85}
{Flower}, D.~R., {Pineau des For{\^ e}ts}, G., \& {Hartquist}, T.~W. 1985,
  \mnras, 216, 775

\bibitem[{{Flower} \& {Pineau des For{\^e}ts}(2010)}]{flower10}
{Flower}, D.~R. \& {Pineau des For{\^e}ts}, G. 2010, \mnras, 912

\bibitem[{{Flower} \& {Pineau des For{\^e}ts}(2013)}]{flower13}
{Flower}, D.~R. \& {Pineau des For{\^e}ts}, G. 2013, \mnras, 436, 2143

\bibitem[{{Flower} \& {Pineau des For{\^e}ts}(2015)}]{flower15}
{Flower}, D.~R. \& {Pineau des For{\^e}ts}, G. 2015, \aap, 578, A63

\bibitem[{{Flower} {et~al.}(2006){Flower}, {Pineau des For{\^e}ts}, \&
  {Walmsley}}]{flower06}
{Flower}, D.~R., {Pineau des For{\^e}ts}, G., \& {Walmsley}, C.~M. 2006, \aap,
  449, 621

\bibitem[{{Flower} \& {Roueff}(1998{\natexlab{a}})}]{flowerroueff98}
{Flower}, D.~R. \& {Roueff}, E. 1998{\natexlab{a}}, Journal of Physics B Atomic
  Molecular Physics, 31, 2935

\bibitem[{{Flower} \& {Roueff}(1998{\natexlab{b}})}]{flower98}
{Flower}, D.~R. \& {Roueff}, E. 1998{\natexlab{b}}, {Journal of Physics B}, 31,
  2935

\bibitem[{{Flower} {et~al.}(1998){Flower}, {Roueff}, \& {Zeippen}}]{flower98a}
{Flower}, D.~R., {Roueff}, E., \& {Zeippen}, C.~J. 1998, Journal of Physics B
  Atomic Molecular Physics, 31, 1105

\bibitem[{{Froebrich} {et~al.}(2015){Froebrich}, {Makin}, {Davis}, {Gledhill},
  {Kim}, {Koo}, {Rowles}, {Eisl{\"o}ffel}, {Nicholas}, {Lee}, {Williamson}, \&
  {Buckner}}]{froebrich15}
{Froebrich}, D., {Makin}, S.~V., {Davis}, C.~J., {et~al.} 2015, \mnras, 454,
  2586

\bibitem[{{Garc{\'\i}a-Bernete} {et~al.}(2022){Garc{\'\i}a-Bernete},
  {Rigopoulou}, {Alonso-Herrero}, {Donnan}, {Roche}, {Pereira-Santaella},
  {Labiano}, {Peralta de Arriba}, {Izumi}, {Ramos Almeida}, {Shimizu},
  {H{\"o}nig}, {Garc{\'\i}a-Burillo}, {Rosario}, {Ward}, {Bellocchi}, {Hicks},
  {Fuller}, \& {Packham}}]{garcia-bernete22}
{Garc{\'\i}a-Bernete}, I., {Rigopoulou}, D., {Alonso-Herrero}, A., {et~al.}
  2022, \aap, 666, L5

\bibitem[{{Godard} {et~al.}(2019){Godard}, {Pineau des For{\^e}ts}, {Lesaffre},
  {Lehmann}, {Gusdorf}, \& {Falgarone}}]{godard19}
{Godard}, B., {Pineau des For{\^e}ts}, G., {Lesaffre}, P., {et~al.} 2019, \aap,
  622, A100

\bibitem[{{Goldader} {et~al.}(1997){Goldader}, {Joseph}, {Doyon}, \&
  {Sanders}}]{Goldader1997}
{Goldader}, J.~D., {Joseph}, R.~D., {Doyon}, R., \& {Sanders}, D.~B. 1997,
  \apj, 474, 104

\bibitem[{{Guillard} {et~al.}(2009){Guillard}, {Boulanger}, {Pineau des
  For{\^e}ts}, \& {Appleton}}]{Guillard2009}
{Guillard}, P., {Boulanger}, F., {Pineau des For{\^e}ts}, G., \& {Appleton},
  P.~N. 2009, \aa, 502, 515

\bibitem[{{Guillard} {et~al.}(2012){Guillard}, {Ogle}, {Emonts}, {Appleton},
  {Morganti}, {Tadhunter}, {Oosterloo}, {Evans}, \& {Evans}}]{Guillard2012}
{Guillard}, P., {Ogle}, P.~M., {Emonts}, B.~H.~C., {et~al.} 2012, \apj, 747, 95

\bibitem[{{Guillet} {et~al.}(2009){Guillet}, {Jones}, \& {Pineau des
  For{\^e}ts}}]{guillet09}
{Guillet}, V., {Jones}, A.~P., \& {Pineau des For{\^e}ts}, G. 2009, \aap, 497,
  145

\bibitem[{{Guillet} {et~al.}(2011){Guillet}, {Pineau des For{\^e}ts}, \&
  {Jones}}]{guillet11}
{Guillet}, V., {Pineau des For{\^e}ts}, G., \& {Jones}, A.~P. 2011, \aap, 527,
  A123

\bibitem[{{Gusdorf} {et~al.}(2011){Gusdorf}, {Giannini}, {Flower}, {Parise},
  {G{\"u}sten}, \& {Kristensen}}]{gusdorf11}
{Gusdorf}, A., {Giannini}, T., {Flower}, D.~R., {et~al.} 2011, \aap, 532, A53

\bibitem[{{Gusdorf} {et~al.}(2008){Gusdorf}, {Pineau des For{\^e}ts}, {Cabrit},
  \& {Flower}}]{gusdorf08b}
{Gusdorf}, A., {Pineau des For{\^e}ts}, G., {Cabrit}, S., \& {Flower}, D.~R.
  2008, \aap, 490, 695

\bibitem[{{Gustafsson} {et~al.}(2010){Gustafsson}, {Ravkilde}, {Kristensen},
  {Cabrit}, {Field}, \& {Pineau des For{\^e}ts}}]{gustafsson10}
{Gustafsson}, M., {Ravkilde}, T., {Kristensen}, L.~E., {et~al.} 2010, \aap,
  513, A5

\bibitem[{{Hollenbach} \& {McKee}(1989)}]{hollenbach89}
{Hollenbach}, D. \& {McKee}, C.~F. 1989, \apj, 342, 306

\bibitem[{{Ingalls} {et~al.}(2011){Ingalls}, {Bania}, {Boulanger}, {Draine},
  {Falgarone}, \& {Hily-Blant}}]{Ingalls2011}
{Ingalls}, J.~G., {Bania}, T.~M., {Boulanger}, F., {et~al.} 2011, \apj, 743,
  174

\bibitem[{{Kaufman} \& {Neufeld}(1996)}]{kaufman96b}
{Kaufman}, M.~J. \& {Neufeld}, D.~A. 1996, \apj, 456, 611

\bibitem[{{Kristensen} {et~al.}(2007){Kristensen}, {Ravkilde}, {Field},
  {Lemaire}, \& {Pineau des For{\^e}ts}}]{kristensen07}
{Kristensen}, L.~E., {Ravkilde}, T.~L., {Field}, D., {Lemaire}, J.~L., \&
  {Pineau des For{\^e}ts}, G. 2007, \aap, 469, 561

\bibitem[{{Kristensen} {et~al.}(2008){Kristensen}, {Ravkilde}, {Pineau des
  For{\^e}ts}, {Cabrit}, {Field}, {Gustafsson}, {Diana}, \&
  {Lemaire}}]{kristensen08}
{Kristensen}, L.~E., {Ravkilde}, T.~L., {Pineau des For{\^e}ts}, G., {et~al.}
  2008, \aap, 477, 203

\bibitem[{{Le Bourlot} {et~al.}(1999){Le Bourlot}, {Pineau des For{\^e}ts}, \&
  {Flower}}]{lebourlot99}
{Le Bourlot}, J., {Pineau des For{\^e}ts}, G., \& {Flower}, D.~R. 1999, \mnras,
  305, 802

\bibitem[{{Lehmann} {et~al.}(2020){Lehmann}, {Godard}, {Pineau des For{\^e}ts},
  \& {Falgarone}}]{lehmann20}
{Lehmann}, A., {Godard}, B., {Pineau des For{\^e}ts}, G., \& {Falgarone}, E.
  2020, \aap, 643, A101

\bibitem[{{Lehmann} {et~al.}(2022){Lehmann}, {Godard}, {Pineau des For{\^e}ts},
  {Vidal-Garc{\'\i}a}, \& {Falgarone}}]{lehmann22}
{Lehmann}, A., {Godard}, B., {Pineau des For{\^e}ts}, G., {Vidal-Garc{\'\i}a},
  A., \& {Falgarone}, E. 2022, \aap, 658, A165

\bibitem[{{Lehmann} \& {Wardle}(2016)}]{lehmann16}
{Lehmann}, A. \& {Wardle}, M. 2016, \mnras, 455, 2066

\bibitem[{{Lesaffre} {et~al.}(2004{\natexlab{a}}){Lesaffre}, {Chi{\`e}ze},
  {Cabrit}, \& {Pineau des For{\^e}ts}}]{lesaffre04a}
{Lesaffre}, P., {Chi{\`e}ze}, J.~P., {Cabrit}, S., \& {Pineau des For{\^e}ts},
  G. 2004{\natexlab{a}}, \aap, 427, 147

\bibitem[{{Lesaffre} {et~al.}(2004{\natexlab{b}}){Lesaffre}, {Chi{\`e}ze},
  {Cabrit}, \& {Pineau des For{\^e}ts}}]{lesaffre04b}
{Lesaffre}, P., {Chi{\`e}ze}, J.~P., {Cabrit}, S., \& {Pineau des For{\^e}ts},
  G. 2004{\natexlab{b}}, \aap, 427, 157

\bibitem[{{Lesaffre} {et~al.}(2013){Lesaffre}, {Pineau des For{\^e}ts},
  {Godard}, {Guillard}, {Boulanger}, \& {Falgarone}}]{lesaffre13}
{Lesaffre}, P., {Pineau des For{\^e}ts}, G., {Godard}, B., {et~al.} 2013, \aap,
  550, A106

\bibitem[{{Lutz} {et~al.}(2003){Lutz}, {Sturm}, {Genzel}, {Spoon}, {Moorwood},
  {Netzer}, \& {Sternberg}}]{Lutz2003}
{Lutz}, D., {Sturm}, E., {Genzel}, R., {et~al.} 2003, \aap, 409, 867

\bibitem[{{Maret} {et~al.}(2009){Maret}, {Bergin}, {Neufeld}, {Green},
  {Watson}, {Harwit}, {Kristensen}, {Melnick}, {Sonnentrucker}, {Tolls},
  {Werner}, {Willacy}, \& {Yuan}}]{maret09}
{Maret}, S., {Bergin}, E.~A., {Neufeld}, D.~A., {et~al.} 2009, \apj, 698, 1244

\bibitem[{{Martin} \& {Mandy}(1995)}]{martin95}
{Martin}, P.~G. \& {Mandy}, M.~E. 1995, \apjl, 455, L89

\bibitem[{{Mathis} {et~al.}(1983){Mathis}, {Mezger}, \& {Panagia}}]{mathis83}
{Mathis}, J.~S., {Mezger}, P.~G., \& {Panagia}, N. 1983, \aap, 128, 212

\bibitem[{{Mouri}(1994)}]{Mouri1994}
{Mouri}, H. 1994, \apj, 427, 777

\bibitem[{{Mullan}(1971)}]{mullan71}
{Mullan}, D.~J. 1971, \mnras, 153, 145

\bibitem[{{Neufeld} {et~al.}(2019){Neufeld}, {DeWitt}, {Lesaffre}, {Cabrit},
  {Gusdorf}, {Tram}, \& {Richter}}]{neufeld19}
{Neufeld}, D.~A., {DeWitt}, C., {Lesaffre}, P., {et~al.} 2019, \apjl, 878, L18

\bibitem[{{Neufeld} {et~al.}(2006){Neufeld}, {Melnick}, {Sonnentrucker},
  {Bergin}, {Green}, {Kim}, {Watson}, {Forrest}, \& {Pipher}}]{neufeld06}
{Neufeld}, D.~A., {Melnick}, G.~J., {Sonnentrucker}, P., {et~al.} 2006, \apj,
  649, 816

\bibitem[{{Nisini} {et~al.}(2010){Nisini}, {Giannini}, {Neufeld}, {Yuan},
  {Antoniucci}, {Bergin}, \& {Melnick}}]{nisini10}
{Nisini}, B., {Giannini}, T., {Neufeld}, D.~A., {et~al.} 2010, \apj, 724, 69

\bibitem[{{Reach} {et~al.}(2019){Reach}, {Tram}, {Richter}, {Gusdorf}, \&
  {DeWitt}}]{reach19}
{Reach}, W.~T., {Tram}, L.~N., {Richter}, M., {Gusdorf}, A., \& {DeWitt}, C.
  2019, \apj, 884, 81

\bibitem[{{Richard} {et~al.}(2022){Richard}, {Lesaffre}, {Falgarone}, \&
  {Lehmann}}]{richard22}
{Richard}, T., {Lesaffre}, P., {Falgarone}, E., \& {Lehmann}, A. 2022, \aap,
  664, A193

\bibitem[{{Rigby} {et~al.}(2023){Rigby}, {Perrin}, {McElwain}, {Kimble},
  {Friedman}, {Lallo}, {Doyon}, {Feinberg}, {Ferruit}, {Glasse}, {Rieke},
  {Rieke}, {Wright}, {Willott}, {Colon}, {Milam}, {Neff}, {Stark}, {Valenti},
  {Abell}, {Abney}, {Abul-Huda}, {Scott Acton}, {Adams}, {Adler}, {Aguilar},
  {Ahmed}, {Albert}, {Alberts}, {Aldridge}, {Allen}, {Altenburg},
  {{\'A}lvarez-M{\'a}rquez}, {Alves de Oliveira}, {Andersen}, {Anderson},
  {Anderson}, {Argyriou}, {Armstrong}, {Arribas}, {Artigau}, {Arvai},
  {Atkinson}, {Bacon}, {Bair}, {Banks}, {Barrientes}, {Barringer}, {Bartosik},
  {Bast}, {Baudoz}, {Beatty}, {Bechtold}, {Beck}, {Bergeron}, {Bergkoetter},
  {Bhatawdekar}, {Birkmann}, {Blazek}, {Blome}, {Boccaletti}, {B{\"o}ker},
  {Boia}, {Bonaventura}, {Bond}, {Bosley}, {Boucarut}, {Bourque}, {Bouwman},
  {Bower}, {Bowers}, {Boyer}, {Bradley}, {Brady}, {Braun}, {Breda},
  {Bresnahan}, {Bright}, {Britt}, {Bromenschenkel}, {Brooks}, {Brooks},
  {Brown}, {Brown}, {Brown}, {Bunker}, {Burger}, {Bushouse}, {Cale}, {Cameron},
  {Cameron}, {Canipe}, {Caplinger}, {Caputo}, {Cara}, {Carey}, {Carniani},
  {Carrasquilla}, {Carruthers}, {Case}, {Catherine}, {Chance}, {Chapman},
  {Charlot}, {Charlow}, {Chayer}, {Chen}, {Cherinka}, {Chichester}, {Chilton},
  {Chonis}, {Clampin}, {Clark}, {Clark}, {Coe}, {Coleman}, {Comber}, {Comeau},
  {Connolly}, {Cooper}, {Cooper}, {Coppock}, {Correnti}, {Cossou}, {Coulais},
  {Coyle}, {Cracraft}, {Curti}, {Cuturic}, {Davis}, {Davis}, {Dean}, {DeLisa},
  {deMeester}, {Dencheva}, {Dencheva}, {DePasquale}, {Deschenes}, {Hunor
  Detre}, {Diaz}, {Dicken}, {DiFelice}, {Dillman}, {Dixon}, {Doggett},
  {Donaldson}, {Douglas}, {DuPrie}, {Dupuis}, {Durning}, {Easmin}, {Eck},
  {Edeani}, {Egami}, {Ehrenwinkler}, {Eisenhamer}, {Eisenhower}, {Elie},
  {Elliott}, {Elliott}, {Ellis}, {Engesser}, {Espinoza}, {Etienne}, {Etxaluze},
  {Falini}, {Feeney}, {Ferry}, {Filippazzo}, {Fincham}, {Fix}, {Flagey},
  {Florian}, {Flynn}, {Fontanella}, {Ford}, {Forshay}, {Fox}, {Franz}, {Fu},
  {Fullerton}, {Galkin}, {Galyer}, {Garc{\'\i}a Mar{\'\i}n}, {Gardner},
  {Gardner}, {Garland}, {Garrett}, {Gasman}, {Gaspar}, {Gaudreau}, {Gauthier},
  {Geers}, {Geithner}, {Gennaro}, {Giardino}, {Girard}, {Giuliano},
  {Glassmire}, {Glauser}, {Glazer}, {Godfrey}, {Golimowski}, {Gollnitz},
  {Gong}, {Gonzaga}, {Gordon}, {Gordon}, {Goudfrooij}, {Greene}, {Greenhouse},
  {Grimaldi}, {Groebner}, {Grundy}, {Guillard}, {Gutman}, {Ha}, {Haderlein},
  {Hagedorn}, {Hainline}, {Haley}, {Hami}, {Hamilton}, {Hammel}, {Hansen},
  {Harkins}, {Harr}, {Hart}, {Hart}, {Hartig}, {Hashimoto}, {Haskins},
  {Hathaway}, {Havey}, {Hayden}, {Hecht}, {Heller-Boyer}, {Henriques}, {Henry},
  {Hermann}, {Hernandez}, {Hesman}, {Hicks}, {Hilbert}, {Hines}, {Hoffman},
  {Holfeltz}, {Holler}, {Hoppa}, {Hott}, {Howard}, {Howard}, {Hunter},
  {Hunter}, {Hurst}, {Husemann}, {Hustak}, {Ilinca Ignat}, {Illingworth},
  {Irish}, {Jackson}, {Jahromi}, {Jakobsen}, {James}, {James}, {Januszewski},
  {Jenkins}, {Jirdeh}, {Johnson}, {Johnson}, {Jones}, {Jones}, {Jones},
  {Jones}, {Jordan}, {Jordan}, {Jurczyk}, {Jurling}, {Kaleida}, {Kalmanson},
  {Kammerer}, {Kang}, {Kao}, {Karakla}, {Kavanagh}, {Kelly}, {Kendrew},
  {Kennedy}, {Kenny}, {Keski-kuha}, {Keyes}, {Kidwell}, {Kinzel}, {Kirk},
  {Kirkpatrick}, {Kirshenblat}, {Klaassen}, {Knapp}, {Scott Knight},
  {Knollenberg}, {Koehler}, {Koekemoer}, {Kovacs}, {Kulp}, {Kumari},
  {Kyprianou}, {La Massa}, {Labador}, {Labiano}, {Lagage}, {Lajoie}, {Lallo},
  {Lam}, {Lamb}, {Lambros}, {Lampenfield}, {Langston}, {Larson}, {Law},
  {Lawrence}, {Lee}, {Leisenring}, {Lepo}, {Leveille}, {Levenson}, {Levine},
  {Levy}, {Lewis}, {Lewis}, {Libralato}, {Lightsey}, {Link}, {Liu}, {Lo},
  {Lockwood}, {Logue}, {Long}, {Long}, {Loomis}, {Lopez-Caniego}, {Lorenzo
  Alvarez}, {Love-Pruitt}, {Lucy}, {Luetzgendorf}, {Maghami}, {Maiolino},
  {Major}, {Malla}, {Malumuth}, {Manjavacas}, {Mannfolk}, {Marrione},
  {Marston}, {Martel}, {Maschmann}, {Masci}, {Masciarelli}, {Maszkiewicz},
  {Mather}, {McKenzie}, {McLean}, {McMaster}, {Melbourne}, {Mel{\'e}ndez},
  {Menzel}, {Merz}, {Meyett}, {Meza}, {Miskey}, {Misselt}, {Moller},
  {Morrison}, {Morse}, {Moseley}, {Mosier}, {Mountain}, {Mueckay}, {Mueller},
  {Mullally}, {Murphy}, {Murray}, {Murray}, {Mustelier}, {Muzerolle},
  {Mycroft}, {Myers}, {Myrick}, {Nanavati}, {Nance}, {Nayak}, {Naylor},
  {Nelan}, {Nickson}, {Nielson}, {Nieto-Santisteban}, {Nikolov},
  {Noriega-Crespo}, {O'Shaughnessy}, {O'Sullivan}, {Ochs}, {Ogle}, {Oleszczuk},
  {Olmsted}, {Osborne}, {Ottens}, {Owens}, {Pacifici}, {Pagan}, {Page}, {Park},
  {Parrish}, {Patapis}, {Paul}, {Pauly}, {Pavlovsky}, {Pedder}, {Peek},
  {Pena-Guerrero}, {Penanen}, {Perez}, {Perna}, {Perriello}, {Phillips},
  {Pietraszkiewicz}, {Pinaud}, {Pirzkal}, {Pitman}, {Piwowar}, {Platais},
  {Player}, {Plesha}, {Pollizi}, {Polster}, {Pontoppidan}, {Porterfield},
  {Proffitt}, {Pueyo}, {Pulliam}, {Quirt}, {Quispe Neira}, {Ramos Alarcon},
  {Ramsay}, {Rapp}, {Rapp}, {Rauscher}, {Ravindranath}, {Rawle}, {Regan},
  {Reichard}, {Reis}, {Ressler}, {Rest}, {Reynolds}, {Rhue}, {Richon},
  {Rickman}, {Ridgaway}, {Ritchie}, {Rix}, {Robberto}, {Robinson}, {Robinson},
  {Robinson}, {Rock}, {Rodriguez}, {Rodriguez Del Pino}, {Roellig}, {Rohrbach},
  {Roman}, {Romelfanger}, {Rose}, {Roteliuk}, {Roth}, {Rothwell}, {Rowlands},
  {Roy}, {Royer}, {Royle}, {Rui}, {Rumler}, {Runnels}, {Russ}, {Rustamkulov},
  {Ryden}, {Ryer}, {Sabata}, {Sabatke}, {Sabbi}, {Samuelson}, {Sapp},
  {Sappington}, {Sargent}, {Sauer}, {Scheithauer}, {Schlawin}, {Schlitz},
  {Schmitz}, {Schneider}, {Schreiber}, {Schulze}, {Schwab}, {Scott}, {Sembach},
  {Shanahan}, {Shaughnessy}, {Shaw}, {Shawger}, {Shay}, {Sheehan}, {Shen},
  {Sherman}, {Shiao}, {Shih}, {Shivaei}, {Sienkiewicz}, {Sing}, {Sirianni},
  {Sivaramakrishnan}, {Skipper}, {Sloan}, {Slocum}, {Slowinski}, {Smith},
  {Smith}, {Smith}, {Smith}, {Snyder}, {Soh}, {Tony Sohn}, {Soto}, {Spencer},
  {Stallcup}, {Stansberry}, {Starr}, {Starr}, {Stewart}, {Stiavelli},
  {Straughn}, {Strickland}, {Stys}, {Summers}, {Sun}, {Sunnquist}, {Swade},
  {Swam}, {Swaters}, {Swoish}, {Taylor}, {Taylor}, {Te Plate}, {Tea}, {Teague},
  {Telfer}, {Temim}, {Thatte}, {Thompson}, {Thompson}, {Thomson}, {Tikkanen},
  {Tippet}, {Todd}, {Toolan}, {Tran}, {Trejo}, {Truong}, {Tsukamoto},
  {Tustain}, {Tyra}, {Ubeda}, {Underwood}, {Uzzo}, {Van Campen}, {Vandal},
  {Vandenbussche}, {Vila}, {Volk}, {Wahlgren}, {Waldman}, {Walker}, {Wander},
  {Warfield}, {Warner}, {Wasiak}, {Watkins}, {Weaver}, {Weilert}, {Weiser},
  {Weiss}, {Weissman}, {Welty}, {West}, {Wheate}, {Wheatley}, {Wheeler},
  {White}, {Whiteaker}, {Whitehouse}, {Whiteleather}, {Whitman}, {Williams},
  {Willmer}, {Willoughby}, {Wilson}, {Wirth}, {Wislowski}, {Wolf}, {Wolfe},
  {Wolff}, {Workman}, {Wright}, {Wu}, {Wu}, {Wymer}, {Yates}, {Yeager},
  {Yeates}, {Yerger}, {Yoon}, {Young}, {Yu}, {Zak}, {Zeidler}, {Zhou},
  {Zielinski}, {Zincke}, \& {Zonak}}]{rigby22}
{Rigby}, J., {Perrin}, M., {McElwain}, M., {et~al.} 2023, \pasp, 135, 048001

\bibitem[{{Rosenthal} {et~al.}(2000){Rosenthal}, {Bertoldi}, \&
  {Drapatz}}]{rosenthal00}
{Rosenthal}, D., {Bertoldi}, F., \& {Drapatz}, S. 2000, \aap, 356, 705

\bibitem[{{Santangelo} {et~al.}(2014){Santangelo}, {Antoniucci}, {Nisini},
  {Codella}, {Bjerkeli}, {Giannini}, {Lorenzani}, {Lundin}, {Cabrit},
  {Calzoletti}, {Liseau}, {Neufeld}, {Tafalla}, \& {van
  Dishoeck}}]{santangelo14}
{Santangelo}, G., {Antoniucci}, S., {Nisini}, B., {et~al.} 2014, \aap, 569, L8

\bibitem[{{Tram} {et~al.}(2018){Tram}, {Lesaffre}, {Cabrit}, {Gusdorf}, \&
  {Nhung}}]{tram18}
{Tram}, L.~N., {Lesaffre}, P., {Cabrit}, S., {Gusdorf}, A., \& {Nhung}, P.~T.
  2018, \mnras, 473, 1472

\bibitem[{{Valentijn} \& {van der Werf}(1999)}]{Valentijn1999}
{Valentijn}, E.~A. \& {van der Werf}, P.~P. 1999, \apjl, 522, L29

\bibitem[{{Verma} {et~al.}(2005){Verma}, {Charmandaris}, {Klaas}, {Lutz}, \&
  {Haas}}]{Verma2005}
{Verma}, A., {Charmandaris}, V., {Klaas}, U., {Lutz}, D., \& {Haas}, M. 2005,
  \ssr, 119, 355

\bibitem[{{Wilgenbus} {et~al.}(2000){Wilgenbus}, {Cabrit}, {Pineau des For{\^
  e}ts}, \& {Flower}}]{wilgenbus00}
{Wilgenbus}, D., {Cabrit}, S., {Pineau des For{\^ e}ts}, G., \& {Flower}, D.~R.
  2000, \aap, 356, 1010

\bibitem[{{Wright} {et~al.}(1993){Wright}, {Geballe}, \& {Graham}}]{Wright1993}
{Wright}, G.~S., {Geballe}, T.~R., \& {Graham}, J.~R. 1993, in Evolution of
  Galaxies and their Environment, ed. J.~M. {Shull} \& H.~A. {Thronson}

\bibitem[{{Yang} {et~al.}(2022){Yang}, {Green}, {Pontoppidan}, {Bergner},
  {Cleeves}, {Evans}, {Garrod}, {Jin}, {Kim}, {Kim}, {Lee}, {Sakai},
  {Shingledecker}, {Shope}, {Tobin}, \& {van Dishoeck}}]{yang22}
{Yang}, Y.-L., {Green}, J.~D., {Pontoppidan}, K.~M., {et~al.} 2022, \apjl, 941,
  L13

\end{thebibliography}

\appendix

\section{The ISM platform}\label{app:ismdb}

The ISM platform\footnote{\url{http://ism.obspm.fr}} is a web portal that contains a series of services developed for the diffusion of state-of-the-art astrochemical models and the preparation and interpretation of observations. Regarding the Paris-Durham shock code, the platform provides access to the numerical code and its previous versions, a full documentation of the physical processes implemented, a tutorial to learn how to run the code locally, and a series of selected references. The platform also provides two analysis tools, IDAT and the Chemistry Analyzer tool, which can be used to study the output of the shock code and identify the processes responsible for the thermochemical evolution of the gas in a simulation. Finally, the platform contains a numerical database (InterStellar Medium DataBase or ISMDB) that provides an easy access to recalculated grid of theoretical models.

On this platform it is possible to ``Search models in ISMDB'' and from there ``Browse models.'' This leads to a page where combinations of input shock parameters can be specified, and once the selection has been made, it is possible to ``Get model.'' The resulting page shows the input parameters as well as some of the resulting quantities (e.g., shock type). The entire model output can be downloaded for further analysis, or the model can be quickly inspected directly through ``Online analysis with IDAT.'' This tool allows the user to select different quantities and plot them against distance through the shock on one or two different y-axes if so desired. An example could be the velocities through the shock as well as the temperature.

\section{Tables with extracted parameters}\label{app:tables}

We here provide example tables of the physical quantities already extracted from the grid (Tables \ref{tab:phys} --  \ref{tab:atom}). These tables are available on CDS in electronic format. These tables include: 
\begin{itemize}
    \item[\ref{tab:phys}] Physical quantities such as peak temperature, density, width, and age of the shock; 
    \item[\ref{tab:coldens}] Column densities of relevant species, particularly H, H$_2$, O, OH, H$_2$, C$^+$, C, and CO; 
    \item[\ref{tab:exc}] Data required for creating H$_2$ excitation diagrams, i.e., ln($N$/$g$) and $E$ for each of the 150 levels; 
    \item[\ref{tab:int}] H$_2$ integrated intensities of the 1000 lines extracted, along with their wavelength; 
    \item[\ref{tab:width}] Width of the H$_2$ emitting zone for the $\varv$ = 0--0 S(1), 1--0 S(1), 0--0 S(9), 1--0 O(5), and 2--1 S(1) lines; 
    \item[\ref{tab:opr}] H$_2$ $o/p$ ratios determined both locally and integrated through the shock; 
    \item[\ref{tab:atom}] Integrated line intensities of 29 transitions arising from C$^+$, Si$^+$, H, C, Si, O, S$^+$, N$^+$, N, and S. 
\end{itemize}

An energy cutoff of 99.9\% was used to define the point at which integrated quantities (e.g., line intensities, column densities) were integrated to (Sect. \ref{sect:grid}). Tests were performed using cutoffs at 95\%, 99\%, 99.9\%, 99.99\%, and 99.999\%. The two lower values (95 and 99\%) did not capture the H$_2$-emitting zone, particularly in strong CJ-type shocks where the temperature exceeds 10$^5$ K. The difference between 99.9\% and 99.99\% cutoffs were on the order of a few percent in terms of H$_2$ integrated line intensities for the $\varv$ = 0--0 S(1), 1--0 S(1), and 2--1 S(1) transitions for most shock conditions. Thus, a threshold of 99.9\% ensured that most of the H$_2$ radiative cooling zone was encompassed. 

\begin{table*}
\caption{Physical parameters, e.g., temperature and size of the shock. \label{tab:phys}}
\begin{center}
\tiny
\begin{tabular}{c c c c c c c c c c c c c c c c c c}
\hline\hline
$n_{\rm H}$ & $\varv_{\rm s}$ & $b$ & $G_0$ & $\zeta_{\rm H2}$ & $X$(PAH) & Type\tablefootmark{a}  & $T_{\rm gas,initial}$ & $T_{\rm gas,max}$ & $\Delta z$ & $\Delta t$ & $N_{\rm H}$ & $n_{\rm H,max}$ \\
(cm$^{-3}$) & (km s$^{-1}$) & &  & (s$^{-1}$) &  &  & (K) & (K) & (AU) & (years) & (cm$^{-2}$) & (cm$^{-3}$) \\ \hline
1.0e+02 & 3.0 & 0.1 & 0.0e+00 & 1.0e-17 & 1.0e-06 & 0 & 34.8 & 496.4 & 1.80e+03 & 5.58e+04 & 5.28e+19 & 2.24e+03 \\
1.0e+02 & 3.0 & 0.3 & 0.0e+00 & 1.0e-17 & 1.0e-06 & 1 & 34.8& 79.9  & 1.53e+05 & 5.35e+05 & 5.06e+20 & 7.10e+02 \\
1.0e+02 & 3.0 & 1.0 & 0.0e+00 & 1.0e-17 & 1.0e-06 & 1 & 34.8 & 34.8 & 6.02e+05 & 1.00e+06 & 9.51e+20 & 1.23e+02 \\
\ldots & \ldots & \ldots & \ldots & \ldots & \ldots & \ldots & \ldots & \ldots & \ldots & \ldots & \ldots & \ldots \\ \hline
\end{tabular}
\vspace{-12pt}
\end{center}
  \tablefoot{This table is only an extract. The full version is available at the CDS. 
\tablefoottext{a}{Resulting shock type: 0 is for J, 1 is for C, 2 is for C$^*$, and 3 is for CJ-type shocks, 99 is for a model that did not converge.}
}
\end{table*}

\begin{table*}
\caption{Column densities of H, H$_2$, O, OH, H$_2$O, C$^+$, C, and CO. \label{tab:coldens}}
\begin{center}
\tiny
\begin{tabular}{c c c c c c c c c c c c c c c c c c}
\hline\hline
$n_{\rm H}$ & $\varv_{\rm s}$ & $b$ & $G_0$ & $\zeta_{\rm H2}$ & $X$(PAH) & Type & $N$(H) & $N$(H$_2$) & $N$(O) & $N$(OH) & $N$(H$_2$O) & \ldots \\
(cm$^{-3}$) & (km s$^{-1}$) & &  & (s$^{-1}$) &  &  & (cm$^{-2}$) & (cm$^{-2}$) & (cm$^{-2}$) & (cm$^{-2}$) & (cm$^{-2}$) & \ldots \\ \hline
1.0e+02 & 3.0 & 0.1 & 0.0e+00 & 1.0e-17 & 1.0e-06 & 0 & 6.78e+16 & 2.64e+20 & 1.52e+08 & 2.64e+10 & 7.63e+09 & \ldots \\
1.0e+02 & 3.0 & 0.3 & 0.0e+00 & 1.0e-17 & 1.0e-06 & 1 & 5.37e+17 & 2.53e+20 & 1.28e+10 & 4.50e+11 & 1.21e+11 & \ldots \\
1.0e+02 & 3.0 & 1.0 & 0.0e+00 & 1.0e-17 & 1.0e-06 & 1 & 7.96e+18 & 4.75e+20 & 4.70e+10 & 1.92e+11 & 5.22e+10 & \ldots \\
\ldots & \ldots & \ldots & \ldots & \ldots & \ldots & \ldots & \ldots & \ldots & \ldots & \ldots & \ldots & \ldots \\ \hline
\end{tabular}
\end{center}
\tablefoot{This table is only an extract. The full version is available at the CDS. 
}
\end{table*}

\begin{table*}
\caption{Values of ln($N/g$), where $N$ is in units of cm$^{-2}$, and $E/k_{\rm B}$ for 150 H$_2$ levels, useful for creating excitation diagrams. \label{tab:exc}}
\begin{center}
\tiny
\begin{tabular}{c c c c c c c c c c c c c c c c c c c}
\hline\hline
$n_{\rm H}$ & $\varv_{\rm s}$ & $b$ & $G_0$ & $\zeta_{\rm H2}$ & $X$(PAH) & Type &  v=0, J=0 & $E$ & v=0, J=1 & $E$ & v=0, J=2 & $E$ &  \ldots \\
(cm$^{-3}$) & (km s$^{-1}$) &  &  & (s$^{-1}$) &  &  &  & (K) & & (K) & & (K)  \\ \hline
1.0e+02 & 3.0 & 0.1 & 0.0e+00 & 1.0e-17 & 1.0e-06 & 0 & 44.62 & 0.00 & 39.92 & 170.50 & 39.02 & 509.85 & \ldots \\
1.0e+02 & 3.0 & 0.3 & 0.0e+00 & 1.0e-17 & 1.0e-06 & 1 & 46.90 & 0.00 & 42.25 & 170.50 & 37.68 & 509.85 & \ldots \\
1.0e+02 & 3.0 & 1.0 & 0.0e+00 & 1.0e-17 & 1.0e-06 & 1 & 47.55 & 0.00 & 42.64 & 170.50 & 29.41 & 509.85 & \ldots \\
\ldots & \ldots & \ldots & \ldots & \ldots & \ldots & \ldots & \ldots & \ldots & \ldots & \ldots & \ldots & \ldots & \ldots \\ \hline
\end{tabular}
\end{center}
\tablefoot{This table is only an extract. The full version is available at the CDS. 
}
\end{table*}

\begin{table*}
\caption{H$_2$ integrated intensities and wavelengths of 1000 transitions. \label{tab:int}}
\begin{center}
\tiny
\begin{tabular}{c c c c c c c c c c c c c c c c c c c}
\hline\hline
$n_{\rm H}$ & $\varv_{\rm s}$ & $b$ & $G_0$ & $\zeta_{\rm H2}$ & $X$(PAH) & Type & v=0,J=2 -- v=0,J=0 & $\lambda$ & v=0,J=3 -- v=0,J=1 & $\lambda$ & \ldots \\
(cm$^{-3}$) & (km s$^{-1}$) &  &  & (s$^{-1}$) &  &  & (erg cm$^{-2}$ s$^{-1}$ sr$^{-1}$) & ($\mu$m) & (erg cm$^{-2}$ s$^{-1}$ sr$^{-1}$) & ($\mu$m) & \ldots \\ \hline
1.0e+02 & 3.0 & 0.1 & 0.0e+00 & 1.0e-17 & 1.0e-06 & 0 & 7.29e-08 & 28.25 & 1.53e-08 & 17.05 & \ldots \\
1.0e+02 & 3.0 & 0.3 & 0.0e+00 & 1.0e-17 & 1.0e-06 & 1 & 1.91e-08 & 28.25 & 1.10e-10 & 17.05 & \ldots \\
1.0e+02 & 3.0 & 1.0 & 0.0e+00 & 1.0e-17 & 1.0e-06 & 1 & 4.91e-12 & 28.25 & 4.07e-11 & 17.05 & \ldots \\
\ldots & \ldots & \ldots & \ldots & \ldots & \ldots & \ldots & \ldots & \ldots & \ldots & \ldots & \ldots \\ \hline
\end{tabular}
\end{center}
\tablefoot{This table is only an extract. The full version is available at the CDS. 
}
\end{table*}

\begin{table*}
\caption{Widths of the region where 80\% of H$_2$ emission is generated for five transitions. \label{tab:width}}
\begin{center}
\tiny
\begin{tabular}{c c c c c c c c c c c c c c c c c c}
\hline\hline
$n_{\rm H}$ & $\varv_{\rm s}$ & $b$ & $G_0$ & $\zeta_{\rm H2}$ & $X$(PAH) & Type & $z$(v=0--0 S(1)) & $z$(v=1--0 S(1)) & $z$(v=0--0 S(9)) & $z$(v=1--0 O(5)) & $z$(v=2--1 S(1)) \\
(cm$^{-3}$) & (km s$^{-1}$) &  &  & (s$^{-1}$) &  &  & (AU) & (AU) & (AU) & (AU) & (AU) \\ \hline
1.0e+02 & 3.0 & 0.1 & 0.0e+00 & 1.0e-17 & 1.0e-06 & 0 & 1.89e+02 & 1.38e+02 & 8.49e+01 & 1.14e+03 & 1.44e+03 \\
1.0e+02 & 3.0 & 0.3 & 0.0e+00 & 1.0e-17 & 1.0e-06 & 1 & 2.50e+04 & 8.36e+04 & 8.65e+04 & 8.65e+04 & 8.65e+04 \\
1.0e+02 & 3.0 & 1.0 & 0.0e+00 & 1.0e-17 & 1.0e-06 & 1 & 1.86e+02 & 1.33e+02 & 8.38e+01 & 1.07e+03 & 1.46e+03 \\
\ldots & \ldots & \ldots & \ldots & \ldots & \ldots & \ldots & \ldots & \ldots & \ldots & \ldots & \ldots \\ \hline
\end{tabular}
\end{center}
\tablefoot{This table is only an extract. The full version is available at the CDS. 
}
\end{table*}

\begin{table*}
\caption{H$_2$ $o/p$ ratios, both local and integrated. \label{tab:opr}}
\begin{center}
\tiny
\begin{tabular}{c c c c c c c c c c c c c c c c c c}
\hline\hline
$n_{\rm H}$ & $\varv_{\rm s}$ & $b$ & $G_0$ & $\zeta_{\rm H2}$ & $X$(PAH) & Type & $o/p_{\rm ini}$ & $o/p_{\rm max}$ & $N_{\rm o}/N_{\rm p}$ & $N_{\rm o}/N_{\rm p}$($\varv=0$, $J$=2--9) & $N_{\rm o}/N_{\rm p}$($\varv=1$, $J$=2--9) \\
(cm$^{-3}$) & (km s$^{-1}$) &  &  & (s$^{-1}$) &   \\ \hline
1.0e+02 & 3.0 & 0.1 & 0.0e+00 & 1.0e-17 & 1.0e-06 & 0 & 0.07 & 0.08 & 0.08 & 0.01 & 3.70  \\
1.0e+02 & 3.0 & 0.3 & 0.0e+00 & 1.0e-17 & 1.0e-06 & 1 & 0.07 & 0.09 & 0.09 & 0.00 & 3.72  \\
1.0e+02 & 3.0 & 1.0 & 0.0e+00 & 1.0e-17 & 1.0e-06 & 1 & 0.07 & 0.07 & 0.07 & 0.32 & 3.70  \\
\ldots & \ldots & \ldots & \ldots & \ldots & \ldots & \ldots & \ldots & \ldots & \ldots & \ldots & \ldots \\ \hline
\end{tabular}
\end{center}
\tablefoot{This table is only an extract. The full version is available at the CDS. 
}
\end{table*}

\begin{table*}
\caption{Integrated intensities of 29 transitions from C$^+$, Si$^+$, H, C, Si, O, S$^+$, N$^+$, N, and S. \label{tab:atom}}
\begin{center}
\tiny
\begin{tabular}{c c c c c c c c c c c c c c c c c c c}
\hline\hline
$n_{\rm H}$ & $\varv_{\rm s}$ & $b$ & $G_0$ & $\zeta_{\rm H2}$ & $X$(PAH) & Type & 
C$^+$(158$\mu$m) & C$^+$(2324.7\AA) & C$^+$(2323.5\AA) & \ldots \\
(cm$^{-3}$) & (km s$^{-1}$) &  &  & (s$^{-1}$) &  &  & (erg cm$^{-2}$ s$^{-1}$ sr$^{-1}$) & (erg cm$^{-2}$ s$^{-1}$ sr$^{-1}$) & (erg cm$^{-2}$ s$^{-1}$ sr$^{-1}$) &  \ldots \\ \hline
1.0e+02 & 3.0 & 0.1 & 0.0e+00 & 1.0e-17 & 1.0e-06 & 0 & 4.77e-11 & 5.37e-54 & 2.89e-54 & \ldots \\
1.0e+02 & 3.0 & 0.3 & 0.0e+00 & 1.0e-17 & 1.0e-06 & 1 & 1.45e-09 & 1.48e-53 & 7.98e-54 & \ldots \\
1.0e+02 & 3.0 & 1.0 & 0.0e+00 & 1.0e-17 & 1.0e-06 & 1 & 2.04e-10 & 5.63e-55 & 3.03e-55 & \ldots \\
\ldots & \ldots & \ldots & \ldots & \ldots & \ldots & \ldots & \ldots & \ldots & \ldots & \ldots \\ \hline
\end{tabular}
\end{center}
\tablefoot{This table is only an extract. The full version is available at the CDS. 
}
\end{table*}

\FloatBarrier

\section{Additional figures}

\subsection{Excitation temperatures}\label{app:tex}

\begin{figure*}
\centering
\includegraphics[width=0.4\textwidth]{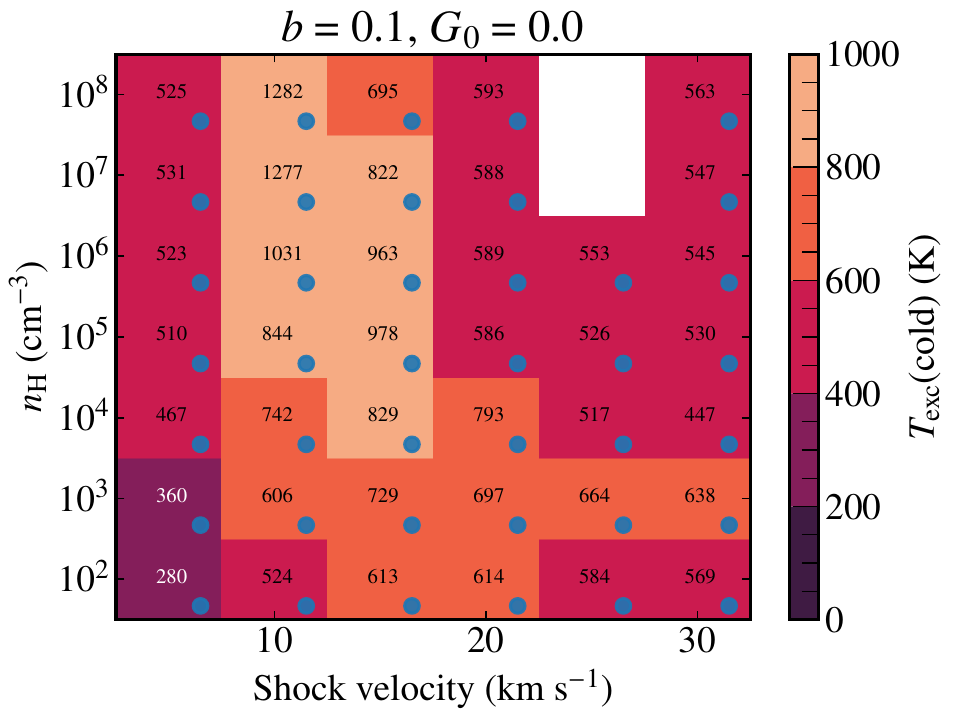}
\includegraphics[width=0.4\textwidth]{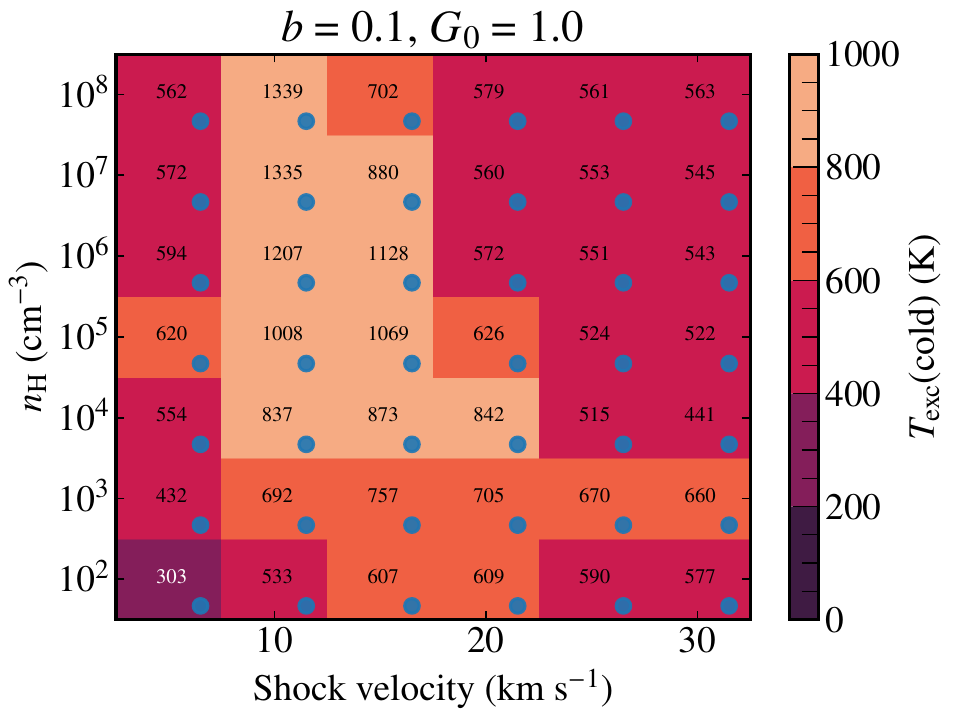}
\includegraphics[width=0.4\textwidth]{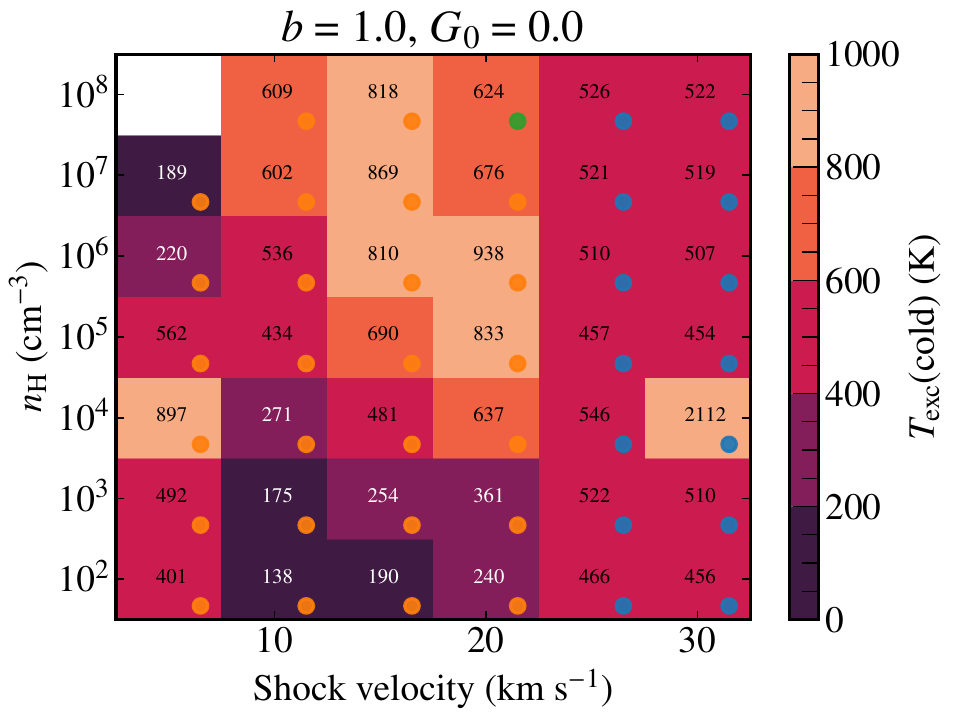}
\includegraphics[width=0.4\textwidth]{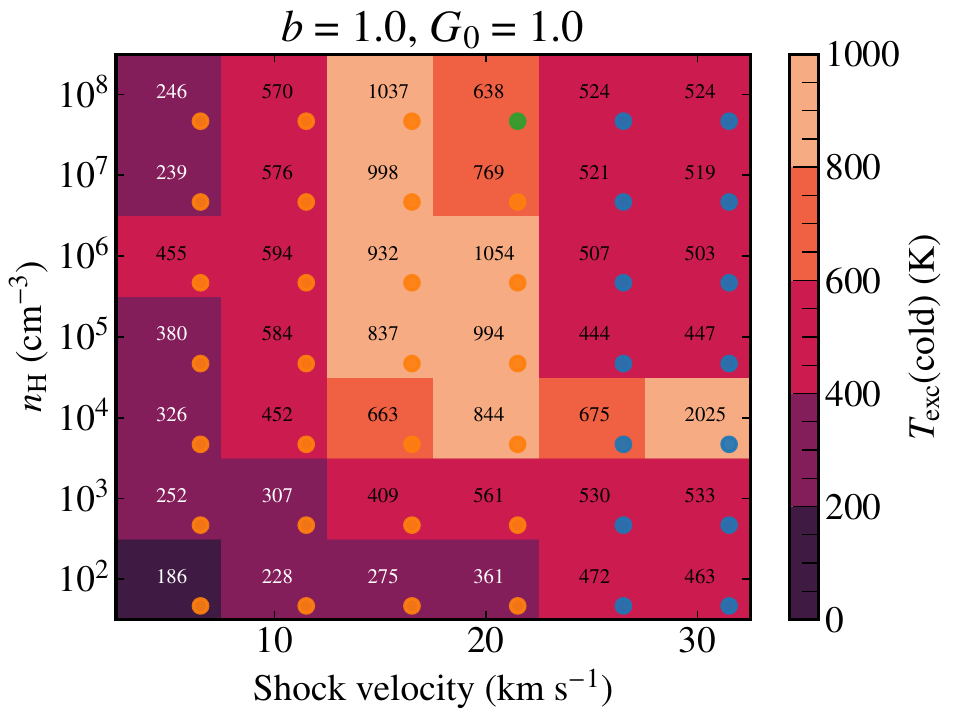}
\caption{Excitation temperature determined from the $\varv$ = 0, $J$ = 3 to 5 level populations (corresponding to the S(1) to S(3) transitions) for $b$ = 0.1 and 1, and $G_0$ = 0 and 1. \label{fig:tcold}}
\end{figure*}

\begin{figure*}
\centering
\includegraphics[width=0.4\textwidth]{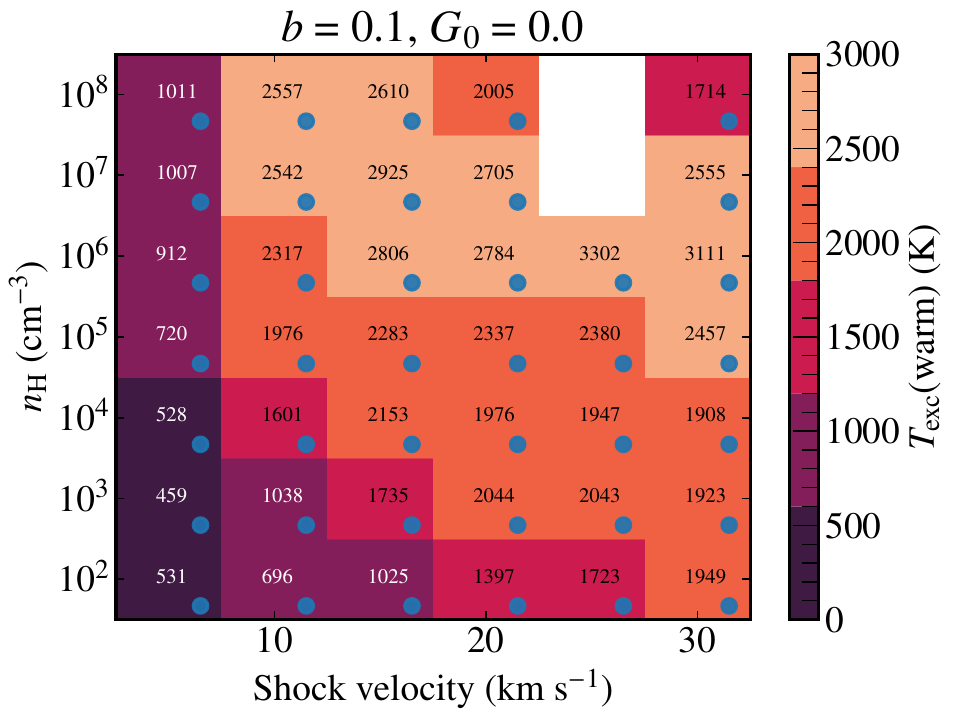}
\includegraphics[width=0.4\textwidth]{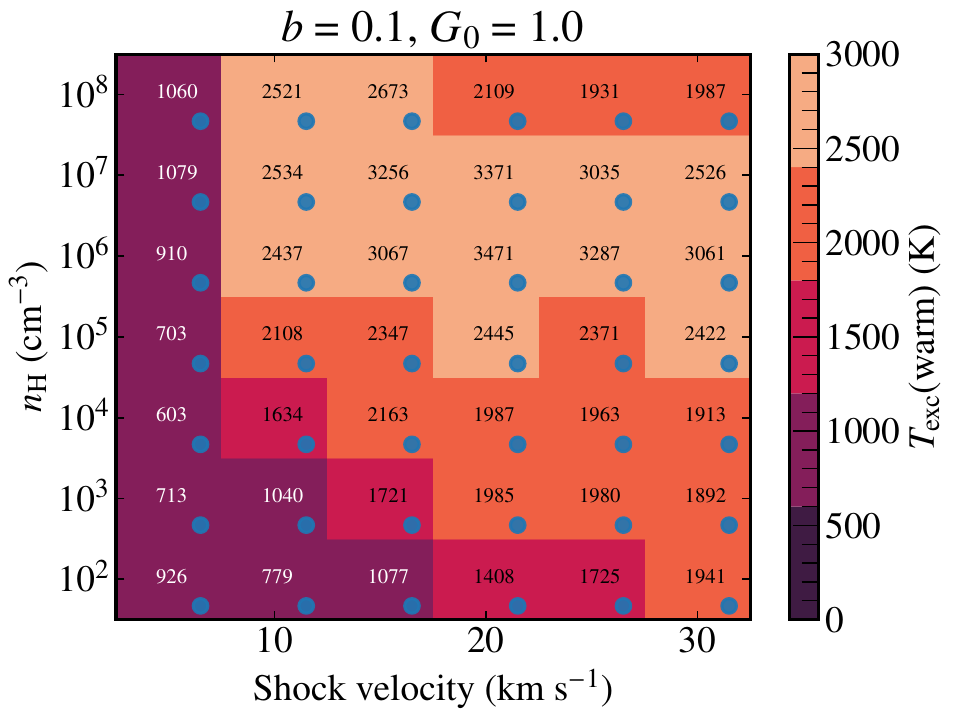}
\includegraphics[width=0.4\textwidth]{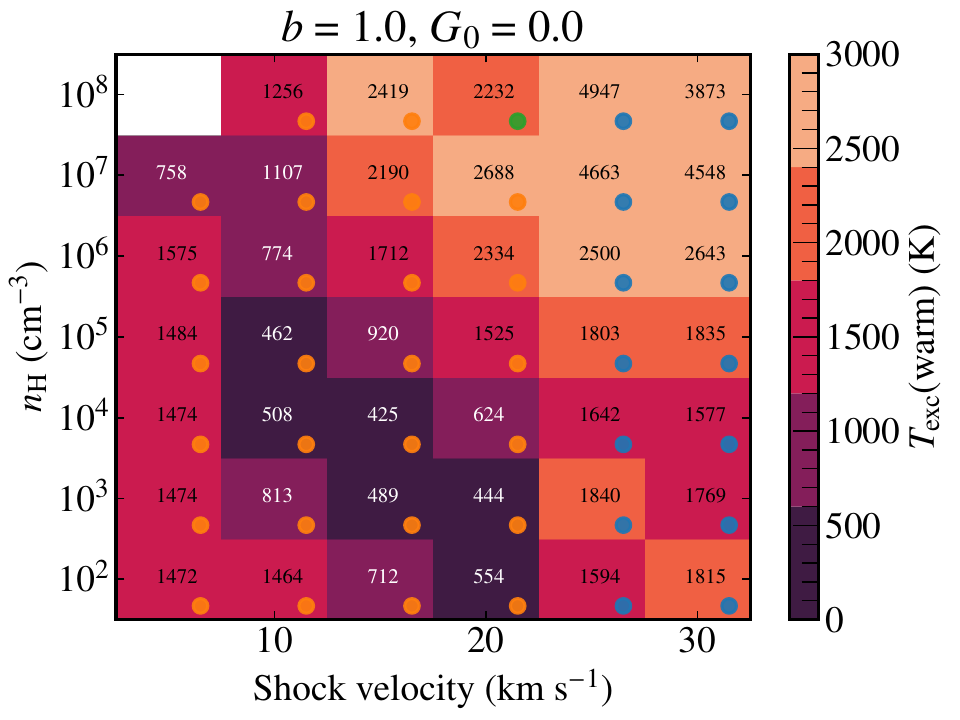}
\includegraphics[width=0.4\textwidth]{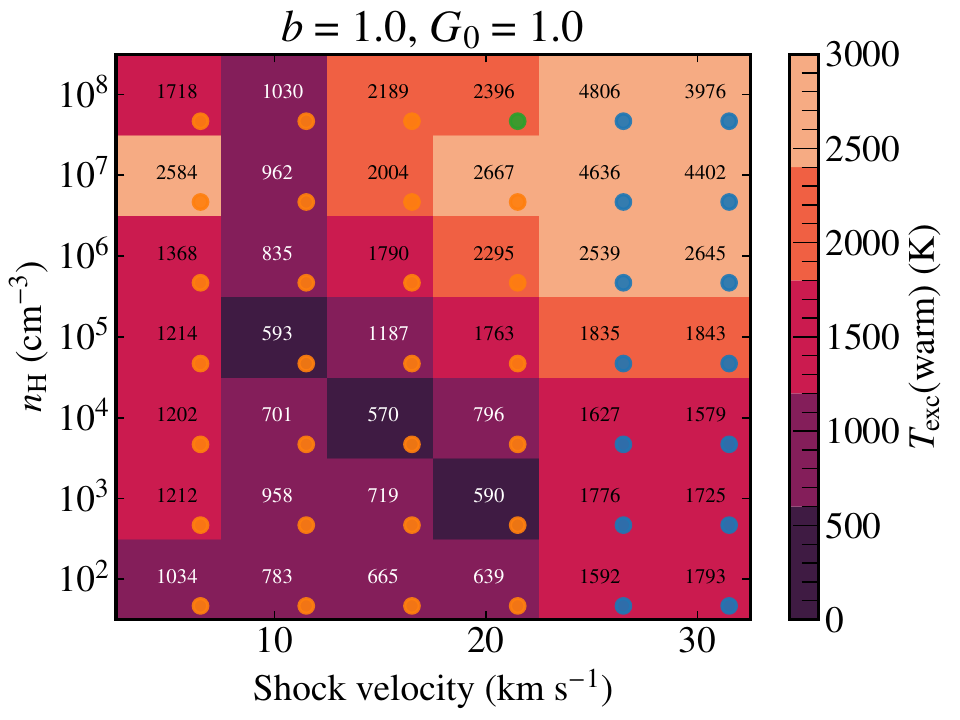}
\caption{Excitation temperature determined from the $\varv$ = 0, $J$ = 6 to 11 level populations (corresponding to the S(4) to S(9) transitions) for $b$ = 0.1 and 1, and $G_0$ = 0 and 1. \label{fig:twarm}}
\end{figure*}

\begin{figure*}
\centering
\includegraphics[width=0.4\textwidth]{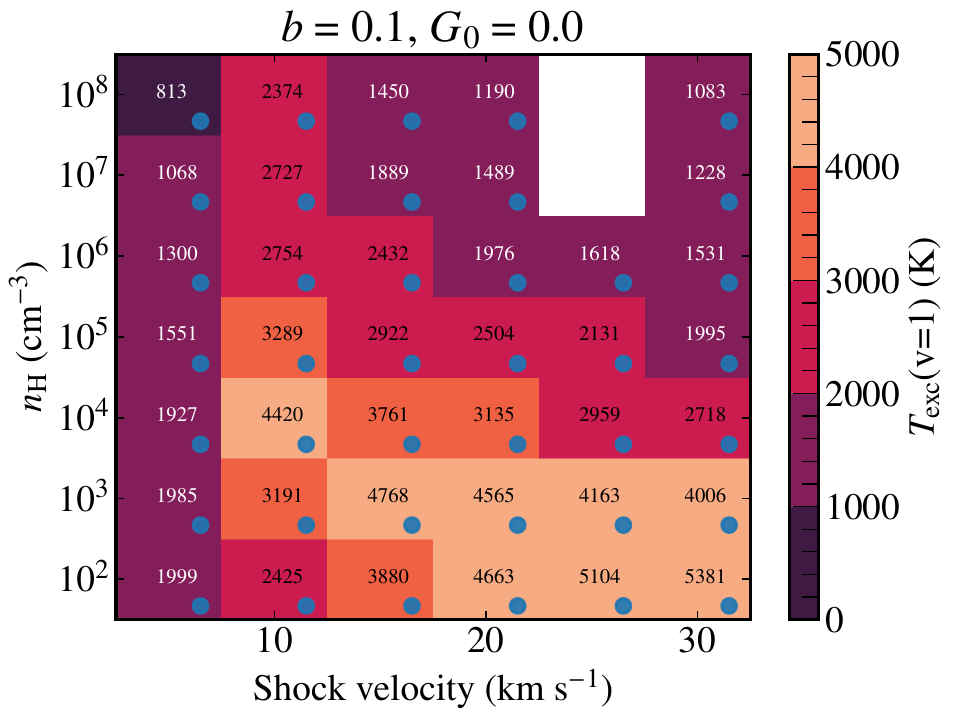}
\includegraphics[width=0.4\textwidth]{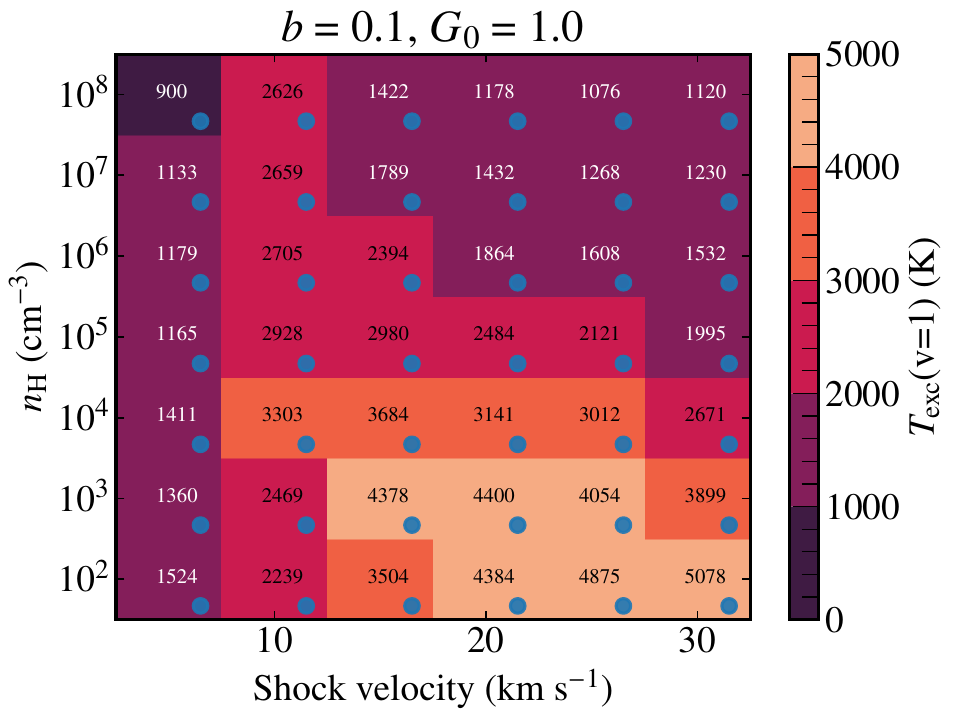}
\includegraphics[width=0.4\textwidth]{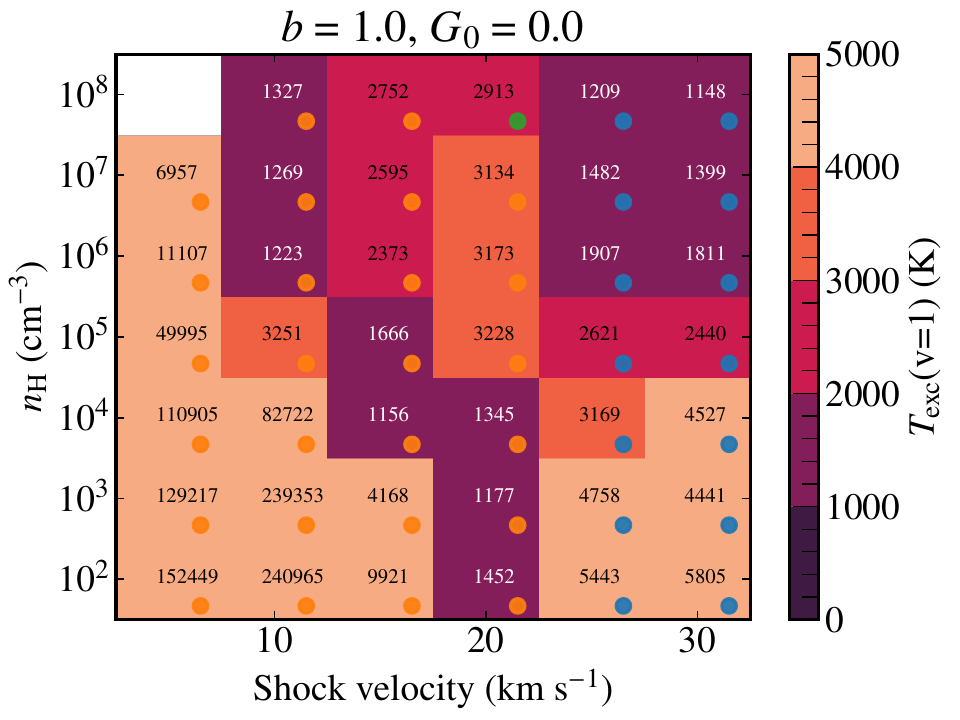}
\includegraphics[width=0.4\textwidth]{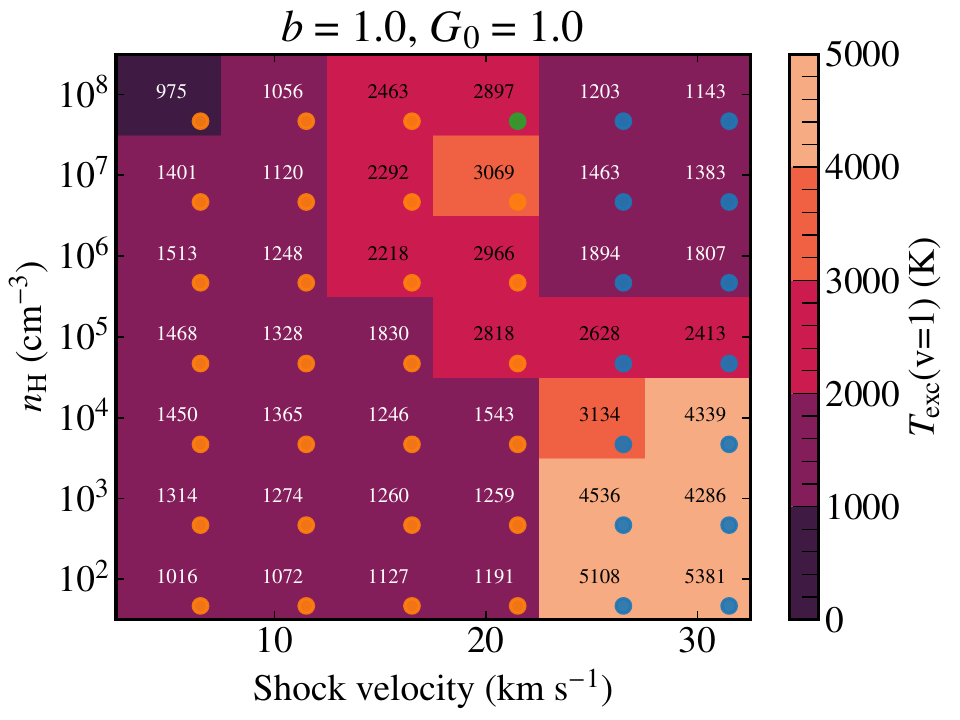}
\caption{Excitation temperature determined from the $\varv$ = 1, $J$ = 0 to 8 level populations for $b$ = 0.1 and 1, and $G_0$ = 0 and 1. \label{fig:tv1}}
\end{figure*}

\begin{figure*}
\centering
\includegraphics[width=0.4\textwidth]{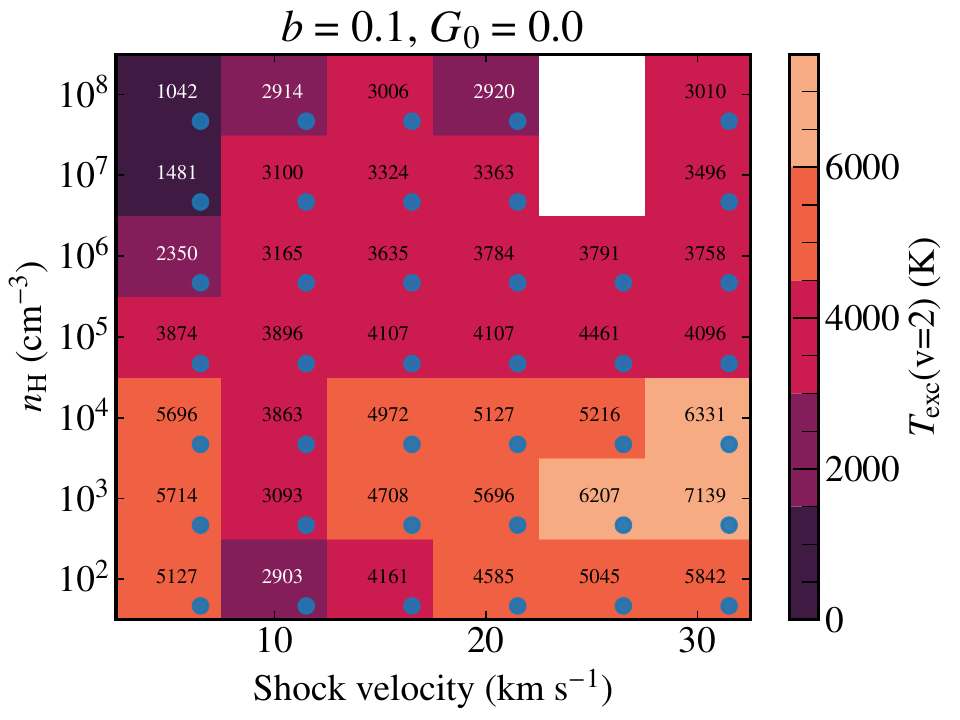}
\includegraphics[width=0.4\textwidth]{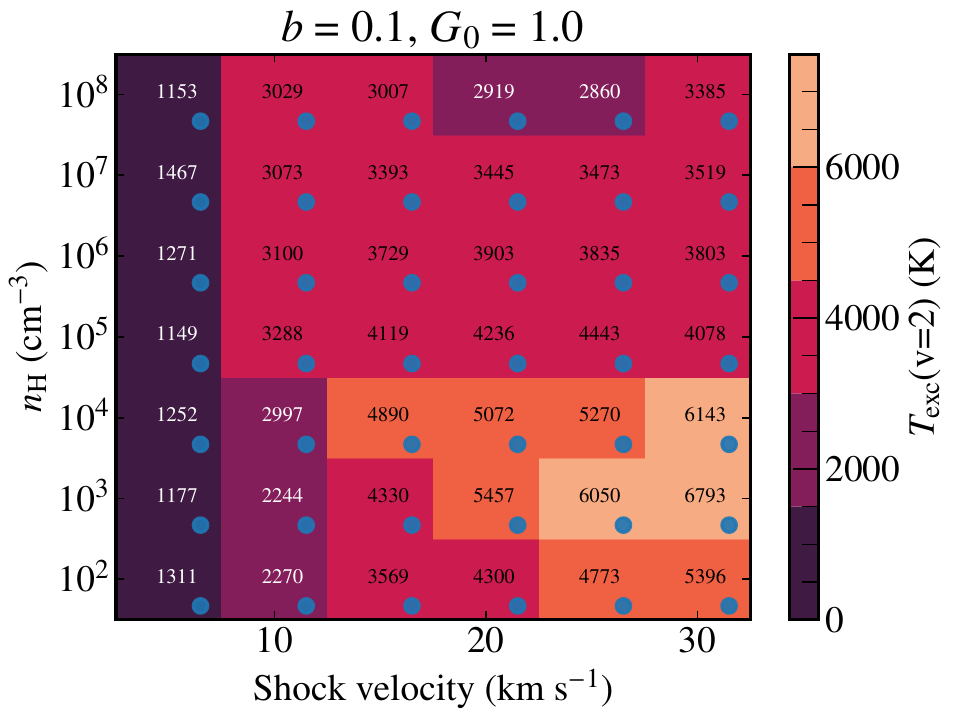}
\includegraphics[width=0.4\textwidth]{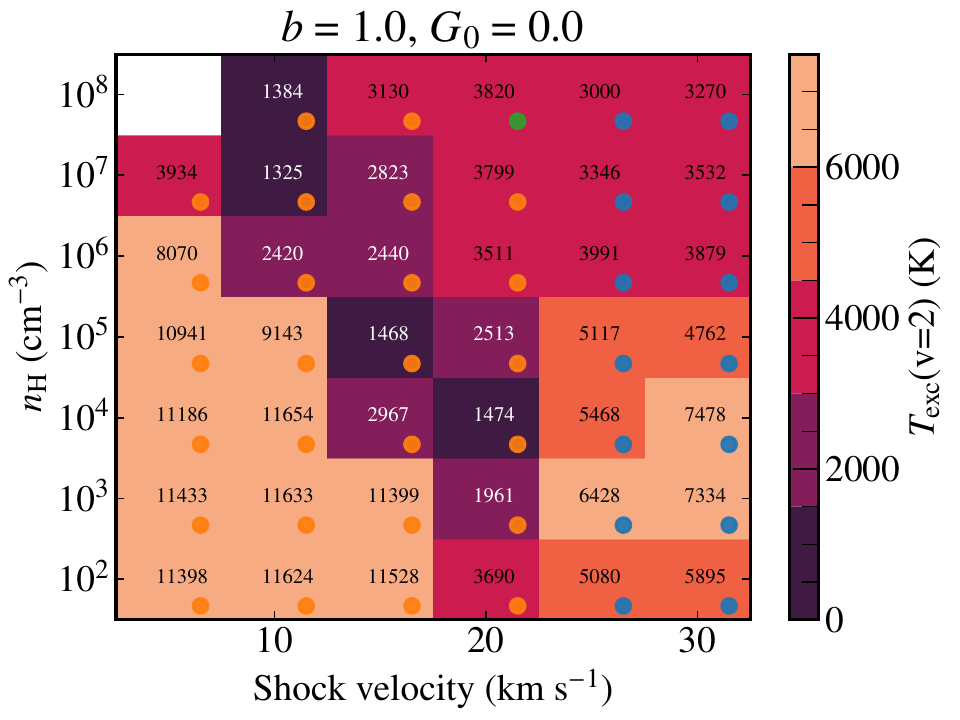}
\includegraphics[width=0.4\textwidth]{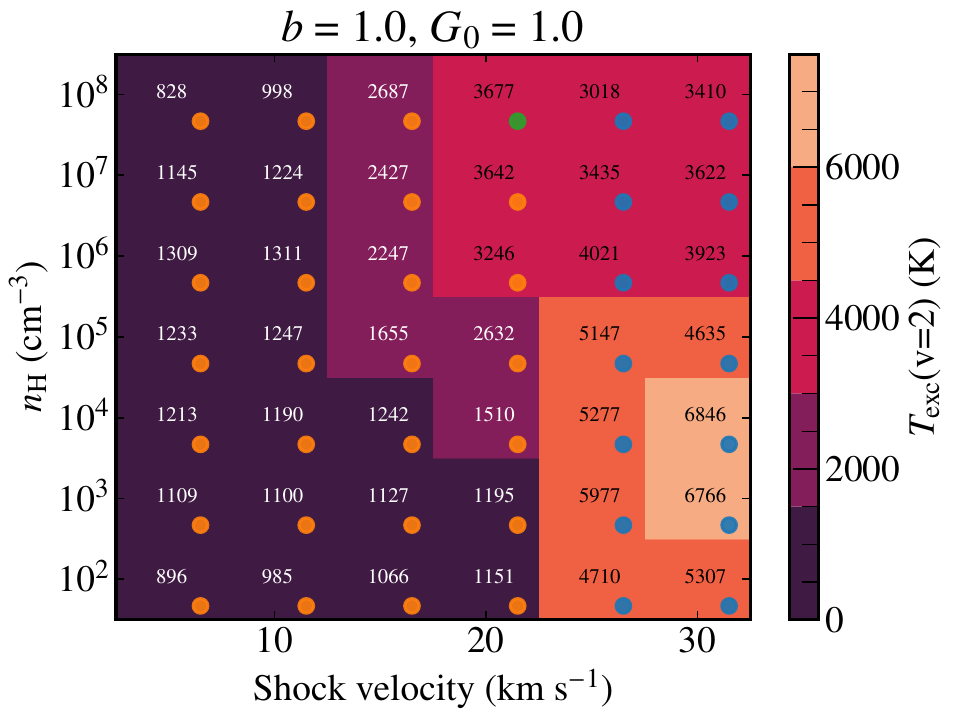}
\caption{Excitation temperature determined from the $\varv$ = 2, $J$ = 0 to 8 level populations for $b$ = 0.1 and 1, and $G_0$ = 0 and 1. \label{fig:tv2}}
\end{figure*}

Excitation temperatures have been extracted and calculated from a subset of the grid. Figures \ref{fig:tcold} and \ref{fig:twarm} show these temperatures calculated from the $\varv$ = 0, $J$ = 3 to 5 levels (S(1) to S(3)) and the $\varv$ = 0, $J$ = 6 to 11 levels (S(4) to S(9)) levels, respectively. The excitation temperatures are shown for $b$ = 0.1 and 1, and $G_0$ = 0 and 1. Figures \ref{fig:tv1} and \ref{fig:tv2} show excitation temperatures for the $\varv$ = 1, $J$ = 0--8 and $\varv$ = 2, $J$ = 0--8 vibrationally excited levels.

\subsection{Cosmic ray ionization rate}

\begin{figure*}
\centering
\includegraphics[width=0.48\textwidth]{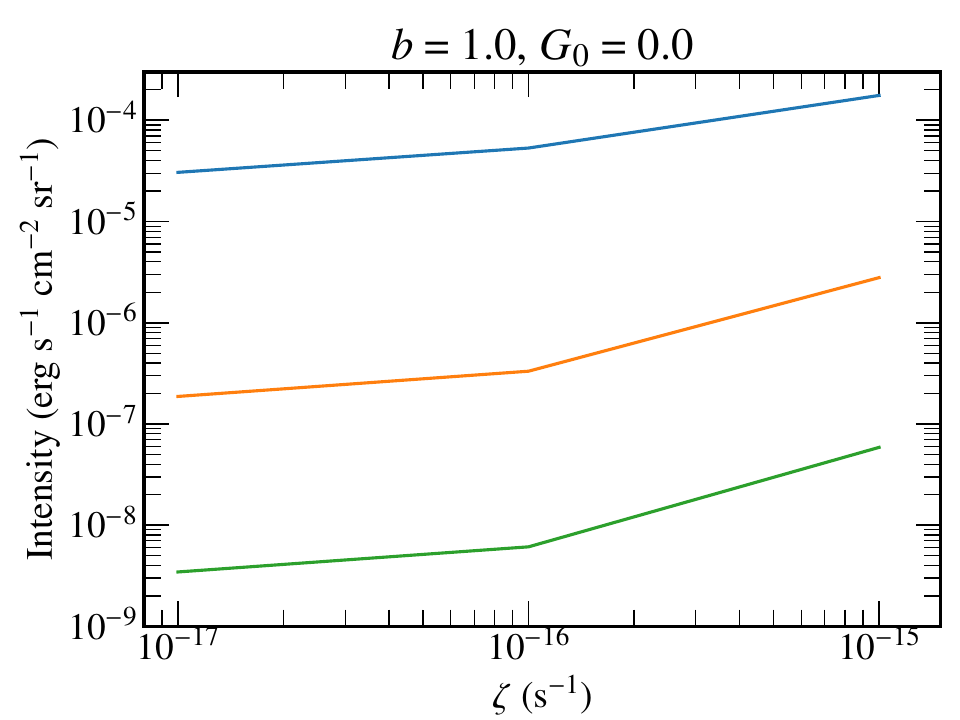}
\includegraphics[width=0.48\textwidth]{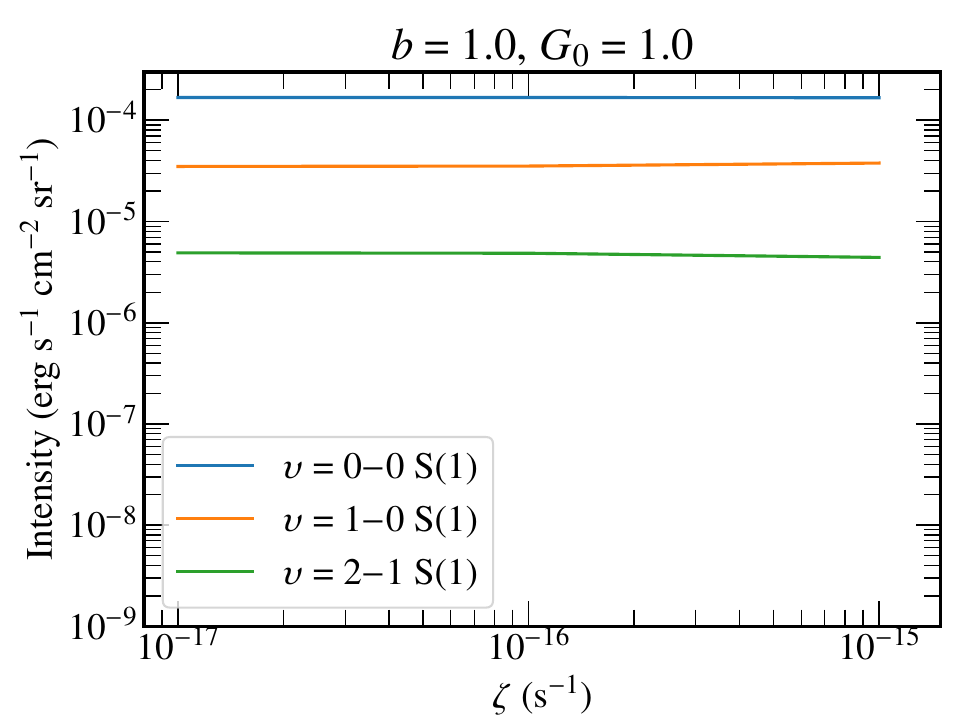}
\caption{Integrated intensity of select H$_2$ lines as a function of the cosmic-ray ionization rate, $\zeta_{\rm H2}$. The panel on the left is for $G_0$ = 0, and on the right for $G_0$ = 1. Both are for a C-type shock with shock velocity of 20 km s$^{-1}$, density of 10$^4$ cm$^{-3}$, and $b$ = 1.0. The PAH abundance is 10$^{-8}$. \label{fig:h2_zeta}}
\end{figure*}

In the model, cosmic rays may ionize H$_2$ and other species. When these species recombine, primarily H$_2$, secondary UV photons are emitted. Direct excitation by cosmic rays is not included. In this manner, cosmic rays serve as an additional source of both ionization and thus energy input. The expectation is that they will impact the H$_2$ emission to a similar degree as an external UV field. Their impact, however, is smaller than that of UV radiation. This is illustrated in Fig. \ref{fig:h2_zeta}, where the integrated line intensity of three representative lines are shown as a function of the cosmic ray ionization rate, $\zeta_{\rm H2}$, for Model B. In this case, the PAH abundance is set to 10$^{-8}$. For no external UV radiation, the integrated intensity increases by $\sim$ one order of magnitude when $\zeta_{\rm H2}$ increases by two orders of magnitude. For $G_0$ = 1, there is practically no change in intensity over the same range of $\zeta_{\rm H2}$, however, the vibrationally excited lines are significantly brighter than for the shocks without an external radiation field. 

\FloatBarrier

\section{Dominant cooling lines}\label{app:cool}

It is natural, when examining such a large grid, to identify the dominant H$_2$ cooling lines, that is, the H$_2$ lines that are most likely to be observed for a given set of input parameters. One way of identifying these lines for the entire grid, is to go through each model and tabulate the lines with integrated intensities that are greater than 25\% of the maximum intensity. This arbitrary cutoff is chosen from the perspective that if the strongest line is detected at 20$\sigma$, then these lines would also be detectable at the 5$\sigma$ level. Next, the lines are sorted according to which ones are present in the largest number of models, i.e., which are typically the dominant cooling lines in a global perspective. The lines that are present in at least 25\% of models are tabulated in Table \ref{tab:bright_lines}. 

Twenty-four lines are present in at least 25\% of models. The lines are either $\varv$ = 0--0 or 1--0 transitions; the higher-excited levels are clearly not sufficiently populated over the majority of the grid. Some of the lines in Table \ref{tab:bright_lines} are observable from the ground, for example, the often bright $\varv$ = 1--0 S(1) line at 2.12 $\mu$m, but the majority of the lines are not (17/24 lines). All lines are, however, observable with the JWST. Eighteen lines are observable with NIRSpec, while seven are observable with MIRI. At 5.06 $\mu$m, the $\varv$ = 0--0 S(8) line is observable with both instruments, and could serve as a cross-calibrator between the two instruments. 

\begin{table}[]
    \caption{Dominant cooling lines of H$_2$. \label{tab:bright_lines}}
    \centering
    \begin{tabular}{l c c}
\hline\hline
Line & Fraction\tablefootmark{a} & Wavelength ($\mu$m) \\ \hline
0-0 S(7) & 0.621 & 5.52 \\
1-0 Q(1)\tablefootmark{b} & 0.605 & 2.41 \\
0-0 S(5) & 0.596 & 6.92 \\
0-0 S(9) & 0.599 & 4.70 \\
1-0 O(3)\tablefootmark{b} & 0.586 & 2.81 \\
0-0 S(11) & 0.529 & 4.19 \\
1-0 S(1)\tablefootmark{c} & 0.493 & 2.12 \\
1-0 S(3)\tablefootmark{d} & 0.492 & 1.96 \\
1-0 S(5) & 0.491 & 1.84 \\
1-0 Q(3)\tablefootmark{c} & 0.451 & 2.43 \\
0-0 S(13) & 0.451 & 3.85 \\
0-0 S(3) & 0.443 & 9.67 \\
1-0 S(2) & 0.418 & 2.04 \\
1-0 S(7) & 0.370 & 1.75 \\
1-0 Q(5)\tablefootmark{d} & 0.367 & 2.46 \\
1-0 O(2) & 0.345 & 2.63 \\
1-0 O(5)\tablefootmark{c} & 0.327 & 3.24 \\
0-0 S(6) & 0.324 & 6.12 \\
0-0 S(4) & 0.319 & 8.03 \\
0-0 S(8) & 0.313 & 5.06 \\
1-0 S(4) & 0.305 & 1.89 \\
0-0 S(15) & 0.305 & 3.63 \\
0-0 S(10) & 0.261 & 4.41 \\
0-0 S(2) & 0.255 & 12.29 \\
\hline
    \end{tabular}
\tablefoot{
\tablefoottext{a}{For each model, the lines with integrated intensities $>$ 25\% of the maximum intensity are recorded. The lines which appear in more than 25\% of models are recorded here as fractions of the total number of models.}
\tablefoottext{b}{Lines share the same upper level, $\varv$ = 1, $J$ = 1.}
\tablefoottext{c}{Lines share the same upper level, $\varv$ = 1, $J$ = 3.}
\tablefoottext{d}{Lines share the same upper level, $\varv$ = 1, $J$ = 5.}
}
\end{table}

\end{document}